\documentclass[a4paper,12pt]{article}
\usepackage{graphicx,amsmath,amsfonts,amssymb}
\usepackage{psfrag}
\usepackage{epsf}
\usepackage{latexsym}

\newlength{\cqfd}
\def\R{\mathbb{R}}

\def\Z{\mathbb{Z}}
\def\sig{\sigma}

\newcommand{\ep}{\epsilon}

\newcommand{\fin}{\hspace*{\cqfd}\Box}

\newcommand{\ve}{\vspace{1ex}}
\newcommand{\vsp}[1]{\vspace{#1 ex}}

\newcommand{\proof}{\vspace{1ex}\noindent{\bf Proof.}\ }

\renewcommand{\ni}{\noindent}

\newtheorem{theorem}{Theorem}

\newtheorem{lemma}{Lemma}

\newtheorem{remark}{Remark}

\begin{document}

\title{{\bf Breathers in inhomogeneous nonlinear lattices~:
an analysis via centre manifold reduction.}}

\author{Guillaume James$^{\, a)}$\thanks{Corresponding author},
Bernardo S\'anchez-Rey$^{\, b)}$
and Jesus Cuevas$^{\, b)}$
\\
~\\
$\,^{a)}$ 
Institut de Math\'ematiques de Toulouse, UMR 5219,\\
D\'epartement de Math\'ematiques, INSA de Toulouse,\\ 
135 avenue de Rangueil,
31077 Toulouse Cedex 4, France.\\
{\small e-mail : Guillaume.James@insa-toulouse.fr}
~\\~\\
$\,^{b)}$
Grupo de F\'isica No Lineal, Universidad de Sevilla,\\ 
Departamento de F\'isica Aplicada I,
Escuela Universitaria Polit\'ecnica,\\
c/. Virgen de \'Africa 7,
41011-Sevilla, Espa\~{n}a.\\
{\small e-mail : bernardo@us.es, jcuevas@us.es}
}

\date{\today}

\maketitle
\begin{abstract}
We consider an infinite chain of particles
linearly coupled to their nearest neighbours
and subject to an anharmonic local potential.
The chain is assumed weakly inhomogeneous, i.e.
coupling constants, particle masses 
and on-site potentials can have small variations
along the chain. We look for small amplitude and
time-periodic solutions, and in particular spatially
localized ones (discrete breathers).
The problem is reformulated as a
nonautonomous recurrence in a space of time-periodic functions, where the dynamics
is considered along the discrete spatial coordinate.
Generalizing to nonautonomous maps a centre manifold 
theorem previously obtained for infinite-dimensional
autonomous maps \cite{james2}, 
we show that small amplitude oscillations
are determined by finite-dimensional nonautonomous mappings, whose dimension depends on the
solutions frequency. We consider the case of two-dimensional
reduced mappings, which occurs for frequencies close to
the edges of the phonon band
(computed for the unperturbed homogeneous chain). 
For an homogeneous chain,
the reduced map is autonomous and reversible, and
bifurcations of reversible homoclinics or heteroclinic solutions 
are found for appropriate parameter values. 
These orbits correspond respectively
to discrete breathers for the infinite chain, or ``dark'' breathers superposed on a 
spatially extended standing wave. Breather existence is shown in some cases for any
value of the coupling constant, which generalizes 
(for small amplitude solutions) an existence result obtained
by MacKay and Aubry at small coupling \cite{ma}.
For an inhomogeneous chain the study of the nonautonomous reduced map
is in general far more involved. Here this problem is considered
when the chain presents a finite number of defects.
For the principal part of the reduced recurrence, using the
assumption of weak inhomogeneity, we show that homoclinics
to $0$ exist when the image of the unstable manifold under
a linear transformation (depending on the sequence) intersects
the stable manifold. This provides a geometrical
understanding of tangent bifurcations of discrete
breathers commonly observed in classes of systems with impurities
as defect strengths are varied. The case of a mass
impurity is studied in detail, and our geometrical analysis is
successfully compared with direct numerical simulations.
\end{abstract}

\section{Introduction}

It is now well established that many nonlinear 
networks of interacting particles sustain 
time-periodic and spatially localized oscillations 
commonly denoted as {\it discrete breathers}.
In spatially periodic systems, breathers 
are also called {\em intrinsically localized modes} \cite{sievtak}
in distinction to Anderson modes of disordered linear systems 
\cite{anderson}.
The properties of discrete breathers have been
analyzed in an important number of numerical works 
(see the reviews \cite{revflach,vazquez,dlms}) and
their existence in periodic systems
has been proved analytically in different contexts, 
see \cite{ma,aubry2,sepmac,ar5,aubkadel,pankovb,flachb,james2}
and their references.
In the context of numerical simulations or experiments 
discrete breathers often denote a larger class of spatially localized
oscillations, such as metastable states, oscillations with a
certain degree of periodicity, or even chaotic oscillations
interacting with a noisy extended background \cite{iva,gersh}.
Nonlinear waves of this type are now actually detected in real 
materials \cite{sato,Swanson_ea,SchwarzEnSie,EdHamm,ura} and
also generated in artificial systems such as
Josephson junction arrays, micromechanical cantilever arrays 
and coupled optical waveguides (see references in \cite{revcam}).
They are thought to play a role in various physical processes
such as the formation of local fluctuational openings in the 
DNA molecule \cite{peyrard,peyrev}, which occurs in 
particular during thermal denaturation experiments.

Beyond spatially periodic systems, it is a fundamental 
and challenging problem to understand breather properties 
in {\it nonlinear and inhomogeneous} media, such
as non-periodic or disordered crystals, amorphous solids and
biological macromolecules.
For example the interplay between nonlinearity and disorder
can provide an alternative interpretation for slow
relaxation processes in glasses \cite{kaubry}.
In quasi-one-dimensional media, moving localized waves 
interacting with impurities \cite{kiselev,cimpur,fori},
extended defects \cite{ting}
or local bends of the lattice (see \cite{ck} and its
references) can remain trapped and release vibrational
energy at specific sites.

The modelling of thermal denaturation of DNA and the analysis of its local
fluctuational openings, also known as denaturation bubbles,
represents another problem where heterogeneity is important. 
In order to describe these phenomena, 
a nonlinear model at the scale of the DNA base pair
has been introduced by Peyrard and Bishop \cite{peyrard} 
and further improved by Dauxois et al \cite{dauxm}. The model
describes the stretching $x_n (t)$ of the H-bonds
between two bases, in the $n$th base pair along a
DNA molecule (a large value of $x_n$ corresponding to
a local opening). Each bond fluctuates in an effective
anharmonic potential $V$ and interacts with its 
nearest-neighbours. The model is described by an
Hamiltonian system, and can be coupled with a
thermostat to study the effect of thermal
noise in denaturation experiments. 
This model accurately describes the
thermal denaturation of real DNA segments 
provided their heterogeneity is taken into account
\cite{campa} (for example, the dissociation energy 
of AT and GC base pairs are different). The
Hamiltonian of the system reads
\begin{equation}
\label{pb}
H=\sum_{n=-\infty}^{+\infty}{\frac{m}{2}{\dot{x}_n}^2+V_{s_n}(x_n)
+\frac{k}{2}\, (1+\rho \, e^{-\beta (x_{n+1}+x_n)})\, (x_{n+1}-x_n)^2},
\end{equation}
where $V_{s_n}(x) = D_{s_n}(1 - e^{-a_{s_n}x})^2$ is a Morse potential
depending on the base pairs sequence $s_n \in \{ AT , GC \}$.
The case $\rho =0$ yields a particular case of
{\it Klein-Gordon lattice}, i.e. the model consists in
a chain of anharmonic oscillators with harmonic nearest-neighbours coupling.
For parameters corresponding to real DNA sequences,
Langevin molecular dynamics of (\ref{pb}) have shown that some locations of discrete breathers 
heavily depend on the sequence and seem to coincide with functional 
sites in DNA \cite{kalo}, but at the present time this conclusion remains
controversial \cite{titus}.

From a mathematical point of view, Albanese and Fr\"ohlich have proved the existence of breathers
for a class of random Hamiltonian
systems describing an infinite array of coupled anharmonic oscillators \cite{alba}
(see also the earlier work \cite{fquasi} of Fr\"ohlich et al concerning
quasiperiodic localized oscillations). These
breather families can be parametrized by the solutions frequencies, which
belong to fat Cantor sets (i.e. with nonzero Lebesgue
measure) of asymptotically full relative measure in the limit of zero amplitude. 
These solutions are nonlinear ``continuations'' of a given Anderson mode from the
limit of zero amplitude, and the gaps in their frequency values originate from
a dense set of resonances present in the system.
For disordered Klein-Gordon lattices,
complementary numerical results on the continuation of breathers
with respect to frequency or the transition between breathers
to Anderson modes are available in \cite{kaubry,archi}.

In addition, the existence of breathers in
inhomogeneous Klein-Gordon lattices (with disordered on-site potentials)
has been proved by Sepulchre and MacKay \cite{sepmac2,sepmac}
for small coupling $k$. The proof is based on the continuation method previously introduced 
by MacKay and Aubry \cite{ma} for an homogeneous chain
(method considerably generalized in \cite{sepmac}).
For $k=0$ the system reduces to an array of uncoupled non-identical
anharmonic oscillators, and the simplest type of discrete breather consists of a single particle oscillating
while the others are at rest. Under a nonresonance condition \cite{sepmac2,sepmac},
this solution can be continued to small values of $k$ (in most cases at fixed frequency)
using the implicit function theorem, yielding a spatially localized solution.

In this paper we provide complementary mathematical tools for studying time-periodic
oscillations (not necessarily spatially localized)
in inhomogeneous infinite lattices. The theory is developed
in a very general framework, and applied to breather {\em bifurcations} in
inhomogeneous Klein-Gordon lattices as lattice parameters and
breather frequencies are varied. We start from a general
Klein-Gordon lattice with Hamiltonian
\begin{equation}
\label{exkg}
H=\sum_{n=-\infty}^{+\infty}{\frac{M_n}{2}{\dot{x}_n}^2+D_n V(A_n x_n)+\frac{K_n}{2}(x_{n+1}-x_n)^2}
\end{equation}
(case $\rho =0$ of (\ref{pb}) with more general inhomogeneities). 
The potential $V$ is assumed sufficiently smooth
in a neighbourhood of $0$ with $V^{\prime}(0)=0$, $V^{\prime\prime}(0)=1$.
The general theory is a priori valid
for small inhomogeneities and small amplitude oscillations. In particular,
in our application to system (\ref{exkg}) we assume
$M_n$, $D_n$, $A_n$, $K_n$ to be close (uniformly in $n$) to positive constants.  
However, considering an example of Klein-Gordon lattice with a mass defect, 
we check using numerical computations that our tools remain applicable up 
to strongly nonlinear regimes, and sometimes for a large inhomogeneity.

Our analysis is based on a centre manifold reduction and the concept of
{\em spatial dynamics}. This concept was introduced by K. Kirchg\"assner \cite{klaus}
for nonlinear elliptic PDE in infinite strips, considered as an (ill-posed)
evolution problem in the unbounded space coordinate, and locally 
reduced to a finite-dimensional ODE on an invariant centre manifold.
This idea was transposed to the context of travelling waves in homogeneous
infinite oscillator chains by Iooss and Kirchg\"assner \cite{ioossK},
and centre manifold reduction has been subsequently applied to the analysis of travelling waves
and pulsating travelling waves in different one-dimensional homogeneous lattices
\cite{iooss,jamessire,ioossjames,sire,pelr,ioossp}. Indeed,
looking for travelling waves in an oscillator chain yields an advance-delay differential equation
(a system of such equations in the case of pulsating travelling waves), which can be
reformulated as an infinite-dimensional evolution problem in the moving frame coordinate,
and locally reduced to a finite-dimensional ODE under appropriate spectral conditions.
 
In reference \cite{james}, one of us has proved the existence of breathers in
Fermi-Pasta-Ulam (FPU) lattices using a similar technique in a discrete context.
The dynamical equations for time-periodic solutions were reformulated as an
infinite-dimensional recurrence relation in a space of time-periodic functions,
and then locally reduced to a finite-dimensional mapping on a centre manifold, where
breathers corresponded to homoclinic orbits to $0$. 
A general centre manifold theorem for infinite-dimensional maps
with unbounded linearized operator has been proved subsequently \cite{james2} and
has been used to analyze breather bifurcations in diatomic FPU lattices 
\cite{jamesnoble,jk} and spin lattices \cite{noblenon}.
More generally, the dynamical equations of many one-dimensional lattices
can be reformulated as infinite-dimensional maps in loop spaces
as one looks for small amplitude time-periodic oscillations
(\cite{james2}, section 6.1).

As shown in the present paper, the centre manifold reduction theorem
readily applies to homogeneous 
Klein-Gordon lattices, where 
$M_n =m$, $D_n =d$, $A_n =a$, $K_n =k$ in (\ref{exkg})
and $m,d,a,k >0$. This reduction result rigorously justifies
(in the weakly nonlinear regime) a formal one-Fourier mode
approximation previously introduced in reference \cite{bountis}.
The equations of motion read
\begin{equation}
\label{KGh}
m\, \frac{d^{2}x_{n}}{dt^{2}}+d a \, V^{\prime}(a\, x_{n})=
k\, (x_{n+1}-2x_{n} + x_{n-1}),\ \ \ n\in \Z .
\end{equation}
Looking for time-periodic solutions (with frequency $\omega$)
and setting $x_n (t)=\tilde{x}_n (\omega \, t)$, (\ref{KGh})
can be formulated as an (ill-posed) recurrence relation 
$(\tilde{x}_{n+1},\tilde{x}_n)=F(\tilde{x}_n , \tilde{x}_{n-1})$
in a space of $2\pi$-periodic functions.
Using the theorem of reference \cite{james2}, 
one can locally reduce the problem
to a finite-dimensional mapping on a centre manifold
whose dimension depends on the frequency $\omega$. 
More precisely, equation (\ref{KGh}) linearized
at $x_n =0$ admits solutions in the form of
linear waves (phonons) with 
$x_n (t)=A\, \cos{(q n-\omega_q t)}$,
whose frequency satisfies the dispersion relation 
\begin{equation}
\label{dispersns}
m\omega_q^2 = a^2 d+2k (1-\cos{q}) .
\end{equation}
The frequencies $\omega_q$ lie in 
a band $[\omega_{min},\omega_{max}]$ with $\omega_{min} >0$.
In the nonlinear case,
the dimension of the centre manifold depends on how many multiples
of $\omega$ belong to (or are close to) the phonon band.
When $\omega \approx \omega_{max}$ or 
$\omega \approx \omega_{min}$ (with no additional resonance), the
centre manifold is two-dimensional if solutions are searched even in time,
which reduces (\ref{KGh}) locally to a two-dimensional 
reversible mapping on the centre manifold.
For appropriate parameter values, this
map admits small amplitude homoclinic solutions to $0$ corresponding to
breather solutions of (\ref{KGh}).  
Breather solutions in this system have been proved
to exist by MacKay and Aubry \cite{ma} for small values of the
coupling parameter $k$. Known regions of breather existence
are considerably extended here, since we prove the existence 
of small amplitude breathers for arbitrary values of $k$ in some cases
and for frequencies close to the phonon band edges (see theorem \ref{existencebr} p.\pageref{existencebr}).
In addition
we prove the existence of ``dark breather'' solutions, which converge 
towards a nonlinear standing wave as $n\rightarrow \pm\infty$ and have a 
much smaller amplitude at the centre of the chain. These solutions
correspond to heteroclinic orbits of the reduced two-dimensional map.

Furthermore, we extend this analysis to
the case when small lattice inhomogeneities are present. 
The dynamical equations of the inhomogeneous system (\ref{exkg}) take the form
\begin{equation}
\label{KG}
M_n\, \frac{d^{2}x_{n}}{dt^{2}}+D_n A_n\, V^{\prime}(A_n\, x_{n})=
K_n\, (x_{n+1}-x_{n}) -K_{n-1}(x_n-x_{n-1}),\ \ \ n\in \Z ,
\end{equation}
and time-periodic solutions can be obtained as orbits of a nonautonomous 
map $({x}_{n+1},{x}_n)=F(\lambda_n ,{x}_n , {x}_{n-1})$, where the
nonconstant lattice parameters are embedded in a
multicomponent parameter $\lambda_n$.
Fixing $M_n=m+m_n$, $D_n=d+d_n$, $A_n=a+a_n$, $K_n=k+k_n$, we consider
the case when constant lattice parameters $m,d,a,k >0$ are perturbed by uniformly small
sequences $(m_n)_{n\in \Z}$, $(d_n)_{n\in \Z}$, $(a_n)_{n\in \Z}$, $(k_n)_{n\in \Z}$.
We prove (see theorem \ref{reductionnh} p.\pageref{reductionnh})
that small amplitude time-periodic solutions with frequencies
close to $\omega_{min}$ or $\omega_{max}$
are determined by a two-dimensional nonautonomous map.
Moreover, we generalize this reduction result in the case
when several multiples of $\omega$ are close to
the band $[\omega_{min},\omega_{max}]$, which yields a
higher-dimensional reduced problem
(see theorem \ref{reductionnhg} p.\pageref{reductionnhg}).

In fact we prove this type of reduction result in a very general framework,
for infinite-dimensional mappings with small nonautonomous perturbations,
considered in a neighbourhood of a non-hyperbolic fixed point, or close
to a bifurcation. The linear autonomous part of the map must 
satisfy a property of spectral separation
(see theorem \ref{rvc} p.\pageref{rvc}), but a large number of
one-dimensional lattices with finite-range coupling
fall within this category. 
We obtain a direct proof of the reduction result
by observing that 
any nonautonomous mapping ${u}_{n+1}=F(\lambda_n ,{u}_n )$
can be seen as a projection of an extended
autonomous mapping, to which the centre manifold
theorem of reference \cite{james2} can be applied under appropriate assumptions. 
The centre manifold of the extended map is infinite-dimensional,
but this case is also covered in \cite{james2}.
The reduced nonautonomous mapping for the original system can
be interpreted as a projection on a finite-dimensional subspace
of the extended autonomous mapping restricted to
the invariant centre manifold.

We use this reduction result to analyse the case when equation
(\ref{KG}) presents a mass defect at a single site, 
all other lattice parameters being independent of $n$.
In that case, the linearized problem admits a spatially
localized mode (usually denoted as an impurity mode), and a
nonlinear continuation of this mode can be computed
\cite{fori}, corresponding to a Lyapunov family of
periodic orbits. Klein-Gordon systems with a coupling defect or a
harmonic impurity in the on-site potential share
similar characteristics \cite{cimpur}, as well as
nonlinear lattices with a different type of nonlinearity \cite{kiv}. 
In addition to this simple localization phenomenon, single impurities can have
more complex effects in a nonlinear system. Indeed
it is a common feature to observe a complex sequence
of tangent bifurcations between (deformations of)
site-centered and bond-centered breathers in some
neighbourhood of the defect 
as the strength of an impurity is varied \cite{ck,kiv}. Using numerical
computations we show some examples of such bifurcations 
in the present paper, as one varies the strength of a mass 
defect in system (\ref{KG}). From a physical point of view
it is quite important to understand how a local change in the 
lattice parameters modifies the set of spatially localized solutions.
For example, this could contribute to 
explain how a mutation at a specific location
of an homogeneous sequence of (artificial) DNA would modify the
structure of fluctuational openings \cite{kalo}. 

This paper provides a qualitative explanation of such tangent
bifurcations, which reveals also very precise
quantitatively when compared with numerical simulations
of the Klein-Gordon model.
According to the previously described reduction theorem,
for a small mass defect of size $\epsilon$, small amplitude breather solutions of
(\ref{KG}) with frequencies below (and close to) 
$\omega_{min}$ are described by a two-dimensional nonautonomous mapping
$v_{n+1}=f(v_n ,\omega ) + \epsilon\, g(n, v_n ,\omega ,\epsilon )$.
Here we only study the principal part of the reduced mapping
as $(v_n ,\omega ,\epsilon )\approx (0,\omega_{min},0)$.
We show that this truncated reduced map admits an homoclinic
orbit to $0$ (corresponding to an approximate breather solution
for the oscillator chain) if, for $\epsilon =0$,
the image of the unstable manifold of $0$ under a certain linear shear 
intersects its stable manifold. The linear shear is $O(\epsilon )$-close to
the identity.
When the on-site potential is soft, these manifolds have very
complicated windings characteristic of homoclinic chaos, hence
the set of their intersections changes in a complex way as the
linear shear varies, or equivalently as one varies
the mass defect. This phenomenon explains the existence of
the above mentioned tangent bifurcations, at least for small defect sizes, 
and for small amplitude breathers with frequencies close to the phonon band.
In addition, we show (by comparison with direct numerical simulations
of the Klein-Gordon model) that this picture remains valid quite far from 
the weakly nonlinear regime. 
Let us note that, to obtain an exact solution of (\ref{KG}) from an orbit of
the truncated map, it would be necessary to control the effect of
higher order terms (with respect to $v_n$, $\omega - \omega_{min}$,
$\epsilon$) present in the full reduced mapping and prove the
persistence of this solution. This problem is not examined here
from the analytical side, but we compare instead numerically computed
solutions of (\ref{KG}) with approximate solutions deduced from
the truncated map. The very good agreement leads us to conjecture
that most of the homoclinic bifurcations existing for the truncated problem
persist for the full reduced system.

Lastly we consider the more general case when system (\ref{KG}) admits a finite
number of defects, i.e. perturbations $m_n$, $d_n$, $a_n$, $k_n$ have a
compact support (as above these perturbations are assumed to be small, of order $\epsilon$). 
We show that the approach developped for a single impurity can be
extended to this case (see lemma \ref{lemint3} p.\pageref{lemint3}),
where the linear shear is replaced by a more general {\it linear} 
near-identity transformation $A_\epsilon$.
The linear transformation $A_\epsilon$ provides a useful tool
for studying breather bifurcations in Klein-Gordon lattices 
with a finite number of impurities, 
as for the single impurity case that we
have analyzed in detail.
By computing the principal part of $A_\epsilon$
as $\epsilon$ is small and frequencies are
close to $\omega_{min}$, 
we show that the effect of the parameter sequence 
on the set of small amplitude breather solutions should mainly depend on
weighted averages of the defects values.

The outline of the paper is as follows. 
Section \ref{rrsi} presents the centre manifold reduction
theory for time-periodic oscillations in
weakly inhomogeneous nonlinear lattices.
We treat the case of Klein-Gordon lattices in detail
in sections \ref{formul} and \ref{redkg}, and formulate
the reduction theory in a much more
general setting in section \ref{redgeneral}. 
Section \ref{epshl} concerns spatially homogeneous Klein-Gordon
lattices. Existence theorems for small amplitude
breather and dark breather solutions are deduced from the
dynamics of two-dimensional reversible maps on 
invariant centre manifolds. The case of weakly
inhomogeneous Klein-Gordon chains
is considered in section \ref{nfail}, where the truncated
reduced map is analyzed for a finite number of defects. 
A geometrical condition for the existence of homoclinic
orbits to $0$ is derived in section \ref{many}, and
some homoclinic
bifurcations are studied in detail in 
section \ref{single} for a single mass defect.
In the latter case, breather solutions are numerically computed 
in section \ref{numerical} and the results are successfully
compared with our analytical findings. 

\section{\label{rrsi}Reduction result for small inhomogeneities}

In this section we consider system (\ref{KG}) in the limit of small
inhomogeneities. We show that all small amplitude time-periodic solutions
are determined by a finite-dimensional nonautonomous map, whose dimension
depends on the frequency domain under consideration. For this purpose we reformulate
(\ref{KG}) as a map in a loop space, perturbed by a small nonautonomous
term (section \ref{formul}). Then we prove 
in section \ref{redgeneral} a
general centre manifold reduction theorem for infinite-dimensional
maps with small 
nonautonomous perturbations. This result is based on the centre manifold theorem
proved in reference \cite{james2} for autonomous systems.
Our general result is applied to the inhomogeneous
Klein-Gordon lattice, which yields the above mentioned reduction
result (section \ref{redkg}).

\subsection{\label{formul}The Klein-Gordon system as a map in a loop space}

We set $x_n (t)=y_n (\omega (k/m)^{1/2} t)$ 
in equation (\ref{KG}),
where
$y_n$ is $2\pi$-periodic in $t$ (hence
$x_n$ is time-periodic with frequency $ \omega (k/m)^{1/2}$).
The constant $a>0$ being fixed, we also define
$ \tilde{V}(x) = a^{-2}\, V(ax)$. Equation (\ref{KG}) becomes
\begin{equation}
\label{dfpu}
\omega^2 (1+\ep_n )\, \frac{d^{2}y_{n}}{dt^{2}}
+\Omega^2 (1+\eta_n )\, \tilde{V}^{\prime}((1+\gamma_n )y_{n})=
y_{n+1}-y_{n} -(1+\kappa_n)\, (y_n -y_{n-1}),\ \ \ n\in \Z
\end{equation}
where $\Omega^2={a^2 d}/{k}$ and 
$1+\ep_n =(1+\frac{m_n}{m})/(1+\frac{k_n}{k})$,
$1+\eta_n =(1+\frac{d_n}{d})(1+\frac{a_n}{a})/(1+\frac{k_n}{k})$,
$\gamma_n = \frac{a_n}{a}$, 
$1+\kappa_n =(1+\frac{k_{n-1}}{k}) /(1+\frac{k_n}{k})$.
The sequences
$(\ep_n)_{n\in \Z}$, $(\eta_n)_{n\in \Z}$, $(\gamma_n)_{n\in \Z}$
$(\kappa_n)_{n\in \Z}$ will be assumed sufficiently
small in $\ell_\infty(\Z )$, where
$\ell_\infty (\Z )$ is the classical Banach
space of bounded sequences on $\Z$, equiped with
the supremum norm.
To simplify the notations, we shall drop the tilde in the sequel
when referring to the renormalized potential $\tilde{V}$.
Moreover we shall use the shorter notations $\{\ep \} $
when referring to sequences $(\ep_n)_{n\in \Z}$.

\ve

To analyze system (\ref{dfpu}) we use the same approach as in
reference \cite{james2} for spatially homogeneous systems.
We reformulate (\ref{dfpu}) as a (nonautonomous) recurrence
relation in a space of $2\pi$-periodic functions of $t$,
and locally reduce the (spatial) dynamics to one on a finite-dimensional
centre manifold. 
We restrict our
attention to the case when $y_{n}$ 
is even in $t$ in order to deal with lower-dimensional problems.
More precisely, we assume $y_{n}\in H^{2}_{\#}$ for all
$n\in\Z$, where
$H^{n}_{\#}
=\{ \, y\in H^{n}_{per}(0,2\pi ),\ y\mbox{ is even}\, \}$
and $H^{n}_{per}(0,2\pi )$
denotes the classical Sobolev space of $2\pi$-periodic
functions
($H^{0}_{per}(0,2\pi )=L^{2}_{per}(0,2\pi )$).

\ve

Since our analysis concerns small amplitude solutions and
small inhomogeneities, the first step consists in studying the
linearized system at $y_n=0$ when 
$\ep_n , \eta_n ,\gamma_n , \kappa_n$ are fixed equal to $0$.
In that case equation (\ref{dfpu}) yields
\begin{equation}
\label{dfpulin}
\omega^2 \, \frac{d^{2}y_{n}}{dt^{2}}
+\Omega^2 y_{n}=
y_{n+1}-2 y_{n} + y_{n-1},\ \ \ n\in \Z .
\end{equation}
Now we rewrite the problem as an infinite-dimensional
linear mapping.
For this purpose we
introduce $Y_{n}=(y_{n-1},y_{n})\in D$,
where $D=H^{2}_{\#}\times H^{2}_{\#}$.
Equation (\ref{dfpulin}) can be written
\begin{equation}
\label{evol1}
Y_{n+1}=A_{\omega}\, Y_{n},
\ \ \ n\in \Z ,
\end{equation}
where
\begin{equation}
\label{defia}
A_{\omega}(z,y)=
\Big( \, y\, ,\,   
\omega^2 \, \frac{d^{2}y}{dt^{2}}
+(\Omega^2 +2) y-z
\, \Big) 
\end{equation}
and equation (\ref{evol1}) holds in
$X=H^{2}_{\#}\times H^{0}_{\#}$.
The operator 
$A_\omega :\, D\subset X\rightarrow X$
is unbounded in $X$ (of domain $D$) and closed
(we omit the additional parameter $\Omega $ in the notation $A_\omega$).

The spectrum of $A_{\omega}$
consists in essential spectrum at the origin and
an infinite number of eigenvalues 
$\sigma_{p}, \sigma_{p}^{-1}$ ($p\geq 0$)
depending on $\omega$, $\Omega$, and
satisfying the dispersion relation
 \begin{equation}
\label{dispers}
\sigma^{2}+(\omega^{2} p^{2}-\Omega^2 -2)\sigma +1=0
\end{equation}
(it follows that $\sigma_p$ is either real or has modulus one).
Equation (\ref{dispers}) is directly obtained by setting
$y_n = \sigma^n\, \cos{(pt)}$ in equation (\ref{dfpulin}).
The invariance $\sigma\rightarrow\sigma^{-1}$ in 
(\ref{dispers}) originates from the
invariance $n\rightarrow -n$ in (\ref{dfpulin}).
In the sequel
we shall note $\sig_{p}$ the solution of (\ref{dispers}) satisfying
$|\sig_{p}|\geq 1$ and $\mbox{Im}\, \sig_{p}\leq 0$.
Clearly $\sig_{p}$ is real negative
for $p$ large enough
and $\lim_{p\rightarrow +\infty}\sig_p = -\infty$. Moreover
$\sig_{p}^{-1}$ accumulates at $\sigma =0$ as $p\rightarrow +\infty$.
It follows that the number of eigenvalues of $A_{\omega}$
on the unit circle is finite for any value of the
parameters $\omega ,\Omega$.

In addition, the eigenvalues $\sigma_{p}, \sigma_{p}^{-1}$
defined by (\ref{dispers}) lie on the unit circle
when $\Omega \leq \omega p \leq (4+\Omega^2)^{1/2}$.
This property as a simple interpretation.
Multiplying (\ref{dispers}) by $\sigma^{-1}$,
setting $\sigma =e^{iq}$ and 
$\omega_q = \omega \, p \, (k/m)^{1/2}$,
one finds the usual dispersion relation (\ref{dispersns}).
Consequently, if $\omega \, p \, (k/m)^{1/2}$ lies
inside the phonon band $[\omega_{min},\omega_{max}]$
for some $p \in \mathbb{N}$, then
$A_{\omega}$ admits a pair of eigenvalues $e^{\pm iq}$
on the unit circle determined by the dispersion relation
(\ref{dispersns}). This condition on $\omega$ is equivalent
to prescribing $\Omega \leq \omega p \leq (4+\Omega^2)^{1/2}$.

\vspace{1ex}

Now let us describe the spectrum of $A_{\omega}$ 
near the unit circle
when $\Omega >0$ is fixed and $\omega$ is varied.
As we shall see, the number of eigenvalues of $A_{\omega}$
on the unit circle changes as $\omega$ crosses
an infinite sequence of decreasing critical values
$\omega_1 > \omega_2  >\ldots >0$. 
Small amplitude solutions of the nonlinear system bifurcating from
$y_n=0$ will be found
near these critical frequencies. 

We begin by studying the evolution of each pair of eigenvalues
$\sig_{p}$, $\sig_{p}^{-1}$ as $\omega$ varies.
Firstly, one can easily check that $\sigma_{0}, \sigma_{0}^{-1}$
are independent of $\omega$, real positive and lie strictly
off the unit circle.

Secondly we consider the case $p\geq 1$.
For $\omega > \sqrt{4+\Omega^2}/p$, $\sigma_{p}, \sigma_{p}^{-1}$
are real negative and lie strictly off the unit circle.
When $\omega$ decreases, they approach the unit
circle and
one has $\sigma_{p}= \sigma_{p}^{-1}=-1$ for 
$\omega = \sqrt{4+\Omega^2}/p$
(this corresponds to a frequency $\omega_q$ at the top of
the phonon band, for a wavenumber $q=\pi$).
At this critical parameter value,
$\sigma_{p}=-1$ is a
double non semi-simple eigenvalue of $A_\omega$.
For $\Omega /p<\omega < \sqrt{4+\Omega^2}/p$, $\sigma_{p}, \sigma_{p}^{-1}$
lie on the unit circle, and approach $+1$ as
$\omega$ decreases. 
One has $\sigma_{p}= \sigma_{p}^{-1}=1$ for 
$\omega = \Omega /p$, and then $+1$ is a
double non semi-simple eigenvalue of $A_\omega$
(this corresponds to a frequency $\omega_q$ at the bottom of
the phonon band, for a wavenumber $q=0$).
For $\omega < \Omega /p$, $\sigma_{p}, \sigma_{p}^{-1}$
are real positive and lie strictly off the unit circle.

Now let us qualitatively describe the evolution of the whole spectrum
of $A_\omega$. 
When $\omega  > \sqrt{4+\Omega^2}$ the spectrum of $A_\omega$
lies strictly off the unit circle
(both inside and outside).
When $\omega$ decreases, the eigenvalues $\sigma_{p}$ approach the unit
circle for all $p\geq 1$. As the first critical value 
$\omega_1  =\sqrt{4+\Omega^2}$ is reached,
the eigenvalues
$\sigma_{1},\sigma_{1}^{-1}$ collide 
and yield a double (non semi-simple) eigenvalue $\sigma_1 =-1$, while
the remaining part of the spectrum is hyperbolic.
When $\omega$ is further decreased, two different situations
occur depending on the value of $\Omega$. 

\begin{figure}[h]
\psfrag{legend1}[0.9]{{\small Case $\omega > \omega_1$}}
\psfrag{legend2}[0.9]{{\small Case $\omega_2 < \omega < \omega_1$}}
\psfrag{legend3}[0.9]{{\small Case $\omega_3 < \omega < \omega_2$}}
\begin{center}
\includegraphics[scale=0.4]{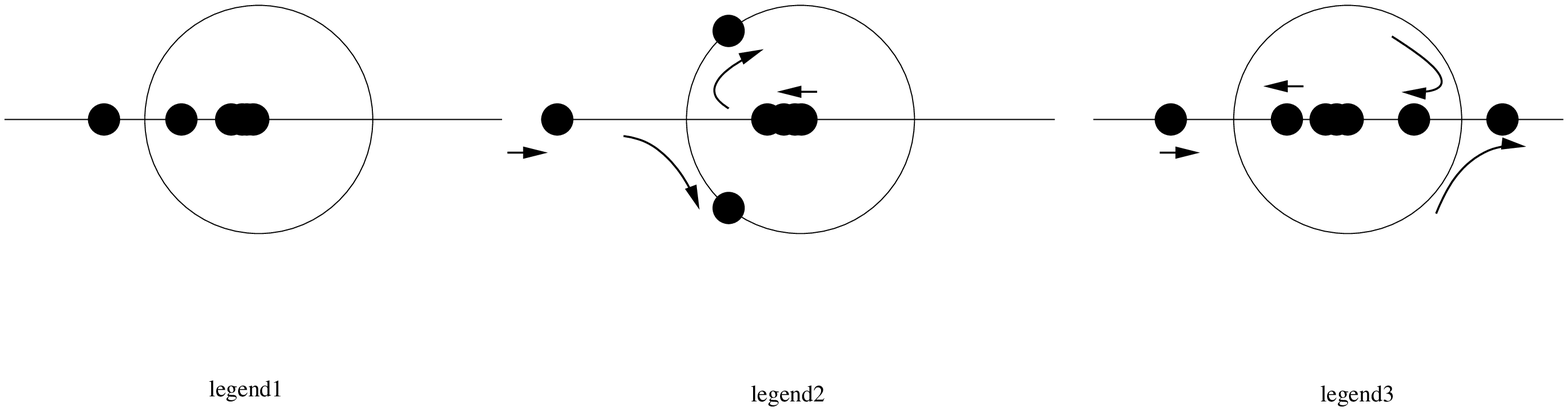}
\end{center}
\caption{\label{bifurc} 
Spectrum of $A_\omega$ near the unit circle as $\omega$ is varied, in the case
$\Omega > 2/\sqrt{3}$. The unbounded part of the spectrum on the negative real
axis is not shown. The arrows indicate how the eigenvalues have moved
from their positions in the previous graph, after $\omega$ has been decreased.} 
\end{figure}

For $\Omega > 2/\sqrt{3}$, $\sigma_{1},\sigma_{1}^{-1}$
are the only eigenvalues on the unit circle for
$\Omega \leq \omega \leq \sqrt{4+\Omega^2}$.
One has $\sigma_{1}= \sigma_{1}^{-1}=1$
at the second critical value
$\omega_2  = \Omega$.
When $\omega$ is further decreased,
$\sigma_{1}, \sigma_{1}^{-1}$
are real positive and lie strictly off the unit circle.
One has $\sigma_{2}= \sigma_{2}^{-1}=-1$ at
the third critical value 
$\omega_3  = \sqrt{4+\Omega^2}/2 < \Omega$.
The situation is sketched in figure \ref{bifurc}.

\begin{figure}[h]
\psfrag{legend1}[0.9]{{\small Case $\omega > \omega_1$}}
\psfrag{legend2}[0.9]{{\small Case $\omega_2 < \omega < \omega_1$}}
\psfrag{legend3}[0.9]{{\small Case $\omega_3 < \omega < \omega_2$}}
\begin{center}
\includegraphics[scale=0.4]{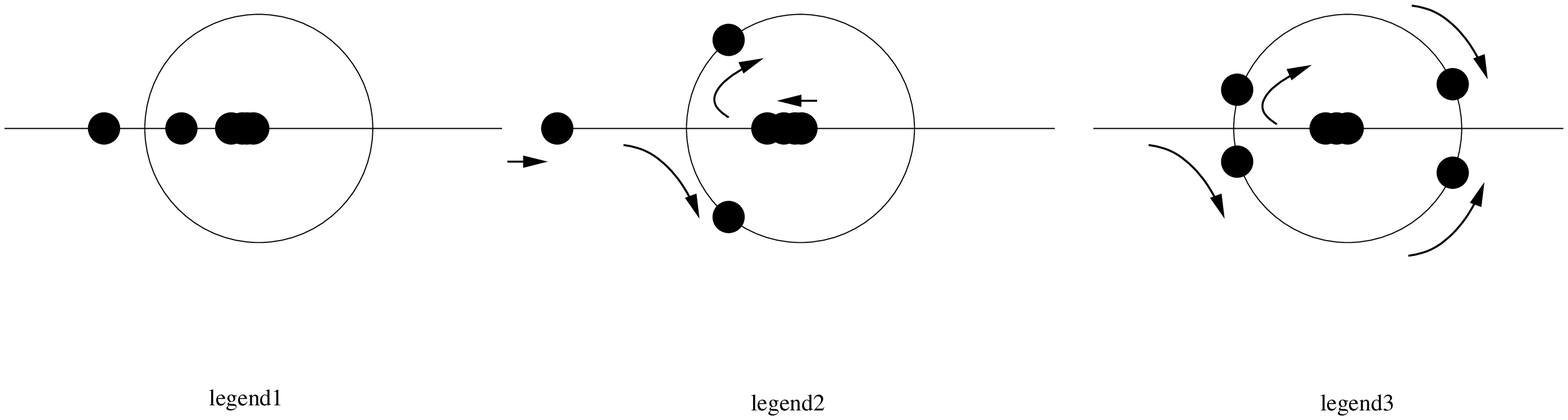}
\end{center}
\caption{\label{bifurc2} 
Spectrum of $A_\omega$ near the unit circle as $\omega$ is varied, in the case
$\Omega < 2/\sqrt{3}$. The unbounded part of the spectrum on the negative real
axis is not shown. The arrows indicate how the eigenvalues have moved
from their positions in the previous graph, after $\omega$ has been decreased.} 
\end{figure}

The case $\Omega < 2/\sqrt{3}$ is different, since
$\sigma_{1},\sigma_{1}^{-1}$
are the only eigenvalues on the unit circle in
the smaller frequency range
$\sqrt{4+\Omega^2}/2 < \omega \leq \sqrt{4+\Omega^2}$.
Indeed one has
$\sigma_{2}= \sigma_{2}^{-1}=-1$ at
second critical value
$\omega_2  = \sqrt{4+\Omega^2}/2 > \Omega$.
For $\omega < \sqrt{4+\Omega^2}/2$ and 
$\omega \approx \sqrt{4+\Omega^2}/2$
the spectrum of $A_\omega$ on the unit circle consists
in two pairs of simple eigenvalues 
$\sigma_{1},\sigma_{1}^{-1},\sigma_{2},\sigma_{2}^{-1}$.
In the interval
$\Omega < \omega < \sqrt{4+\Omega^2}/2$ other eigenvalues
may collide at $-1$ depending on the value of $\Omega$.
The situation is sketched in figure \ref{bifurc2}.

In what follows we restrict our attention to
the neighbourhood of critical frequencies
$\omega \approx \omega_2 $ with $\Omega > 2/\sqrt{3}$,
and $\omega \approx \omega_1 $.
This leads us to consider the small parameter $\mu$
defined by $\omega^2=\omega_i^2+\mu$.
As $\omega$ equals one of the critical frequencies $\omega_1 ,\omega_2$, 
the spectrum of $A_\omega$ on the unit circle only consists in 
a double eigenvalue $-1$ or $+1$. 
For $\omega \approx \omega_1$ or $\omega_2$,
the above spectral analysis shows that the fixed point $Y=0$ of
(\ref{evol1}) is hyperbolic when $\omega >\omega_1 $
or $\omega <\omega_2 $.
In this case, when nonlinear effects will be taken into account, we shall see that
the stable and unstable manifolds $W^{s}(0)$, $W^{u}(0)$ may
intersect depending on the local properties of the anharmonic
potential $V$,
leading to the existence of
homoclinic orbits to $Y=0$.

\ve

System (\ref{dfpu}) will be analyzed in the limit of small amplitude
solutions and for small parameters $\mu$, 
$\{\ep \} ,\{\eta  \},\{\gamma \} ,\{\kappa \}$.
The parameter space will be denoted as
$E=\R\times {(\ell_\infty (\Z ))}^4$.
All parameters are embedded in a multicomponent parameter
$\{ \lambda \} =(\mu ,\{\ep \} ,\{\eta  \},\{\gamma \} ,\{\kappa \})\in E$.
In addition we denote by
$\tau_n$ the index shift in $\ell_\infty (\Z )$, i.e.
$\{ \tau_n\, \{\ep \} \}_k =\epsilon_{n+k}$.

Equation (\ref{dfpu}) can be rewritten
in the form of a nonautonomous mapping in a function space.
More precisely we have
\begin{equation}
\label{nl}
Y_{n+1}=L\, Y_n+N(Y_{n},\lambda_n ), \ \ \ 
n\in \mathbb{Z},
\end{equation}
where 
$Y_{n}=(y_{n-1},y_{n}) =(z_n,y_n) \in D$,
$\lambda_n =(\mu ,\ep_n ,\eta_n ,\gamma_n ,\kappa_n )\in \R^5$,
$L=A_{\omega_i}$ (for $i=1$ or $2$)
and $N(z,y,\lambda_n )=(\, 0\, ,\, N_2(z,y,\lambda_n )\, ) $,
$$
N_2(z,y,\lambda_n )=
(\omega_i^2 \ep_n +\mu  (1+\ep_n ))\, \frac{d^{2}y}{dt^{2}}
+\Omega^2 [(1+\eta_n)(1+\gamma_n)-1] y
+\kappa_n (y - z)
+W(y,\eta_n ,\gamma_n ) ,
$$
\begin{equation}
\label{defwnh}
W(y,\eta ,\gamma )=
\Omega^2 \,
(1+\eta )\, \big( \, V^\prime [(1+\gamma)\, y]- (1+\gamma)\, y \, \big).
\end{equation}
Equation (\ref{nl}) holds in the Hilbert space $X$.
The potential $V$ is assumed sufficiently smooth
($C^{p+1}$, with $p\geq 5$) in a neighbourhood of $0$.
It follows that $N\, :\, D\times \R^5\rightarrow X$
is $C^k$ ($k={p-2}\geq 3$) in a
neighbourhood of $(Y,\lambda )=0$. The operator $N$ consists in
higher order terms as $(Y,\lambda_n )\approx 0$, i.e.
we have $N(0,\lambda )=0$, $D_{Y}N(0,0)=0$.

We note that
(\ref{nl}) is invariant under
the symmetry $T\, Y=Y(\cdot+\pi )$. Moreover
the usual invariance under 
index shifts $\{ Y\} \rightarrow \tau_1 \{ Y \}  $
is broken by the inhomogeneity of
the lattice, and replaced by the invariance
$(\{ Y\},\{\lambda \} )\rightarrow (\tau_1 \{ Y \},\tau_1 \{  \lambda \})$.

In the next section we prove a general 
centre manifold reduction theorem for maps 
having the form (\ref{nl}), under appropriate
spectral conditions on $L$ and for small 
nonautonomous perturbations $\{ \lambda \} \in E$. 
This analysis relies on the reduction results
proved in reference \cite{james2} for autonomous
maps. To simplify the proof, problem (\ref{nl}) will
be considered as a projection of a suitable autonomous
mapping to which the centre manifold theorem can be
directly applied.

\subsection{\label{redgeneral} Centre manifold reduction for nonautonomous perturbations of 
infinite-dimensional maps}

In this section we reformulate the situation of section \ref{formul}
in a general framework, and prove a local centre manifold reduction result
for problems of this type. This level of generality is relevant for
nonlinear lattices, because the dynamical equations of many one-dimensional lattices
can be reformulated as infinite-dimensional maps in loop spaces
as one looks for small amplitude time-periodic oscillations.
Indeed, if the coupling between sites has finite range
(i.e. $x_n$ is coupled to $x_k$ for $|n-k|\leq p$), then in general $x_{n+p}$ can
be obtained locally as a function of
$x_{n+p-1},\ldots ,x_{n-p}$ using the implicit function theorem
(for some examples see e.g. \cite{james2}, section 6.1, or \cite{jk}).

To work in a general setting, let us
consider a Hilbert space $X$ and a closed linear operator
$L : D\subset X\rightarrow X$ of domain $D$, $L$ being
in general unbounded. 
We equip $D$ with the scalar product
${\langle u,v \rangle}_{D}=
{\langle Lu,Lv \rangle}_{X}+
{\langle u,v \rangle}_{X}$,
hence $D$ is a Hilbert space continuously embedded in $X$.

We denote by
${\cal U}\times {\cal V}$ a neighbourhood of $0$ in 
$D\times {\R}^{p}$ and consider a nonlinear
map $N\in C^{k}({\cal U}\times {\cal V},X)$ ($k\geq 2$),
where $N(Y,\lambda )$
satisfies $N(0,\lambda )=0$, $D_{Y}N(0,0)=0$.
We look for
sequences $(Y_{n})_{n\in \Z}$ in ${\cal U}$ satisfying
\begin{equation}
\label{nlgen}
\forall n\in\Z, \ \ \ 
Y_{n+1}=L\, Y_{n}+N(Y_{n},\lambda_n )\mbox{\ \ \ in } X,
\end{equation}
where $\{ \lambda \}=(\lambda_{n})_{n\in \Z}$ is a bounded sequence in ${\cal V}$
treated as a parameter.
In what follows we shall note $E=\ell_\infty (\Z , \mathbb{R}^p)$
the Banach space in which $\{ \lambda \}$ lies.
Notice that $Y=0$ is a fixed point of (\ref{nlgen}).

We assume that $L$ has
the property of {\it spectral separation},
i.e. $L$ satisfies the assumption (H) described
below (in what follows we note $\sigma (T)$ the spectrum of 
a linear operator $T$).

\vspace{1ex}

\noindent
{\bf Assumption (H)}: 
{\em The operator $L$ has nonempty hyperbolic ($|z|\neq 1$)
and central ($|z|=1$) spectral parts. Moreover,
there exists an annulus 
$\mathcal{A}=\{ \, z\in\mathbb{C}\, ,\,  r\leq |z|\leq R\, \}$ 
($r<1<R$) 
such that the only part of
the spectrum of $L$ in $\mathcal{A}$ lies
on the unit circle.} 

\vspace{1ex}

\begin{figure}[!h]
\begin{center}
\includegraphics[width=5cm]{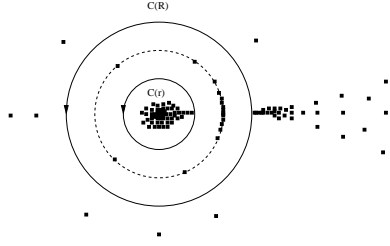} 
\end{center}
\caption{\label{splittingfig} 
Spectrum of $L$ (dots), unit circle (dashed)
and oriented circles $\mathcal{C}(r)$, $\mathcal{C}(R)$.
}
\end{figure}

The situation corresponding to assumption (H) is sketched 
in figure \ref{splittingfig}.
Under assumption (H), the hyperbolic part $\sigma_h$
of $\sigma (L)$
is isolated from its central part $\sigma_c$.
In particular 
this allows one to split $X$ into two 
subspaces $X_c ,X_h$ invariant under $L$,
corresponding to $\sigma_c , \sigma_h$
respectively. More precisely, $L_{h}=L_{|X_{h}}$
and $L_{c}=L_{|X_{c}}$ satisfy $\sigma (L_{h})=\sigma_h$
and $\sigma (L_{c})=\sigma_{c}$.

The invariant subspace $X_{c}$ is called
centre subspace, and $X_h$ is the hyperbolic
subspace. 
The subspace $X_{c}$ is finite-dimensional when
the spectrum of $L$ on the unit circle consists in a
finite number of eigenvalues with finite multiplicities
(we do not need this assumption for the reduction theorem
constructed here).

The spectral projection $\pi_{c}$
on the centre subspace 
can be defined in the following way (see e.g. \cite{kato})
$$
\pi_{c}= \frac{1}{2i\pi}\int_{\mathcal{C}(R)}{(zI-L)^{-1}\, dz}
-
\frac{1}{2i\pi}\int_{\mathcal{C}(r)}{(zI-L)^{-1}\, dz},
$$
where $\mathcal{C}(r)$ denotes the circle of centre $z=0$
and radius $r$ (see figure \ref{splittingfig}).
One has 
$\pi_{c} \in \mathcal{L}(X,D)$, $X_{c}=\pi_{c}X\subset D$ and
$\pi_{c}\, L=L\, \pi_{c}$, where $ \mathcal{L}(X,D)$ denotes
the set of bounded operators from $X$ into $D$.
In the sequel we note $\pi_{h}=I-\pi_{c}$ and
$D_{h}=\pi_{h}\, D$.

\begin{remark}
\label{remreduc}
Let us
consider the situation of section \ref{formul}
and the linear operator $L$ of equation (\ref{nl}).
In the case $\omega = \omega_2 $ and
$\Omega > 2/\sqrt{3}$,
the spectrum of $L$ on the unit circle 
consists in a double non semi-simple
eigenvalue $+1$. Moreover, for
$\omega = \omega_1 $
the spectrum of $L$ on the unit circle 
consists in a double non semi-simple
eigenvalue $-1$.
In both cases the associated invariant
subspace $X_{c}$ is spanned by
$V_{z}=(\cos t ,0)$, $V_{y}=(0,\cos t )$ and
we have
$\pi_{c}\, Y
=\frac{1}{\pi}(\int_{0}^{2\pi}{Y(t)\cos{t}\, dt})\, \cos{t}$.
In addition the unit circle is isolated from the remainder
of the spectrum, since the latter is discrete and only
accumulates at the origin and
at $-\infty$ on the real axis. It follows that $L$
satisfies assumption (H).
\end{remark}

Now we state the centre manifold reduction theorem
in the general case.
In the sequel
we note $Y^c=\pi_c\, Y$, $Y^h=\pi_h\, Y$.

\begin{theorem}
\label{rvc}
Assume that $L$ has the property of spectral separation,
i.e. satisfies assumption (H).
There exists a neighbourhood $\Omega\times \Lambda $ 
of $0$ in $D\times E$
and a map
$\psi\in C^{k}(X_{c} \times \Lambda ,D_{h})$
(with $\psi (0,\{ \lambda \} )=0$, $D_{Y^c}\psi (0,0)=0$) 
such that for all $\{\lambda \} \in \Lambda $ the 
following holds.

\ni
i) If $\{ Y \}$ is a solution of (\ref{nlgen}) 
such that $Y_{n}\in \Omega $ for all $n\in\Z$, then
$Y^h_{n}=\psi(Y^{c}_{n},\tau_n \{\lambda \})$
for all $n\in\Z$ and 
$Y^{c}_{n}$ satisfies the 
nonautonomous recurrence
relation in $X_{c}$
\begin{equation}
\label{red}
\forall n\in\Z, \ \ \
Y^{c}_{n+1}=f_n(Y^{c}_{n},\{\lambda \} ),
\end{equation}
where $f_n\in C^{k}((X_{c}\cap \Omega ) \times \Lambda ,X_{c})$
is defined by
$$
f_n(.,\{\lambda \} )=
\pi_{c}\, (L+N(.,\lambda_n ))\circ (I+\psi (.,\tau_n \{\lambda \})).
$$

\ni
ii) Conversely, if $\{ Y^{c} \}$ is a solution of
(\ref{red}) such that $Y^{c}_{n}\in \Omega $ for all $n\in\Z$,
then $Y_{n}=Y^{c}_{n}+\psi(Y^{c}_{n},\tau_n \{\lambda \} )$ 
satisfies (\ref{nlgen}).

\ni
iii) 
If $L+N(.,\lambda )$ commutes with
a linear isometry $T\in \mathcal{L}(X)\cap \mathcal{L}(D)$
then $T\psi (.,\{ \lambda \} )=\psi (. ,\{ \lambda \} )\circ T$
and $Tf_n(.,\{\lambda \} )=f_n(., \{ \lambda \} )\circ T$.
\end{theorem}

Properties i), ii) reduce the local study of
(\ref{nlgen}) to that of the 
nonautonomous recurrence relation
(\ref{red}) in the subspace $X_{c}$.
Note that the dependency of $\psi$ and the reduced map
$f_n$ with respect to sequences $\{ \lambda \}$
is nonlocal.

\ve

In what follows we give a simple proof of theorem \ref{rvc}
which relies on the fact that the nonautonomous mapping
(\ref{nlgen}) can be seen as a projection of an extended
autonomous mapping, to which the centre manifold
theorem proved in \cite{james2} can be applied. 
This procedure will explain why the result of
theorem \ref{rvc} can be seen as a centre manifold reduction, since
the reduction function $\psi$ will appear as one component of
the function having the centre manifold as its graph. 
The reduced nonautonomous mapping (\ref{red}) will be interpreted as
a projection of the extended autonomous mapping restricted to
the invariant centre manifold.

\ve

Theorem \ref{rvc} has been proved in reference \cite{james2} in
the case of an autonomous mapping, 
when the sequence $\{\lambda \}$ is absent or replaced by a
simple parameter $\lambda \in \mathbb{R}^p$.

To recover this autonomous case we introduce the additional variable
$S_n = \tau_n \{\lambda \} \in E$. Note that for any fixed $n\in \mathbb{Z}$,
$S_n$ denotes a bounded sequence in $\mathbb{R}^p$ 
(to simplify the notations we use the symbol $S_n$ instead of  
$\{ S_n \}$). Given a sequence $ \{\lambda \} \in E$ we also note
$ \delta_0 \{\lambda \} = \lambda_0$. Equation (\ref{nlgen}) can be
rewritten
\begin{equation}
\label{nlaut}
Y_{n+1}=L\, Y_{n}+N(Y_{n},\delta_0 S_n ), \ \ \
S_{n+1}=\tau_1 S_n,
\end{equation}
which consists in an autonomous mapping in $X \times E$.

In what follows we apply the theory of reference \cite{james2}
to system (\ref{nlaut}). As we shall see the corresponding
centre manifold will be infinite-dimensional due to the second
component of (\ref{nlaut}). The case of infinite-dimensional
centre manifolds has been treated in reference \cite{james2},
with the counterpart that theory is restricted to
maps in Hilbert spaces. Consequently the first step is to
search for $ S_n $ in a suitable Hilbert space instead of
the Banach space $E$.
For this purpose we consider the space of sequences
$$
{h}_{-1}  
= \{ \, \{u \} \, / \, u_k \in \mathbb{C}^p , \, \|  \{u \} \|_{-1} <+\infty \, \}, 
$$
where $\|  \{u \} \|_{-1}^2 =  \sum_{k\in \mathbb{Z}}{(1+k^2)^{-1} \, \| u_k \|^2 }$.
The space ${h}_{-1}$ defines a Hilbert space equiped with
the scalar product
$\langle \{u\} , \{v\} \rangle = \sum_{k\in \mathbb{Z}}{(1+k^2)^{-1} \,  u_k \cdot {v}_k }$,
where $\cdot$ denotes the usual scalar product on $ \mathbb{C}^p$ and
$ \| \, \|$ the associated norm.
For all $n\in \mathbb{Z}$ we now search for $S_n$ in the space 
$H={h}_{-1} \cap  (\mathbb{R}^p)^{\mathbb{Z}}$ consisting of
real sequences in ${h}_{-1}$.
Note that $E \subset H$, the embedding being continuous.

Since sequences in $H$ may be unbounded and $N(Y,.)$ is defined on a neighbourhood $\mathcal{V}$
of $\lambda =0$ in $\mathbb{R}^p$, we replace (\ref{nlaut}) by a locally equivalent problem
\begin{equation}
\label{ncut}
(Y_{n+1},S_{n+1})=F(Y_n , S_n)
\end{equation}
where 
$$
F(Y , S)=
\big( \, 
L\, Y +N(Y ,\, \gamma(\delta_0 S )\, ) \, ,\, 
\tau_1 S \, \big) ,
$$
$\gamma \, : \, \mathbb{R}^p \rightarrow \mathcal{V}$ is a $C^\infty$ cut-off function
satisfying $\| \gamma (x) \| \leq \| x\| $,
$\gamma (x)=x$ for $\| x \| < r$, 
$\gamma (x)=0$ for $\| x \| > 2 r$, 
$r$ being chosen small enough
(with $B(0,2r)\subset  \mathcal{V}$).

\ve

Problem (\ref{ncut}) consists in an autonomous mapping in $X\times H$.
In order to apply the centre manifold theorem of reference \cite{james2}
we need to study the spectrum of $DF(0)=L \times \tau_1$. One has clearly
$\sigma (DF(0))=\sigma(L)\, \cup \sigma(\tau_1 )$, where
$\sigma(\tau_1 )$ is determined in the following lemma.

\begin{lemma}
\label{spectretau}
The spectrum $\sigma(\tau_1 )$
of $\tau_1 \, : \, H \rightarrow H$ consists of the unit circle.
\end{lemma}

\proof
Consider the complexification ${h}_{-1}$ of $H$.
Given a sequence $\{ f \} \in {h}_{-1}$ and $z\in \mathbb{C}$,
we look for $\{ u \} \in {h}_{-1}$ satisfying
\begin{equation}
\label{spectral}
(z I - \tau_1 ) \{ u \}  = \{ f \}.
\end{equation}
Equation (\ref{spectral}) can be solved in a simple manner using 
Fourier series.
Recall that the periodic Sobolev space $H^1_{per}(0,2\pi )$ 
can be defined as the set of
functions in 
$L^2 (\mathbb{R}/2\pi \mathbb{Z} , \mathbb{C}^p )$
whose Fourier coefficients form a sequence in $h_1$, where
$$
{h}_1 
= \{ \, \{u \} \, / \, u_k \in \mathbb{C}^p , \,  \sum_{k\in \mathbb{Z}}{(1+k^2) \, \| u_k \|^2 }
<+\infty \, \}. 
$$
In the same way its dual space $H^{-1}_{per}(0,2\pi )$
is isomorphic to $h_{-1}$, where
the isomorphism $C\, : H^{-1}_{per}(0,2\pi ) \rightarrow h_{-1}$
is again given by 
$C_n(T)=\frac{1}{2\pi}\langle T,e^{-int}\rangle$ for all $T\in H^{-1}_{per}(0,2\pi )$.
In addition one has the useful property $\tau_1 C(T)=C(e^{-it}\, T)$
for all $T\in H^{-1}_{per}(0,2\pi )$.
Now return to equation (\ref{spectral}) and
consider $T=C^{-1}(\{ u \})$ and $S=C^{-1}(\{ f \})$. 
One obtains the equivalent problem in $ H^{-1}_{per}(0,2\pi )$
\begin{equation}
\label{specequiv}
(z-e^{-it})\, T=S.
\end{equation}
If $|z|\neq 1$ then (\ref{specequiv}) has the unique solution
$T=(z-e^{-it})^{-1} S$, hence $z\notin \sigma (\tau_1 )$.
If $z=e^{i\theta}$ is chosen on the unit circle, $T=2\pi \delta_{-\theta}$
is a solution for $S=0$, corresponding to an eigenvector 
$\{u \} = \{ e^{in\theta}   \}$ of $\tau_1$.

$\fin$

As it follows from lemma \ref{spectretau},
$\sigma (DF(0))$ consists of the union of 
$\sigma(L)$ with the unit circle. Consequently
$DF(0)$ has the property of spectral separation,
i.e. the hyperbolic part of its spectrum is
isolated from the unit circle. 
Moreover the centre subspace of $DF(0)$ is
simply $X_c \times H$.
With these spectral properties at hand, we now
apply the centre manifold theorem
of reference \cite{james2} which states
the following.

\begin{theorem}
\label{rvcc}
There exists a neighbourhood $\Omega\times \tilde\Lambda $ 
of $(Y,S)=0$ in $D\times H$
and a map
$\psi\in C^{k}( X_{c} \times H ,D_{h})$
(with $\psi (0,0)=0$, $D\psi (0,0)=0$) 
such that the manifold
$$
\mathcal{M} =\{ \, (Y,S) \in D \times H \, /\, 
Y=Y^c+\psi (Y^{c},S ), Y^{c}\in X_{c} \,  \}
$$
has the following properties.

\ve

\noindent
i) $\mathcal{M}$ is locally invariant under $F$, i.e. if 
$(Y,S)\in \mathcal{M} \cap (\Omega\times \tilde\Lambda )$
then $F(Y,S) \in \mathcal{M}$.

\ve

\ni
ii) If $\{(Y ,S )\} $ is a solution of (\ref{ncut}) 
such that $(Y_{n},S_n)\in \Omega \times \tilde\Lambda $ for all $n\in\Z$, then
$(Y_{n},S_n)\in\mathcal{M}$ for all $n\in\Z$ 
(i.e. $Y^h_n =\psi (Y^{c}_n,S_n )$)
and $(Y^{c}_{n},S_n)$ satisfies the recurrence
relation in $X_{c}\times H$
\begin{equation}
\label{redcut}
Y^{c}_{n+1}=\tilde{f}(Y^{c}_{n},S_n), \ \ \ 
S_{n+1}=\tau_1 S_n,
\end{equation}
where 
$$
\tilde{f}(Y^{c} ,S )=
\pi_{c}\, [L+N(.,\gamma (\delta_0 S) )] (Y^c+\psi (Y^c ,S)).
$$

\ve

\ni
iii) 
Conversely, given a solution $\{(Y^c ,S )\} $ of
(\ref{redcut}) such that 
$(Y^{c}_{n},S_n) \in \Omega \times \tilde\Lambda $ for all $n\in\Z$,
consider $Y_{n}=Y^{c}_{n}+\psi(Y^{c}_{n},S_n )$.
Then $(Y_n,S_n)$ defines a solution of (\ref{ncut})
lying on $\mathcal{M}$.

\ve

\ni
iv) 
If $L+N(.,\lambda )$ commutes with
a linear isometry $T\in \mathcal{L}(X)\cap \mathcal{L}(D)$
then $T\psi (Y^c ,S )=\psi (TY^c ,S )$
and $T\tilde{f}(Y^{c} ,S )=\tilde{f}(T Y^{c} ,S )$.
\end{theorem}

The manifold $\mathcal{M}$ is called a local $C^k$ centre manifold
for (\ref{ncut}). It is locally invariant under $F$ (as stated by
property i) ) and the linear isometries of (\ref{ncut}).
Property iv) expresses the invariance of $\mathcal{M}$ under
the linear isometry $T\times I$ of (\ref{ncut}).

\ve

Now the proof of theorem \ref{rvc} follows directly from
theorem \ref{rvcc}. Since $E$ is continuously embedded in $H$,
$\psi$ defines a $C^{k}$ map from $X_{c} \times E$ into $D_{h}$.
In theorem \ref{rvc} we choose $\Lambda$ as a ball of centre
$0$ in $E$ such that $\Lambda \subset \tilde\Lambda $ and $\gamma = I$ on $\Lambda$.
Then problems (\ref{nlgen}) and (\ref{ncut}) are equivalent
for all $\{ \lambda \} \in \Lambda$, with
$S_n = \tau_n \{\lambda \}$, and properties i)-ii)-iii) 
of theorem \ref{rvc} are directly deduced from properties
ii)-iii)-iv) of theorem \ref{rvcc}.
In addition, since $(0,\tau_n \{\lambda \})$  
is a solution of (\ref{ncut}) for all $\{\lambda \} \in \Lambda$ it
follows $\psi (0,\tau_n \{\lambda \} )=0$
(by property ii) of theorem \ref{rvcc}), and consequently
$\psi (0,\{\lambda \} )=0$. 

\subsection{\label{redkg}Application to the Klein-Gordon lattice}

\subsubsection{Reduction result}

In this section we apply the reduction theorem \ref{rvc} to
the inhomogeneous Klein-Gordon lattice considered in section \ref{formul}.
We recall that the inhomogeneous system (\ref{dfpu}) has been
reformulated as a nonautonomous map in a loop space given by expression (\ref{nl}).
All parameters 
(sequences of heterogeneities and frequency shift $\mu$)
are embedded in the multicomponent parameter
$\{ \lambda \} 
=(\mu ,\{\ep \} ,\{\eta  \},\{\gamma \} ,\{\kappa \})\in E=\R\times {(\ell_\infty (\Z ))}^4$.
The problem has exactly the general form (\ref{nlgen}) (in a particular 
case when the first component of $\{ \lambda \}$ is constant) and
consequently the reduction theorem \ref{rvc} can be applied to (\ref{nl}).
This yields the following result for the original system (\ref{dfpu}).

\begin{theorem}
\label{reductionnh}
Fix $\omega^{2}=\omega_c^2 +\mu$ in equation (\ref{dfpu}),
where $\omega_c  =\sqrt{4+\Omega^2}$ or
$\omega_c = \Omega$ (in that case we further assume
$\Omega > 2/\sqrt{3}$).
There exist neighbourhoods
${\cal U}$, ${\cal V}$ and $\mathcal{W}$
of $0$ in $H^{2}_{\#}$, $E$
and $\R$ respectively,
and a $C^k$ map $\phi \, :\, \R^{2}\times E \rightarrow H^{2}_{\#}$
(with $\phi (0,\{\lambda \} )=0$, $D\phi (0,0)=0$) such that
the following holds for all $\{ \lambda \}
\in \mathcal{V}$.

\vspace{1ex}

\noindent
i) All solutions  of (\ref{dfpu}) such that $y_{n}\in {\cal U}$
for all $n\in \Z$ have the form
$$
y_{n}(t)=
\beta_{n}\cos{t}+H_n(t),
$$
where
$H_n =
\phi (\beta_{n-1},\beta_{n},\tau_n \{ \lambda \} )$.
For $\omega_c = \Omega$,
$\beta_{n}$ satisfies a recurrence relation
\begin{equation}
\label{redequanh}
\beta_{n+1}-2\beta_{n}+\beta_{n-1}=R_n(\beta_{n-1},\beta_{n}, \{\lambda \} )
\end{equation}
where $R_n \, : \, \mathcal{W}^2 \times {\cal V} \rightarrow \mathbb{R}$
is $C^k$. The principal part of $R_n$ reads
\begin{equation}
\label{princrn}
R_n(\alpha ,\beta , \{\lambda \}
)=
(\Omega^2\eta_n (1+\gamma_n )-(\Omega^2 +\mu ) \ep_n +\Omega^2 \gamma_n -\mu )\, \beta
+\kappa_n (\beta -\alpha )
+B\, \beta^{3}+\mbox{h.o.t.},
\end{equation}
\begin{equation}
\label{defbnh}
B=\frac{\Omega^2}{8}\, (\, V^{(4)}(0)
-\frac{5}{3} (V^{(3)}(0))^2\, ).
\end{equation}
For $\omega_c  =\sqrt{4+\Omega^2}$ one has
\begin{equation}
\label{redequanh2}
\beta_{n+1}+2\beta_{n}+\beta_{n-1}=R_n(\beta_{n-1},\beta_{n}, \{\lambda \}),
\end{equation}
with 
\begin{equation}
\label{princrn2}
R_n(\alpha ,\beta , \{\lambda \}) =
(\Omega^2\eta_n (1+\gamma_n )-(4+\Omega^2 +\mu ) \ep_n +\Omega^2 \gamma_n -\mu )\, \beta
+\kappa_n (\beta -\alpha )
+\tilde{B}\, \beta^{3}+\mbox{h.o.t.},
\end{equation}
\begin{equation}
\label{defb2}
\tilde{B}=
\frac{\Omega^2}{8}\, 
(\,
V^{(4)}(0)
+{(V^{(3)}(0))}^{2}
(
\frac{\Omega^2}{16+3\Omega^2}-2
)\,
).
\end{equation}
In both cases, higher order terms in $R_n$
are $O( \| (\alpha ,\beta ) \|^3 \, \| \{ \lambda \} \|_E  + \| (\alpha ,\beta ) \|^5   )$
and non-local in $\{ \lambda \}$.

\noindent
ii) If $\beta_{n}$ is a solution of problem (\ref{redequanh}) or (\ref{redequanh2})
(respectively for $\omega_c = \Omega$ or $\omega_c  =\sqrt{4+\Omega^2}$),
such that $\beta_{n}\in {\cal W}$
for all $n\in \Z$, then
$y_{n}(t)=
\beta_{n}\cos{t}+\phi (\beta_{n-1},\beta_{n}, \tau_n \{\lambda \} )$
satisfies equation (\ref{dfpu}).

\noindent
iii) The functions $\phi$ and $R_n$ have the following
symmetries
$$
\phi (-\alpha,-\beta,\{\lambda \} )=
T\phi (\alpha,\beta,\{\lambda \} ),\ \ \
R_n(-\alpha,-\beta,\{\lambda \} )=-R_n(\alpha,\beta,\{\lambda \} ),
$$
where $T$ denotes the half period time shift
$[T\phi (.)](t)=[\phi (.) ] (t+\pi )$.
\end{theorem}

It is straightforward to check that
system (\ref{dfpu}) has the reduction properties i) and ii) described above
since the equivalent system (\ref{nl}) satisfies properties
i) and ii) of theorem \ref{rvc} (see remark \ref {remreduc} p.\pageref{remreduc}).
However there remains to compute the explicit forms
(\ref{princrn}) and (\ref{princrn2})
of the recurrence relations.  
These expressions do not simply correspond to 
the two-dimensional mapping (\ref{red}) rewritten as a second order
recurrence relation.
In addition we rewrite (\ref{red}) in {\em normal form}, i.e. 
we perform a polynomial change of variables which simplifies
(\ref{red}) by keeping only its essential terms.
These computations will be the object of the next three sections.
Property iii) is equivalent to property iii) of
theorem \ref{rvc}, where the symmetry $T$ is the
half period time shift which satisfies $T_{|X_c}=-I$.

A possible way of computing the reduced recurrence relation
would be to consider the equivalent autonomous mapping (\ref{nlaut})
and use a classical computation scheme for centre manifolds of
autonomous systems (see e.g. \cite{vander} for a description of the
method).
The first step consists in computing the Taylor expansion of the
reduction function $\psi$ up to a given order. This can be done
using a nonlocal equation for $Y_n$ (obtained by
expressing $Y_n$ in (\ref{nlaut}) as a function of $N(Y_n , \delta_0 S_n)$) and
computing the Taylor coefficients of $\psi$ by
induction (see \cite{vander}).
The second step is to compute 
the reduced recurrence relation (\ref{redcut}) which is
completely determined by $\psi$.

In the next three sections we shall use a different method yielding
simpler computations.
Firstly we compute the recurrence relations 
(\ref{princrn}) and (\ref{princrn2}) in the autonomous case 
$\{\ep \} =\{\eta  \}=\{\gamma \} =\{\kappa \}=0$, using the method of reference \cite{james2}.
Then, using a symmetry argument, we deduce how the leading order part of
the reduced equation is modified by the nonautonomous terms of 
(\ref{nl}).

\vspace{1ex}

To end this section we point out a generalization of theorem \ref{reductionnh}.
As it follows from the analysis of
section \ref{formul},
the dimension of the centre space $X_{c}$
of $A_{\omega}$ is twice the number of multiples
of $\omega$ lying within the band
$[\Omega ,(4+\Omega^2)^{1/2}]$.
More precisely, if $\Omega \leq \omega p \leq (4+\Omega^2)^{1/2}$
for $p\in \{ p_0 , \ldots , p_1 \}$, with no additional multiples
entering the band, then the centre space
is spanned by the corresponding Fourier modes 
$(\cos (pt) ,0)$, $(0,\cos (pt) )$. As above the following 
reduction result follows from theorem \ref{rvc}.

\begin{theorem}
\label{reductionnhg}
Consider $\omega_c >0$ such that
$\omega_c \, p \in [\Omega ,(4+\Omega^2)^{1/2}]$
for all integers $p\in \{ p_0 , \ldots , p_1 \}$, 
with no additional multiples in this interval. 
Fix $\omega^{2}=\omega_c^2 +\mu$ in equation (\ref{dfpu}) and
note $N=p_1 - p_0 +1$. Consider the subspace $H_c$ of
$H^{2}_{\#}$ spanned by the $N$ Fourier modes
$ \cos({p_0t}), \ldots , \cos({p_1t})$ and its complementary
subspace $H_c^\perp$ consisting of orthogonal Fourier modes.
There exist neighbourhoods
${\cal U}$, ${\cal V}$
of $0$ in $H^{2}_{\#}$, $E$ respectively,
and a $C^k$ map $\phi \, :\, \R^{2N}\times E \rightarrow H_c^\perp$
(with $\phi (0,\{\lambda \} )=0$, $D\phi (0,0)=0$), such that
for all $\{ \lambda \} \in \mathcal{V}$,
all solutions of (\ref{dfpu}) such that $y_{n}\in {\cal U}$
for all $n\in \Z$ have the form
\begin{equation}
\label{redform}
y_{n}(t)=
\sum_{p=p_0}^{p_1}{[\, \beta_{n}^{(p)}\, \cos{(pt)}\, ]}+
\phi (\beta_{n-1}^{(p_0)},\beta_{n}^{(p_0)},\ldots ,\beta_{n-1}^{(p_1)},\beta_{n}^{(p_1)},\tau_n \{ \lambda \} ).
\end{equation}
Moreover, all small amplitude solutions of (\ref{dfpu})
are determined by a finite-dimensional recurrence relation
obtained by projecting (\ref{dfpu}) on $H_c$ and
using the ansatz (\ref{redform}).
\end{theorem}

In the following sections \ref{lower}, \ref{upper} and \ref{inh}, we 
compute the explicit forms of the reduced recurrence relations
given in theorem \ref{reductionnh}.

\subsubsection{\label{lower}Homogeneous case near the lower phonon band edge}

In this section we restrict our attention to the case
when $\Omega > 2/\sqrt{3}$ and
$\omega \approx\omega_2  =\Omega$.
We consider the autonomous case when
$\{\ep \} =\{\eta  \}=\{\gamma \} =\{\kappa \}=0$.
Equation (\ref{nl}) now reads
\begin{equation}
\label{nlfpu1} 
Y_{n+1}=L\, Y_{n}+N(Y_{n},\mu ),
\ \ \ n\in \Z
\end{equation}
where $L=A_\Omega$ is given by (\ref{defia}) and
$$
N((z,y),\mu )=
\Big( \, 0\, ,\, 
\mu\, \frac{d^{2}y}{dt^{2}}+W(y)
\, \Big) ,
$$
with
$W(y)=\Omega^2(V^\prime (y)-y)$.
System (\ref{nlfpu1}) is a reformulation of
the equations of motion for the 
homogenous Klein-Gordon lattice
\begin{equation}
\label{homkg}
\omega^2 \, \frac{d^{2}y_{n}}{dt^{2}}
+\Omega^2 \, V^{\prime}(y_{n})=
y_{n+1}-2y_{n}+y_{n-1},\ \ \ n\in \Z .
\end{equation}
As in the nonautonomous case 
(\ref{nl}), system (\ref{nlfpu1}) is invariant under
the symmetry $T\, Y=Y(\cdot+\pi )$.
Moreover, the invariance
$y_{n}\rightarrow y_{-n}$ of (\ref{homkg})
implies that (\ref{nlfpu1}) is reversible with respect to
the symmetry $R(z,y)=(y,z)$, i.e. if $Y_{n}$ is a solution
then also $RY_{-n}$. In other words, if $Y$ and 
$[L+N(.,\mu )](RY )$ are in some neighbourhood of $0$ in $D$ one has
$(L+N(.,\mu )\circ R)^{2}Y=Y$. Lastly,
due to the existence of the additional symmetry $T$, it is worthwhile
to notice that $TR$ defines an
other reversibility symmetry.

In what follows we use the notations introduced in section \ref{redgeneral}.
We recall that 
the spectrum of $L=A_\Omega$
on the unit circle 
consists in a double non semi-simple
eigenvalue $+1$, and the associated two-dimensional invariant
subspace $X_{c}$ 
is spanned by the vectors
$V_{z}=(\cos t ,0)$, $V_{y}=(0,\cos t )$, with
$$
L_{|X_{c}}=
\left(  
\begin{array}{cc}  
0 &1 \\  
-1 &2 \\  
\end{array}  
\right) 
$$
in the basis $(V_{z},V_{y})$.
For $\mu $ in some neighbourhood $\Lambda$ of $0$,
(\ref{nlfpu1}) admits  
a $C^k$ two-dimensional
local centre manifold $\mathcal{M}_{\mu}\subset D$ 
(which can be written as a graph over $X_{c}$),
{\em locally} invariant under $L+N(.,\mu )$
(see \cite{james2}, theorem 1 p. 32).
One can write
\begin{equation}
\label{cm}
{\cal M_{\mu}}=\{ \, Y\in D\, /\, 
Y=aV_{z}+bV_{y}
+\psi (a,b,\mu ), (a,b)\in\R^{2} \,  \} ,
\end{equation}
where $\psi\in C^{k}(\R^{2}\times \Lambda ,{D}_h)$
and $\psi(a,b ,\mu)=O(\|(a,b) \|^2+\|(a,b) \||\mu|)$. 
Moreover, ${\cal M_{\mu}}$ is invariant under $T$ and $R$
(see \cite{james2}, theorem 2 p. 34 and section 5.2).

In the sequel we use the notations
$P^{\ast}(y)
=\frac{1}{\pi}\int_{0}^{2\pi}{y(t)\cos{t}\, dt}$,
$P_{c}\, y=P^{\ast}(y)\, \cos{t}$
and $H^{2}_{h}=\{ \, y\in H^{2}_{\#}\, /\, P^{\ast}(y)=0\, \}$.
The spectral projection $\pi_{c}$ on $X_{c}$ reads
$\pi_{c}(z,y)=(P_{c}\, z,P_{c}\, y)$ and we have
${D}_h=H^{2}_{h}\times H^{2}_{h}$.

Since ${\cal M_{\mu}}$ is invariant under $R$ and
$V_{z}$, $V_{y}$ are exchanged by $R$,
we have the symmetry property
$R\psi (a,b,\mu )=\psi (b,a,\mu )$.
Consequently, the function $\psi $ has the form
\begin{equation}
\label{varcomp}
\psi (a,b,\mu )=
(\varphi (b,a,\mu ),\varphi (a,b,\mu ))
\end{equation}
with $\varphi\in C^{k}(\R^{2}\times \Lambda ,H^{2}_{h})$.
Since ${\cal M_{\mu}}$ is invariant under $T$ and
$T_{|X_{c}}=-I$ we have in addition
\begin{equation}
\label{id0}
T\varphi (a,b,\mu )=\varphi (-a,-b,\mu ).
\end{equation}

For $\mu\approx 0$, the centre manifold
$\mathcal{M}_{\mu}$ contains all solutions $Y_n$ 
of (\ref{nlfpu1}) staying 
in a sufficiently small 
neighbourhood of
$Y=0$ in $D$
for all $n\in\mathbb{Z}$. 
Their coordinates $(a_n,b_n)$ on $\mathcal{M}_{\mu}$
are thus given by a two-dimensional mapping
which determines {\it all} small amplitude solutions
when $\mu\approx 0$.
The reduced mapping is given by
\begin{equation}
\label{red11}
\left(
\begin{array}{c}
a_{n+1} \\
b_{n+1}
\end{array}
\right)
=
f_\mu
\left(
\begin{array}{c}
a_{n} \\
b_{n}
\end{array}
\right)
\end{equation}
where
\begin{equation}
\label{deffm}
f_\mu
\left(
\begin{array}{c}
a \\
b
\end{array}
\right)
=
\left(
\begin{array}{c}
b, \\
-a+2b+r(a,b,\mu )
\end{array}
\right) ,
\end{equation}
\begin{equation}
\label{deff1}
r(a,b,\mu )=-\mu\, b
+P^{\ast}W( b\cos{t}+\varphi (a,b,\mu )) .
\end{equation}
One obtains equation (\ref{red11}) 
using the fact that 
$$
z_n=a_n\, \cos{t}+\varphi (b_n,a_n,\mu ),
\ \ \
y_n=b_n\, \cos{t}+\varphi (a_n,b_n,\mu )
$$
for $Y_n =(z_n ,y_n)\in \mathcal{M}_{\mu}$
and applying $P^{\ast}$ to
equation (\ref{nlfpu1})
(one has $P^{\ast}\varphi =0$ and
$P^{\ast}\circ\frac{d^{2}}{dt^{2}}=-P^{\ast}$ on $H^{2}_{\#}$).

Since the reduced mapping inherits the symmetries of (\ref{nlfpu1}) \cite{james2},
$f_\mu$ commutes 
with $T_{|X_{c}}=-I$ and thus
$r(-a,-b,\mu )=-r(a,b,\mu )$.
Moreover, $f_\mu$ is reversible with respect to 
the symmetry $R(a,b)=(b,a)$, i.e.
$(f_\mu\circ R)^2=I$. This
yields the identity
$$
r(a,b,\mu )=r(-a+2b+r(a,b,\mu ),b,\mu ).
$$
This imposes the following structure for the
Taylor expansion of $r$ at
$(a,b,\mu )=0$
\begin{equation}
\label{defr1}
r(a,b,\mu )=
-b\mu +c_{1} b^{3}+c_{2} ab^{2}-\frac{1}{2}c_{2} a^{2}b
+O(|b|\, (|a|+|b|)^4+|b|\, (|a|+|b|)^2 |\mu |),
\end{equation}
where coefficients $c_1 ,c_2$ have to be determined.
Note that $r(a,0,\mu )=0$ (see \cite{james2} p.53 for details).

For determining the unknown coefficients of (\ref{defr1}), we
first compute the leading order terms in the Taylor expansion
of $\psi$ at $(a,b,\mu )=0$.
This can be done using the fact that 
$\mathcal{M}_{\mu}$ 
is locally invariant under $L+N(.,\mu )$
(see \cite{james2}, theorem 1 p. 32).
For $(a,b)\approx 0$, this yields
\begin{equation}
\label{eqinvman}
\pi_h \, [L+N(.,\mu )]\, (\, (a,b)\, \cos{t}+\psi (a,b,\mu )\, )
=\psi (f_\mu (a,b) ,\mu ) 
\end{equation}
or equivalently
\begin{equation}
\label{id11}
\varphi (
-a+2b+r(a,b,\mu )
,
b
,\mu
)
=
\varphi (a,b,\mu ),
\end{equation}
\begin{equation}
\label{id21}
\begin{array}{l}
\varphi (
b
,
-a+2b+r(a,b,\mu )
,\mu
)
=\\
(\Omega^2\frac{d^{2}}{dt^{2}}+2+\Omega^2 )\varphi (a,b,\mu )-\varphi (b,a,\mu )
+(1-P_{c})\, [\mu\frac{d^{2}}{dt^{2}}+W]\, ( b\cos{t}+\varphi (a,b,\mu )).
\end{array}
\end{equation}
Thanks to the symmetry property (\ref{id11}),
the Taylor expansion of $\varphi$ at order $2$ takes the form
\begin{equation}
\label{dev1}
\varphi (a,b,\mu )=
\varphi_{011}b\mu 
-\frac{1}{2}\varphi_{110}a^{2} +
\varphi_{110}ab +
\varphi_{020} b^{2} +\mbox{h.o.t}.
\end{equation}
By an identification procedure
we now compute the coefficients $\varphi_{pqr}$ 
in (\ref{dev1}), using
(\ref{id21}) and the expansion
\begin{equation}
\label{exp}
W(y)=\Omega^2(\,
\frac{1}{2}V^{(3)}(0)\, y^2
+
\frac{1}{6}V^{(4)}(0)\, y^3
+O(y^4)\, ).
\end{equation}
Identification at order $b\mu $ gives
$$
(\frac{d^{2}}{dt^{2}}+1)\varphi_{011}=0,
$$
hence $\varphi_{011}=0$ since $\varphi_{011}\in H^{2}_{h}$.
Identification at order $ab$ leads to
\begin{equation}
\label{ab1}
\varphi_{020}=
-\frac{1}{4}
(\Omega^2\frac{d^{2}}{dt^{2}}+2+\Omega^2 )\varphi_{110}
\end{equation}
and identification at order $b^{2}$ yields
\begin{equation}
\label{b21}
-\varphi_{110}+
(\Omega^2\frac{d^{2}}{dt^{2}}-2+\Omega^2 )
\varphi_{020}=
-\frac{1}{2}\Omega^2 V^{(3)}(0)\,
\cos^2{t}.
\end{equation}
Reporting (\ref{ab1}) in (\ref{b21}) gives
$$
(\frac{d^{2}}{dt^{2}}+1)^2\varphi_{110}=
\frac{2}{\Omega^2}
V^{(3)}(0)\, \cos^2{t}
$$
and consequently
$$\varphi_{110}=\frac{1}{\Omega^2}V^{(3)}(0)\,(1+\frac{1}{9}\cos{(2t)}),
\ \ \
\varphi_{020}=
\frac{1}{2}V^{(3)}(0)\,
(-\frac{1}{2}-\frac{1}{\Omega^2}+
(\frac{1}{6}-\frac{1}{9\Omega^2})
\cos{(2t)}).$$
As a conclusion, we obtain
\begin{equation}
\label{phi1}
\begin{array}{rcl}
\varphi (a,b,\mu )&=&
\frac{1}{\Omega^2}V^{(3)}(0)\,(1+\frac{1}{9}\cos{(2t)})
\, (ab -\frac{1}{2}a^{2})\\
& &
+\frac{1}{2}V^{(3)}(0)\,
(-\frac{1}{2}-\frac{1}{\Omega^2}+
(\frac{1}{6}-\frac{1}{9\Omega^2})
\cos{(2t)})\, b^{2} 
+\mbox{h.o.t}.
\end{array}
\end{equation}

We now compute the two-dimensional mapping giving
the coordinates $(a_{n},b_{n})$ of the solutions
on ${\cal M_{\mu}}$.
Equation (\ref{red11}) 
can be written
\begin{equation}
\label{red21}
a_{n+1}=b_{n}, \ \ \
b_{n+1}-2b_{n}+b_{n-1}=r(b_{n-1},b_{n},\mu ).
\end{equation}
Using (\ref{deff1}), (\ref{phi1}) and (\ref{exp}) yields
in equation (\ref{defr1})
\begin{equation}
\label{defc1}
c_{1}=
\frac{1}{8}\Omega^2 V^{(4)}(0)
-
(\frac{19}{9}+\frac{5}{6}\Omega^2 )
\frac{1}{4}(V^{(3)}(0))^2
, \ \ \
c_{2} =
\frac{19}{18}(V^{(3)}(0))^2.
\end{equation}

Lastly, one can write (\ref{red21}) in normal form
using the change of variables
$b_{n}=\beta_{n}-\frac{c_{2}}{12}\beta_{n}^{3}$. 
The normal form of (\ref{red21}) at order $3$ reads
\begin{equation}
\label{redfinp1}
\beta_{n+1}-2\beta_{n}+\beta_{n-1}=
-\mu\, \beta_{n}+B\, \beta_{n}^{3}+\mbox{h.o.t.}
\end{equation}
with
\begin{equation}
\label{defb1}
B=c_1+\frac{c_2}{2}=\frac{\Omega^2}{8}\, (\, V^{(4)}(0)
-\frac{5}{3} (V^{(3)}(0))^2\, ).
\end{equation}
This yields the explicit form (\ref{redequanh})
of the reduced recurrence relation in the autonomous case
$\{\ep \} =\{\eta  \}=\{\gamma \} =\{\kappa \}=0$.

\subsubsection{\label{upper}Homogeneous case near the upper phonon band edge}

In this section we consider the case
$\omega \approx\omega_1  =\sqrt{4+\Omega^2}$, in
the autonomous case when
$\{\ep \} =\{\eta  \}=\{\gamma \} =\{\kappa \}=0$.
Equation (\ref{nl}) takes the form (\ref{nlfpu1}),
where $L=A_{\omega_1}$ is given by (\ref{defia}).
The spectrum of $L$ on the unit circle 
consists in a double non semi-simple
eigenvalue $-1$, and the centre space $X_{c}$ 
is again spanned by
$V_{z}=(\cos t ,0)$, $V_{y}=(0,\cos t )$.

For $\mu = \omega^2 - \omega_1^2$ in some neighbourhood $\Lambda$ of $0$,
there exists
a smooth two-dimensional
local centre manifold $\mathcal{M}_{\mu}\subset D$ 
locally invariant under $L+N(.,\mu )$, $T$, $R$
and having the form (\ref{cm}).
The function $\psi$ having the centre manifold as its graph has
the form (\ref{varcomp}) and shares the property
(\ref{id0}).
For $\mu\approx 0$, the centre manifold
$\mathcal{M}_{\mu}$ contains all solutions $Y_n$ 
of (\ref{nlfpu1}) staying 
in a sufficiently small 
neighbourhood of
$Y=0$ in $D$
for all $n\in\mathbb{Z}$. 
Their coordinates $(a_n,b_n)$ on $\mathcal{M}_{\mu}$
are then given by a two-dimensional mapping,
which determines {\it all} small amplitude solutions
when $\mu\approx 0$.

The operator $L$ has the following structure
in the basis $(V_{z},V_{y})$
$$
L_{|X_{c}}=
\left(  
\begin{array}{cc}  
0 &1 \\  
-1 &-2 \\  
\end{array}  
\right) 
$$
and the reduced mapping is given by
\begin{equation}
\label{red12}
\left(
\begin{array}{c}
a_{n+1} \\
b_{n+1}
\end{array}
\right)
=
f_\mu
\left(
\begin{array}{c}
a_{n} \\
b_{n}
\end{array}
\right)
\end{equation}
where
\begin{equation}
\label{deffm2}
f_\mu
\left(
\begin{array}{c}
a \\
b
\end{array}
\right)
=
\left(
\begin{array}{c}
b, \\
-a-2b+r(a,b,\mu )
\end{array}
\right) 
\end{equation}
and $r$ is defined by (\ref{deff1}).
Since the reduced mapping inherits the symmetries of (\ref{nlfpu1}),
$f_\mu$ commutes 
with $T_{|X_{c}}=-I$
hence
$r(-a,-b,\mu )=-r(a,b,\mu )$.
Moreover, (\ref{red12}) is reversible with respect to 
the symmetry $R(a,b)=(b,a)$, which 
yields the identity
$$
r(a,b,\mu )=r(-a-2b+r(a,b,\mu ),b,\mu ) .
$$
This implies $r(a,0,\mu )=0$ and
\begin{equation}
\label{defr12}
r(a,b,\mu )=
-b\mu +c_{1} b^{3}+c_{2} ab^{2}+\frac{1}{2}c_{2} a^{2}b+\mbox{h.o.t},
\end{equation}
where the coefficients $c_1 ,c_2$ have to be determined.

For this purpose, we
first compute the leading order terms in the Taylor expansion
of $\psi$ at $(a,b,\mu )=0$, using the fact that 
$\mathcal{M}_{\mu}$ 
is locally invariant under $L+N(.,\mu )$.
Equation (\ref{eqinvman}) yields
\begin{equation}
\label{id12}
\varphi (
-a-2b+r(a,b,\mu )
,
b
,\mu
)
=
\varphi (a,b,\mu ),
\end{equation}
\begin{equation}
\label{id22}
\begin{array}{l}
\varphi (
b
,
-a-2b+r(a,b,\mu )
,\mu
)
=\\
(\omega_1^2\frac{d^{2}}{dt^{2}}+2+\Omega^2 )\varphi (a,b,\mu )-\varphi (b,a,\mu )
+(1-P_{c})\, [\mu\frac{d^{2}}{dt^{2}}+W]\, ( b\cos{t}+\varphi (a,b,\mu )).
\end{array}
\end{equation}
The Taylor expansion of $\varphi$ at order $2$ takes the following
form (due to the symmetry property (\ref{id12}))
\begin{equation}
\label{dev2}
\varphi (a,b,\mu )=
\varphi_{011}b\mu 
+\frac{1}{2}\varphi_{110}a^{2} +
\varphi_{110}ab +
\varphi_{020} b^{2} +\mbox{h.o.t}.
\end{equation}
By an identification procedure
we now compute the coefficients $\varphi_{pqr}$ 
in (\ref{dev2}), using
(\ref{id22}) and the expansion (\ref{exp}).
Identification at order $b\mu $ gives
$$
(\frac{d^{2}}{dt^{2}}+1)\varphi_{011}=0,
$$
hence $\varphi_{011}=0$ since $\varphi_{011}\in H^{2}_{h}$.
Identification at order $ab$ leads to
\begin{equation}
\label{ab2}
\varphi_{020}=
\frac{1}{4}
(\omega_1^2\frac{d^{2}}{dt^{2}}+2+\Omega^2 )\varphi_{110}
\end{equation}
and identification at order $b^{2}$ yields
\begin{equation}
\label{b22}
\varphi_{110}+
(\omega_1^2\frac{d^{2}}{dt^{2}}-2+\Omega^2 )
\varphi_{020}=
-\frac{1}{2}\Omega^2 V^{(3)}(0)\,
\cos^2{t}.
\end{equation}
Reporting (\ref{ab2}) in (\ref{b22}) gives
$$
(\omega_1^2\frac{d^{2}}{dt^{2}}+\Omega^2)^2\varphi_{110}=
- 2\Omega^2 V^{(3)}(0)\cos^2{t}
$$
and consequently
$$\varphi_{110}=
-  V^{(3)}(0)
(\frac{1}{\Omega^2}+\frac{\Omega^2}{{(16+3\Omega^2)}^{2}}\cos{(2t)}),
$$
$$
\varphi_{020}=
-\frac{1}{4}\Omega^2 V^{(3)}(0)
(\frac{1}{\Omega^2}+\frac{2}{\Omega^4}+
(\frac{2}{{(16+3\Omega^2)}^{2}}-\frac{1}{16+3\Omega^2})
\cos{(2t)})
.$$
As a conclusion, we obtain
\begin{equation}
\label{phi2}
\begin{array}{rcl}
\varphi (a,b,\mu )&=&
-  V^{(3)}(0)
(\frac{1}{\Omega^2}+\frac{\Omega^2}{{(16+3\Omega^2)}^{2}}\cos{(2t)})
\, (ab +\frac{1}{2}a^{2})\\
& &
-\frac{1}{4}\Omega^2 V^{(3)}(0)
(\frac{1}{\Omega^2}+\frac{2}{\Omega^4}+
(\frac{2}{{(16+3\Omega^2)}^{2}}-\frac{1}{16+3\Omega^2})
\cos{(2t)})
\, b^{2} 
+\mbox{h.o.t}.
\end{array}
\end{equation}

We now compute the two-dimensional mapping giving
the coordinates $(a_{n},b_{n})$ of the solutions
on ${\cal M_{\mu}}$.
Equation (\ref{red12}) 
can be written
\begin{equation}
\label{red22}
a_{n+1}=b_{n}, \ \ \
b_{n+1}+2b_{n}+b_{n-1}=r(b_{n-1},b_{n},\mu ).
\end{equation}
Using (\ref{deff1}), (\ref{phi2}) and (\ref{exp}) yields
in equation (\ref{defr12})
\begin{equation}
\label{defc2}
c_{1}=
\frac{1}{8}\Omega^2 
\big[
V^{(4)}(0)
-{(V^{(3)}(0))}^{2}\Omega^2
(
\frac{2}{\Omega^2}+\frac{4}{\Omega^4}+
\frac{2}{{(16+3\Omega^2)}^{2}}-\frac{1}{16+3\Omega^2}
)
\big],
\end{equation}
\begin{equation}
\label{defc3}
c_{2} =
-\Omega^4 {(V^{(3)}(0))}^2
\big[\frac{1}{\Omega^4}+\frac{1}{2}\frac{1}{{(16+3\Omega^2)}^{2}}\big] .
\end{equation}

The transformation
$b_{n}=\beta_{n}-\frac{c_{2}}{12}\beta_{n}^{3}$ yields
the normal form of (\ref{red22}) of order $3$
\begin{equation}
\label{redfinp2}
\beta_{n+1}+2\beta_{n}+\beta_{n-1}=
-\mu\, \beta_{n}+\tilde{B}\, \beta_{n}^{3}+\mbox{h.o.t.}
\end{equation}
with $\tilde{B}$ defined by (\ref{defb2}).
This yields the explicit form (\ref{redequanh2})
of the reduced recurrence relation in the autonomous case
$\{\ep \} =\{\eta  \}=\{\gamma \} =\{\kappa \}=0$.

\subsubsection{\label{inh}Inhomogeneous cases}

Using the normal form computations performed in the above sections
for the autonomous case, one can obtain by perturbation the
principal part (\ref{redequanh}) (or (\ref{redequanh2})) of
the normal form for $\omega \approx \omega_2$ (or  $\omega \approx \omega_1$)
in the nonautonomous case.
In what follows this computation is
described for $\omega \approx \omega_2$, the treatment for $\omega \approx \omega_1$
being completely similar.

Theorem \ref{reductionnh} is obtained by applying the reduction theorem \ref{rvc}
to the first order system (\ref{nl}).
According to theorem \ref{rvc}-i), small amplitude solutions $Y_n =(z_n, y_n)$
of (\ref{nl}) have the following form for small $\{ \lambda \} \in E$
\begin{equation}
\label{formeyn}
Y_n = (a_n , b_n )\, \cos{t} +  \Psi (a_n,b_n,\tau_n\{\lambda \} ),
\end{equation}
where $\Psi (a,b,\{\lambda \} )=\psi ((a,b)\, \cos{t},\{\lambda \} ) \in D_h$
and $\psi$ denotes the reduction function of theorem \ref{rvc}.
In the sequel we shall note $\Psi = (\Psi_1 ,\Psi_2)$.

Let us compute the explicit form of the reduced map (\ref{red}). For this
purpose, one has to use the ansatz (\ref{formeyn}) in equation (\ref{nl})
and project the latter
on the Fourier mode $\cos{t}$.
Setting $F_n (a,b,\{ \lambda \} )\, \cos{t} =f_n ((a,b)\, \cos{t},\{\lambda \} )$,
the reduced map (\ref{red}) becomes
\begin{equation}
\label{red1nh}
\left(
\begin{array}{c}
a_{n+1} \\
b_{n+1}
\end{array}
\right)
=
F_n (.,\{ \lambda \} )
\left(
\begin{array}{c}
a_{n} \\
b_{n}
\end{array}
\right) ,
\end{equation}
\begin{equation}
\label{deffnh}
F_n (.,\{\lambda \} )
\left(
\begin{array}{c}
a \\
b
\end{array}
\right)
=
\left(
\begin{array}{c}
b, \\
-a+2b+r_n (a,b,\{\lambda \} )
\end{array}
\right) ,
\end{equation}
where (recall $\{ \lambda \} =(\mu ,\{\ep \} ,\{\eta  \},\{\gamma \} ,\{\kappa \})$)
\begin{equation}
\label{defrnh}
\begin{array}{rcl}
r_n (a,b,\{\lambda \} ) 
&=&
-(\Omega^2 \ep_n +\mu  (1+\ep_n ))\, b
+\Omega^2 [(1+\eta_n)(1+\gamma_n)-1] b
+\kappa_n (b - a) \\
&+&P^{\ast}W( b \cos{t} +  \Psi_2 (a ,b ,\tau_n\{\lambda \} ),\eta_n ,\gamma_n )
\end{array}
\end{equation}
and the function $W$ is defined by (\ref{defwnh}).

Since $f_n (.,\{\lambda \})$ commutes with $T$ and $T_{|X_c}=-I$, 
the map $F_n (.,\{\lambda \})$ commutes with $-I$.
We have consequently
\begin{equation}
\label{rinc}
\begin{array}{rcl}
r_n (a,b,\{\lambda \} ) 
&=&
-(\Omega^2 \ep_n +\mu  (1+\ep_n ))\, b
+\Omega^2 [(1+\eta_n)(1+\gamma_n)-1] b
+\kappa_n (b - a) \\
&+&c_1 \, b^3 + c_2 \, ab^2 + c_3 \, a^2 b    \\
&+& O( \| (a,b) \|^3 \, \| \{ \lambda \} \|_E  + \| (a,b) \|^5   ),
\end{array}
\end{equation}
where the coefficients $c_i$ need to be determined.
Now, since $r_n [a,b, (\mu ,0 ,0,0 ,0)  ] = r(a,b,\mu )$ in the 
homogeneous case (see section \ref{lower}), we have
$c_3 = -\frac{1}{2}c_{2}$ and $c_1$, $c_2$ are defined by (\ref{defc1}).
Consequently we have computed the principal part of
the reduced equation (\ref{red1nh}) in the nonautonomous case.

To obtain the normal form of (\ref{red1nh}) of order three we
now define $P(\beta )=\beta-\frac{c_{2}}{12}\beta^{3}$
and consider as in section \ref{lower}
$$
a_n = P(\alpha_n), \ \ \
b_{n}=P(\beta_{n}).
$$
This yields the normal form of (\ref{red1nh}) of order $3$
given in equation (\ref{redequanh}).

Moreover, the small amplitude solutions of
(\ref{dfpu}) have the form 
$$
y_n = (\beta_n -\frac{c_{2}}{12}\beta_n^{3}) \, \cos{t} +  \Psi_2 ( P(\beta_{n-1}) , P(\beta_{n})   ,\tau_n\{\lambda \} ),
$$
therefore the reduction function $\phi$ of theorem \ref{reductionnh}
is given by
$\phi (\alpha,\beta,\{\lambda \} )= 
-\frac{c_{2}}{12}\beta^{3}\cos{t}+
\Psi_2 ( P(\alpha ) , P(\beta )   , \{\lambda \} )
$. Note that the reduction function $\phi$ has a component along
the Fourier mode $\cos{t}$ after the normal form transformation.
 
\section{\label{epshl}Exact periodic solutions for an homogeneous lattice}

Here we consider the case of the homogeneous Klein-Gordon lattice (\ref{KGh}), which
leads us to system (\ref{dfpu}) with
$\{\ep \} =\{\eta  \}=\{\gamma \} =\{\kappa \}=0$.
Breather solutions have been proved
to exist by MacKay and Aubry \cite{ma}
for system (\ref{KGh}) with
small values of the
coupling parameter $k$ and nonresonant breather frequencies. 
Here we prove the existence of small amplitude breathers
for arbitrary values of $k$ in some cases and frequencies
close to the phonon band edges (see theorems \ref{existence1}-i),
\ref{existence2}-i) and \ref{existencebr} below). We also prove the existence
of dark breather solutions, which converge towards a nonlinear
standing wave as $n\rightarrow \pm\infty$ and have a much smaller
amplitude at the centre of the chain.

\ve

Let us start with the case $\omega \approx \Omega$ and $\Omega > 2/\sqrt{3}$
in (\ref{dfpu}).
By theorem \ref{reductionnh}, small amplitude solutions
of (\ref{dfpu}) in $H^{2}_{\#}$ are determined by
the recurrence relation (\ref{redequanh}). This recurrence 
becomes autonomous for an homogeneous lattice and 
takes the form (\ref{redfinp1}).
It is important to note that
the invariance $n\rightarrow -n$ of (\ref{dfpu})
in the homogeneous case is inherited by (\ref{redfinp1})
(see \cite{james2}, section 5.2 and theorem 2).
This invariance implies that the two-dimensional map
$(\beta_{n-1},\beta_n )\mapsto (\beta_{n},\beta_{n+1} )$
is reversible. 
Bifurcations of small amplitude homoclinic and heteroclinic
solutions have been studied in \cite{james2} (section 6.2.3)
for this class of maps. This yields the following result
for the recurrence relation (\ref{redfinp1}).

\begin{lemma}
\label{exislemma1}
Assume $\Omega > 2/\sqrt{3}$ and
$B=
\frac{\Omega^2}{8}\, (\, V^{(4)}(0)
-\frac{5}{3} (V^{(3)}(0))^2\, )
\neq 0$.
For $\mu\approx 0$, the recurrence relation
(\ref{redfinp1}) has the following solutions.

\noindent
i)
For $\mu <0$ and $B<0$,
(\ref{redfinp1}) has at least two homoclinic
solutions $\beta^{1}_{n}$, $\beta^{2}_{n}$
(and also $- \beta^{1}_{n}$, $- \beta^{2}_{n}$)
such that
$\lim_{n\rightarrow\pm\infty}{\beta^{i}_{n}}=0$.
These solutions have the symmetries
$\beta^{1}_{-n+1}=\beta^{1}_{n}$,
$\beta^{2}_{-n}=\beta^{2}_{n}$ and satisfy
$0<\beta^{i}_{n}\leq C\, |\mu |^{1/2}
\, \sigma_{1}^{-|n|}$,
with $\sigma_{1}=1+O(|\mu |^{1/2})>1$.

\noindent
ii) If $\mu$ and $B$ have the same sign,
(\ref{redfinp1}) has two symmetric
fixed points $\pm \beta^{\ast}=O(|\mu |^{1/2})$.

\noindent
iii) For $\mu >0$ and $B>0$,
(\ref{redfinp1}) has at least two heteroclinic
solutions $\beta^{3}_{n}$,
$\beta^{4}_{n}$ (and also $- \beta^{3}_{n}$, $-\beta^{4}_{n}$)
such that
$\lim_{n\rightarrow \pm\infty}{\beta^{i}_{n}}=\pm \beta^{\ast}$.
These solutions have the symmetries
$\beta^{3}_{-n+1}=-\beta^{3}_{n}$ and
$\beta^{4}_{-n}=- \beta^{4}_{n}$.
Moreover,
$\beta^{3}_{n}$, $\beta^{4}_{n}$ are
$O(\mu^{1/2})$ as $n\rightarrow \pm\infty$
and $O(\mu )$ for bounded values of $n$.
\end{lemma}

Note that for $B>0$ and
$\mu <0$ ($\mu\approx 0$),
(\ref{redfinp1})
has no small amplitude homoclinic solution to $0$.

\vspace{1ex}

For $\mu <0$, typical plots of the stable and unstable manifolds of 
the fixed point $\beta_n =0$ are shown in figure
\ref{splitting2} page \pageref{splitting2} (nonintersecting case $B>0$) and
in figures \ref{fig7} and \ref{fig9} pages \pageref{fig7} and \pageref{fig9}
(intersecting case $B<0$).

\vspace{1ex}

Theorem \ref{reductionnh} ensures that
each solution $\beta^{i}_{n}$ in lemma \ref{exislemma1}
corresponds to a
solution $y^{i}_{n}$ of (\ref{dfpu}) given by
\begin{equation}
\label{defiy}
y^{i}_{n}(t)=
\beta^{i}_{n}\cos{t}+\phi (\beta^{i}_{n-1},\beta^{i}_{n},(\mu ,0,0,0,0))
\end{equation}
with $\omega^{2}=\Omega^2+\mu $ in (\ref{dfpu}).
The following result follows
(the symmetries of $y^{i}_{n}$ are due to the
symmetries of $\beta^{i}_{n}$
described in lemma \ref{exislemma1}).

\begin{theorem}
\label{existence1}
Fix $\{\ep \} =\{\eta  \}=\{\gamma \} =\{\kappa \}=0$
in equation (\ref{dfpu}).
Assume $\Omega > 2/\sqrt{3}$ and
$b=
V^{(4)}(0)
-\frac{5}{3} (V^{(3)}(0))^2
\neq 0$.
For $\omega\approx\Omega$,
problem (\ref{dfpu}) has the following
solutions with $y_{n}\in H^{2}_{\#}$
for all $n\in \Z$.

\noindent
i)
For $\omega <\Omega$ and $b<0$,
(\ref{dfpu}) has at least two homoclinic
solutions $y^{1}_{n}$, $y^{2}_{n}$
(and also $y^{1}_{n}(t+\pi )$, $y^{2}_{n}(t+\pi )$)
such that
$\lim_{n\rightarrow\pm\infty}{\| y^{i}_{n}\|_{H^{2}_{\#}}}=0$.
These solutions
satisfy $y^{1}_{-n+1}=y^{1}_{n}$,
$y^{2}_{-n}=y^{2}_{n}$ and have the form
\begin{equation}
\label{gf1}
\textstyle{
y^{i}_{n}=\beta^{i}_{n}\,\cos t +O(|\omega -\Omega|),
}
\end{equation}
where
$0<\beta^{i}_{n}\leq C\, |\omega -\Omega|^{1/2}
\, \sigma_{1}^{-|n|}$
and $\sigma_{1}=1+O({|\omega -\Omega|}^{1/2})>1$.
Solutions $y^{1}_{n}$, $y^{2}_{n}$ correspond to
small amplitude breathers with a slow exponential decay
as $n\rightarrow \pm\infty$.

\noindent
ii) If $\omega -\Omega$ and $b$ have the same sign,
(\ref{dfpu}) admits a solution $y^{0}\in H^{2}_{\#}$ independent of $n$,
corresponding to collective in-phase oscillations.
It has the form
$y^0(t)=\beta^{\ast}\,\cos t +O(|\omega -\Omega|)$
and $\beta^{\ast}=O(|\omega -\Omega|^{1/2})$.

\noindent
iii) For $\omega >\Omega$ and $b>0$,
(\ref{dfpu}) has at least two heteroclinic
solutions $y^{3}_{n}$,
$y^{4}_{n}$ (and also $y^{3}_{n}(t+\pi )$, $y^{4}_{n}(t+\pi )$)
such that
$\lim_{n\rightarrow -\infty}
{\| y^{i}_{n}-y^{0}(t+\pi )\|_{H^{2}_{\#}}}=0$ and
$\lim_{n\rightarrow +\infty}
{\| y^{i}_{n}-y^{0}\|_{H^{2}_{\#}}}=0$.
These solutions satisfy $y^{3}_{-n+1}(t)=y^{3}_{n}(t+\pi )$
and $y^{4}_{-n}(t)=y^{4}_{n}(t+\pi )$.
Moreover, their norms
$\| y^{3}_{n}\|_{H^{2}_{\#}}$,
$\| y^{4}_{n}\|_{H^{2}_{\#}}$ are
$O((\omega -\Omega )^{1/2})$ as $n\rightarrow \pm\infty$
and $O((\omega -\Omega ) )$ for bounded values of $n$.
Solutions $y^{3}_{n}$, $y^{4}_{n}$
correspond to small amplitude dark breathers.
\end{theorem}

In addition,
note that for $b>0$ there exists no
small amplitude discrete breather $y_{n}\in H^{2}_{\#}$
with $\omega <\Omega$ and
$\omega \approx \Omega$
(since (\ref{redfinp1}) has no small amplitude solution
homoclinic to $0$).

\ve

Now we consider the case $\omega \approx \omega_c$ with
$\omega_c  =\sqrt{4+\Omega^2}$. In that case, equation 
(\ref{dfpu}) can be locally reduced to the recurrence relation
(\ref{redequanh2}), which becomes again autonomous if
$\{\ep \} =\{\eta  \}=\{\gamma \} =\{\kappa \}=0$
and has the invariance $n\rightarrow -n$.
This class of recurrence relations has been studied
in \cite{james2} (section 6.2.3, lemma 7) to which we
refer for details. In addition one can note that the
recurrence (\ref{redfinp2}) can be recast in the form
(\ref{redfinp1}) by setting $\beta_n = (-1)^n \tilde{\beta}_n$.
The following result for
the recurrence relation (\ref{redfinp2}) follows.

\begin{lemma}
\label{exislemma2}
Assume 
$\tilde{B}=
\frac{\Omega^2}{8}\, 
(\,
V^{(4)}(0)
+{(V^{(3)}(0))}^{2}
(
\frac{\Omega^2}{16+3\Omega^2}-2
)\, ) \neq 0$.
For $\mu\approx 0$, the recurrence relation
(\ref{redfinp2}) has the following solutions.

\noindent
i)
For $\mu >0$ and $\tilde{B}>0$,
(\ref{redfinp2}) has at least two homoclinic
solutions $\beta^{1}_{n}$, $\beta^{2}_{n}$
(and also $- \beta^{1}_{n}$, $- \beta^{2}_{n}$)
such that
$\lim_{n\rightarrow\pm\infty}{\beta^{i}_{n}}=0$.
These solutions have the symmetries
$\beta^{1}_{-n+1}=-\beta^{1}_{n}$,
$\beta^{2}_{-n}=\beta^{2}_{n}$ and satisfy
$0<(-1)^n \beta^{i}_{n}\leq C\, \mu^{1/2}
\, |\sigma_{1}|^{-|n|}$,
with $|\sigma_{1}|=1+O(|\mu |^{1/2})>1$.

\noindent
ii) If $\mu$ and $\tilde{B}$ have the same sign,
(\ref{redfinp2}) has a period $2$ solution
$\beta^{0}_{n}=(-1)^{n}\beta^{\ast}$,
with $\beta^{\ast}=O( |\mu |^{1/2})$.  

\noindent
iii) For $\mu <0$ and $\tilde{B}<0$,
(\ref{redfinp2}) has at least two heteroclinic
solutions $\beta^{3}_{n}$,
$\beta^{4}_{n}$ (and also $- \beta^{3}_{n}$, $-\beta^{4}_{n}$)
such that
$\lim_{n\rightarrow \pm\infty}
{| \beta^{i}_{n}\mp \beta^{0}_{n} |}=0$.
These solutions have the symmetries
$\beta^{3}_{-n+1}=\beta^{3}_{n}$ and
$\beta^{4}_{-n}=- \beta^{4}_{n}$.
Moreover,
$\beta^{3}_{n}$, $\beta^{4}_{n}$ are
$O(|\mu |^{1/2})$ as $n\rightarrow \pm\infty$
and $O(|\mu |)$ for bounded values of $n$.
\end{lemma}

In addition, for $\tilde{B}<0$ and
$\mu >0$ ($\mu\approx 0$) problem (\ref{redfinp2})
has no small amplitude homoclinic solution to $0$.
As above, the solutions of the reduced recurrence relation
provided by lemma \ref{exislemma2}
yield the following solutions of (\ref{dfpu}).

\begin{theorem}
\label{existence2}
Fix $\{\ep \} =\{\eta  \}=\{\gamma \} =\{\kappa \}=0$
in equation (\ref{dfpu}).
Assume
$\tilde{b}=
V^{(4)}(0)
+{(V^{(3)}(0))}^{2}
(
\frac{\Omega^2}{16+3\Omega^2}-2
) \neq 0$.
For $\omega \approx \omega_c =\sqrt{4+\Omega^2}$,
problem (\ref{dfpu}) has the following
solutions with $y_{n}\in H^{2}_{\#}$
for all $n\in \Z$.

\noindent
i)
For $\omega > \omega_c$ and $\tilde{b}>0$,
(\ref{dfpu}) has at least two homoclinic
solutions $y^{1}_{n}$, $y^{2}_{n}$
(and also $y^{1}_{n}(t+\pi )$, $y^{2}_{n}(t+\pi )$)
such that
$\lim_{n\rightarrow\pm\infty}{\| y^{i}_{n}\|_{H^{2}_{\#}}}=0$.
These solutions
satisfy $y^{1}_{-n+1}(t)=y^{1}_{n}(t+\pi )$,
$y^{2}_{-n}=y^{2}_{n}$ and have the form
\begin{equation}
\label{gf2}
\textstyle{
y^{i}_{n}=\beta^{i}_{n}\,\cos t +O(|\omega -\omega_c |),
}
\end{equation}
where
$0<(-1)^n \beta^{i}_{n}\leq C\, (\omega -\omega_c )^{1/2}
\, |\sigma_{1}|^{-|n|}$
and $|\sigma_{1}|=1+O({(\omega -\omega_c )}^{1/2})>1$.
Solutions $y^{1}_{n}$, $y^{2}_{n}$ correspond to
small amplitude breathers with a slow exponential decay
as $n\rightarrow \pm\infty$.

\noindent
ii) If $\omega -\omega_c $ and $\tilde{b}$ have the same sign,
(\ref{dfpu}) admits a solution 
$y^{0}_{n}$ being
$2$-periodic in $n$, corresponding to 
out-of-phase oscillations.
It has the form $y^{0}_{n}(t)=y(t+n\pi )$
with $y(t)=\beta^{\ast}\,\cos t +O(|\omega -\omega_{c}|)$ 
($y\in H^{2}_{\#}$)
and $\beta^{\ast}=O(|\omega -\omega_{c}|^{1/2})$. 

\noindent
iii) For $\omega <\omega_c$ and $\tilde{b}<0$,
(\ref{dfpu}) has at least two heteroclinic
solutions $y^{3}_{n}$,
$y^{4}_{n}$ (and also $y^{3}_{n}(t+\pi )$, $y^{4}_{n}(t+\pi )$)
such that
$\lim_{n\rightarrow -\infty}
{\| y^{i}_{n}-y^{0}(t+\pi )\|_{H^{2}_{\#}}}=0$ and
$\lim_{n\rightarrow +\infty}
{\| y^{i}_{n}-y^{0}\|_{H^{2}_{\#}}}=0$.
These solutions satisfy $y^{3}_{-n+1}=y^{3}_{n}$
and $y^{4}_{-n}(t)=y^{4}_{n}(t+\pi )$.
Moreover, their norms
$\| y^{3}_{n}\|_{H^{2}_{\#}}$,
$\| y^{4}_{n}\|_{H^{2}_{\#}}$ are
$O(|\omega -\omega_c |^{1/2})$ as $n\rightarrow \pm\infty$
and $O(|\omega -\omega_c |)$ for bounded values of $n$.
Solutions $y^{3}_{n}$, $y^{4}_{n}$
correspond to small amplitude dark breathers.
\end{theorem}

In addition, for $\tilde{b}<0$ there exists no
small amplitude discrete breather $y_{n}\in H^{2}_{\#}$
with $\omega >\omega_c$ and
$\omega \approx \omega_c$.

\vspace{1ex}

It is worthwhile mentioning that approximate breather solutions
of (\ref{dfpu}) can be obtained in the form of modulated plane
waves, using multiscale expansions (see \cite{gian} and references
therein), where the error can be controlled over finite time intervals.
The envelope of a modulated wave satisfies the nonlinear Schr\"odinger (NLS) 
equation, and does not propagate along the chain when a plane wave with wavenumber
$q=0$ or $q=\pi$ is modulated (its group velocity vanishes).
In these two cases the NLS equation is focusing 
(i.e. time-periodic and spatially localized solutions exist)
when $b<0$ and $\tilde{b}>0$ respectively, which coincides
(according to theorems \ref{existence1} and \ref{existence2})
with the parameter values for which exact breathers exist.

In addition, as shown in reference \cite{flach}
the condition $b<0$ leads to the instability of nonlinear 
standing waves with wavenumber $q=0$.
If periodic boundary conditions are considered,
these standing waves become unstable 
above a critical energy via a tangent bifurcation.
When the lattice period tends to infinity,
the energy threshold goes to $0$ and
bifurcating solutions are slowly spatially modulated.
The same result has been obtained for standing waves with $q=\pi$
when $V$ is even and $\tilde{b}>0$.

\vspace{1ex}

In what follows we reformulate the results with respect to the
unscaled original system (\ref{KGh}). 
For conciseness we only
describe breather bifurcations, but conditions for dark breather
bifurcations are easily deduced from theorems \ref{existence1} and \ref{existence2}.
We express the condition $\tilde{b} >0$ of theorem \ref{existence2}
in a different way using the relation
$$
\tilde{b}=b-\frac{16}{3} \, \frac{(V^{(3)}(0))^2}{16+3\Omega^2}.
$$
In addition, as the rescaled potential
$\tilde{V}$ of (\ref{dfpu}) is replaced by the original potential
$V$ of (\ref{KGh}), coefficients $b$ and $\tilde{b}$ are simply
replaced by $h=a^{-2} b$ and $\tilde{h}=a^{-2} \tilde{b}$.

\begin{theorem}
\label{existencebr}
Consider the Klein-Gordon lattice (\ref{KGh}), where the
on-site potential $V$ satisfies $V^\prime (0)=0$, 
$V^{\prime\prime} (0)=1$ and $m,d,a,k >0$.
Assume
$h=V^{(4)}(0)-\frac{5}{3} (V^{(3)}(0))^2 \neq 0$ and 
note $\Omega^2 = a^2 d /k$, $\omega_{min}^2=a^2 d /m$,
$\omega_{max}^2=(a^2 d +4k) /m$ and
$H=V^{(4)}(0)-2(V^{(3)}(0))^2 $.

\vspace{1ex}

\noindent
i) If $h<0$ and $\Omega^2 > 4/3$, system (\ref{KGh}) admits two
families of breather solutions $x^{1}_{n}$, $x^{2}_{n}$
parametrized by their frequency $\omega $
(in addition to phase shift), where $\omega\approx\omega_{min}$
and $\omega <\omega_{min}$. These solutions satisfy
$x^{1}_{-n+1}=x^{1}_{n}$ and $x^{2}_{-n}=x^{2}_{n}$
and decay exponentially as $n\rightarrow \pm \infty$.
As $\omega\rightarrow \omega_{min}$,
the amplitude of oscillations and the exponential rate 
of decay are $O(|\omega -\omega_{min}|^{1/2})$.
The breather profile is a slow modulation of a linear
mode with wavenumber $q=0$.

\vspace{1ex}

\noindent
ii) If $h>0$ and $\Omega^2 > -16 H / (3h)$,
system (\ref{KGh}) admits two
families of breather solutions $x^{1}_{n}$, $x^{2}_{n}$
parametrized by their frequency $\omega $
(in addition to phase shift), where $\omega\approx\omega_{max}$
and $\omega >\omega_{max}$. These solutions satisfy
$x^{1}_{-n+1}(t)=x^{1}_{n}(t+\pi / \omega )$,
$x^{2}_{-n}=x^{2}_{n}$
and decay exponentially as $n\rightarrow \pm \infty$.
As $\omega\rightarrow \omega_{max}$,
the amplitude of oscillations and the exponential rate 
of decay are $O(|\omega -\omega_{max}|^{1/2})$.
The breather profile is a slow modulation of a linear
mode with wavenumber $q=\pi$.
\end{theorem} 

To interpret the conditions on the on-site potential $V$ in
properties i) and ii), it is interesting to note that $V$ is
soft for $h<0$ and hard for $h>0$ near the origin
(i.e. the period of small oscillations in
this potential respectively increases or decreases with amplitude).
The condition on $\Omega$ in property i) corresponds to 
a nonresonance condition, i.e. it ensures that no multiple of
$\omega$ lies in the phonon band $[\omega_{min},\omega_{max}]$
for $\omega \approx \omega_{min}$.
The condition on $\Omega$ in property ii) is of different nature
and is equivalent (with the condition $h>0$) to fixing $\tilde{h}>0$.

Discrete breathers were known to exist in Klein-Gordon lattices
for small coupling $k$ after the work of MacKay and Aubry \cite{ma}.
Theorem \ref{existencebr} considerably enlarges the domain of
breather existence, with the limitation that it
only provides small amplitude solutions. In particular,
it is interesting to note that small amplitude breathers of property ii)
exist for all values of $k$ if $H>0$.

\section{\label{nfail}Normal form analysis for inhomogeneous lattices}

According to theorem \ref{reductionnh},
small amplitude solutions of
(\ref{dfpu}) in $H^{2}_{\#}$ are described
(for small inhomogeneities and frequencies close
to the phonon band edges) by finite-dimensional nonautonomous
recurrence relations. In what follows we only consider
the case $\omega \approx \Omega$, the situation when
$\omega \approx \sqrt{4+\Omega^2}$ yielding
similar phenomena. 
At leading order, the reduced recurrence relation
(\ref{redequanh}) can be approximated by
\begin{equation}
\label{redequapp}
\beta_{n+1}-2\beta_{n}+\beta_{n-1}=
(\Omega^2\eta_n (1+\gamma_n )-(\Omega^2 +\mu ) \ep_n +\Omega^2 \gamma_n -\mu )\, \beta_n
+\kappa_n (\beta_n -\beta_{n-1} )
+B\, \beta_n^{3}.
\end{equation}
Different kinds of techniques
can be employed to obtain homoclinic solutions of (\ref{redequapp}). 
One can use variational methods
for asymptotically periodic sequences \cite{pankov} (see also \cite{wein}
in the homogeneous case), 
or proceed by perturbation near an
uncoupled limit (also denoted anti-continuous or anti-integrable limit)
where $\Omega$ and $B$ are large (see \cite{aa,aubint,alfimov} and 
section 9 of \cite{ma}).
Existence results of localized solutions 
are also available in \cite{afnls} for disordered defect sequences.
Another approach is to start from a known uniformly hyperbolic homoclinic solution
in the homogeneous case, which persists for small inhomogeneities
by the implicit function theorem, and obtain estimates for
defect sizes allowing persistence (see the technique developed by
Bishnani and MacKay \cite{bim}). Interesting related results 
on the structural stability of discrete dynamical systems under
nonautonomous perturbations can be
found in reference \cite{franks}.

With a different point of view, 
we develop here a dynamical system technique, valid for
a finite number of defects, which allows to analyze
{\em bifurcations} of homoclinic solutions as defects
are varied (see sections \ref{single}
and \ref{many}).
For an isolated defect we put in evidence, near critical defect values,
bifurcations of new homoclinic solutions (having no counterpart in the
homogeneous system) or disappearance of homoclinic solutions 
existing in the homogeneous case. Our method is also generalized
to a finite number of defects, with the counterpart that (\ref{redequapp})
is modified by suitable higher order terms depending on the defect sequence
(this procedure only provides approximate solutions of (\ref{redequapp})).
However this does not constitute a strong limitation since the full reduced
equation (\ref{redequanh}) is itself a higher order perturbation of (\ref{redequapp}).
Note that equation (\ref{redequapp}) is valid (according to theorem \ref{reductionnh})
for small defect sizes and $\mu \approx 0$, where the parameter $\mu$
determines for $\mu <0$ the (weak) degree of hyperbolicity of the
fixed point $\beta_n =0$ in the homogeneous case. Our analysis does not
impose conditions on the relative sizes of these parameters. 

In order to obtain exact breather solutions of
(\ref{dfpu}) via theorem \ref{reductionnh}, it would be necessary to
proceed in two steps. The first step is the one described above, where
exact or approximate homoclinic solutions are obtained for the
truncated problem (\ref{redequapp}).
The second step is to show
that these solutions 
persist for the complete equation
(\ref{redequanh}) as higher order terms are added.
For this purpose a typical procedure would be to solve
(\ref{redequanh}) using the contraction mapping theorem
in the neighbourhood of an exact or approximate
solution of (\ref{redequapp}).

In this paper
we shall not examine the persistence of solutions
for the complete equation (\ref{redequanh}).
Instead we shall later compare 
approximate solutions $y_n (t)\approx \beta_n \, \cos{t}$ 
(deduced from (\ref{redequapp}) and theorem \ref{reductionnh})
to numerically computed
solutions of the original problem (\ref{dfpu})
(see section \ref{numerical}).
This will allow us to study the validity of approximation (\ref{redequapp})
far from the small amplitude limit and as
inhomogeneities become larger.

Note that other interesting bifurcations can exist when
impurities act at a purely nonlinear level
(see \cite{sukh,kevr} for some examples in spatially discrete
or continuous systems). This would correspond to the
situation when the on-site potential in (\ref{exkg})
has an harmonic part independent on $n$, whereas
higher order terms are inhomogeneous. The subsequent analysis
of the reduced recurrence relation would be quite different,
and in particular the method developed in sections
\ref{single} and \ref{many} (based on a linear deformation of the
unstable manifold) would not apply.

\subsection{\label{single}Case of a single mass defect}

We start with the simplest case when the coefficients of (\ref{redequapp}) are
constant, except at $n=0$ where their value changes.
To fix the idea we consider the case of a single mass defect in equation (\ref{KG}),
i.e. $D_n =d$, $K_n =k$, $A_n=a$, $M_n = 1+m_n$, $m_n = m_0 \delta_{n0}$.
The case when all lattice parameters are allowed to vary over a
finite number of sites will be considered in section \ref{many}. 

\ve

For equation (\ref{dfpu}) the above assumption
yields $\eta_n =\gamma_n =\kappa_n =0$ and $\epsilon_n = m_0 \delta_{n0}$.
Equation (\ref{redequapp}) reads (recall $\omega^2 = \Omega^2 +\mu$)
\begin{equation}
\label{redequapp2}
\beta_{n+1}-2\beta_{n}+\beta_{n-1}=
-(\omega^2 m_0 \delta_{n0} +\mu )\, \beta_n
+B\, \beta_n^{3}.
\end{equation}
Setting $\beta_{n-1}=\alpha_n$ and
$U_n =(\alpha_n ,\beta_n)^T$, equation (\ref{redequapp2}) can be rewritten
\begin{equation}
\label{red11bis}
U_{n+1}=G_{\omega}(U_n)-\omega^2 m_0 \delta_{n0}
\left(
\begin{array}{c}
0 \\
\beta_n
\end{array}
\right)
\end{equation}
where 
\begin{equation}
\label{defmap}
G_{\omega}(U_n)
=
\left(
\begin{array}{c}
\beta_n  \\
-\alpha_n +2\beta_n+(\Omega^2 -\omega^2 ) \beta_n +B \beta_n^3
\end{array}
\right) .
\end{equation}
One has in particular
\begin{equation}
\label{defmap0}
U_1 =
A(\omega ,m_0)
G_{\omega}(U_0)
\end{equation}
where the linear transformation
\begin{equation}
\label{defA}
A(\omega ,m_0)=
\left(
\begin{array}{cc}
1 & 0  \\
-\omega^2 m_0 & 1
\end{array}
\right)
\end{equation}
corresponds to a linear shear.
Note that the axis $\alpha =0$ consists of
fixed points of $ A(\omega ,m_0)$.

It is worthwhile to notice that the map
$G_{\omega}$ is reversible under the symmetry
$R\, : \, (\alpha ,\beta )\mapsto (\beta ,\alpha )$, i.e.
$G_{\omega} \circ R = R \, G_{\omega}^{-1}$.
In other words, if $U_n$ is a solution of
(\ref{red11bis}) for $m_0 =0$ then $R\, U_{-n}$ is also solution.
This property is due to the fact that equation (\ref{redequapp2}) 
with $m_0 =0$
has the invariances $n\rightarrow n+1$ and $n\rightarrow -n$.
Obviously the latter invariance still exists for 
$m_0 \neq 0$. Consequently, for all $m_0 \in \mathbb{R}$,
if $U_n$ is a solution of
(\ref{red11bis}) then $R\, U_{-n+1}$ is also solution.

Now we shall use a geometrical argument
to find homoclinic orbits to $0$ for equation (\ref{redequapp2}).
In the sequel we consider
the stable manifold $W^s(0)$ of the fixed point
$(\alpha ,\beta ) =0$ of $G_\omega$,
and its unstable manifold $W^u(0)$, both existing for
$\omega < \Omega$. The following result follows immediately.

\begin{lemma}
\label{lemint}
For $0<\omega < \Omega$, equation (\ref{redequapp2}) possesses
an homoclinic orbit to $0$ if and only if $W^s(0)$ and $ A(\omega ,m_0)W^u(0)$ intersect.
\end{lemma}

In addition it is useful to notice that $W^s(0)$ and $W^u(0)$ are exchanged
by the reversibility symmetry $R$.

\subsubsection{\label{linc}Linear case}

As a simple illustration, consider the linear case when $V$ is harmonic, in which
$B=0$.
Equation (\ref{redequapp2}) reads
\begin{equation}
\label{redequalin}
\beta_{n+1}-2\beta_{n}+\beta_{n-1}=
-(\omega^2 m_0 \delta_{n0} +\mu )\, \beta_n.
\end{equation}
In that case, $W^s(0)$ and $W^u(0)$ correspond respectively to the stable and the unstable
eigenspace of a linear mapping in $\mathbb{R}^2$.
The situation is sketched in figure \ref{splitting} below.
The corresponding stable eigenvalue
$\sigma \in (0,1)$ is given by
\begin{equation}
\label{decay}
\sigma = 1-\frac{\mu}{2}-\frac{1}{2}(\mu^2 -4\mu )^{1/2}, \ \ \
\mu =\omega^2 -\Omega^2 <0 ,
\end{equation}
and $W^s(0)$ is the line $\beta = \sigma \alpha$,
$W^u(0)$ being the line $\beta = \sigma^{-1} \alpha$.
For fixed $\omega < \Omega$, $ A(\omega ,m_0)$ maps the unstable eigenspace on the stable eigenspace
if and only if $m_0 > 0$ (mass is increased at the defect) and
\begin{equation}
\label{condlin}
m_0 = m_l (\omega ),
\end{equation}
where
\begin{equation}
\label{masslin}
m_l (\omega ) = \frac{1}{\omega^2}(\sigma^{-1} -\sigma ).
\end{equation}

\begin{figure}[!h]
\begin{center}
\includegraphics[width=6.0cm]{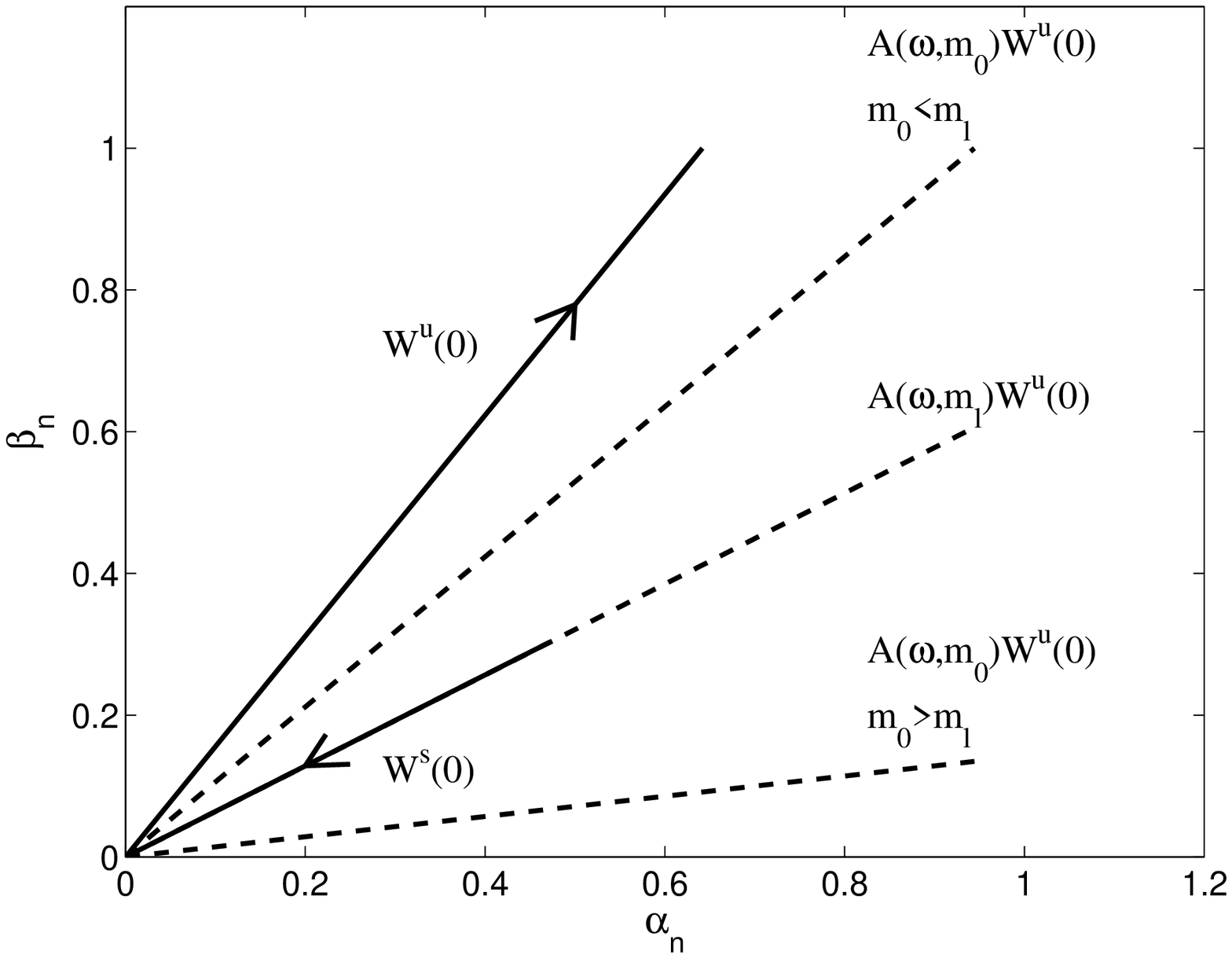}
\includegraphics[width=6.0cm]{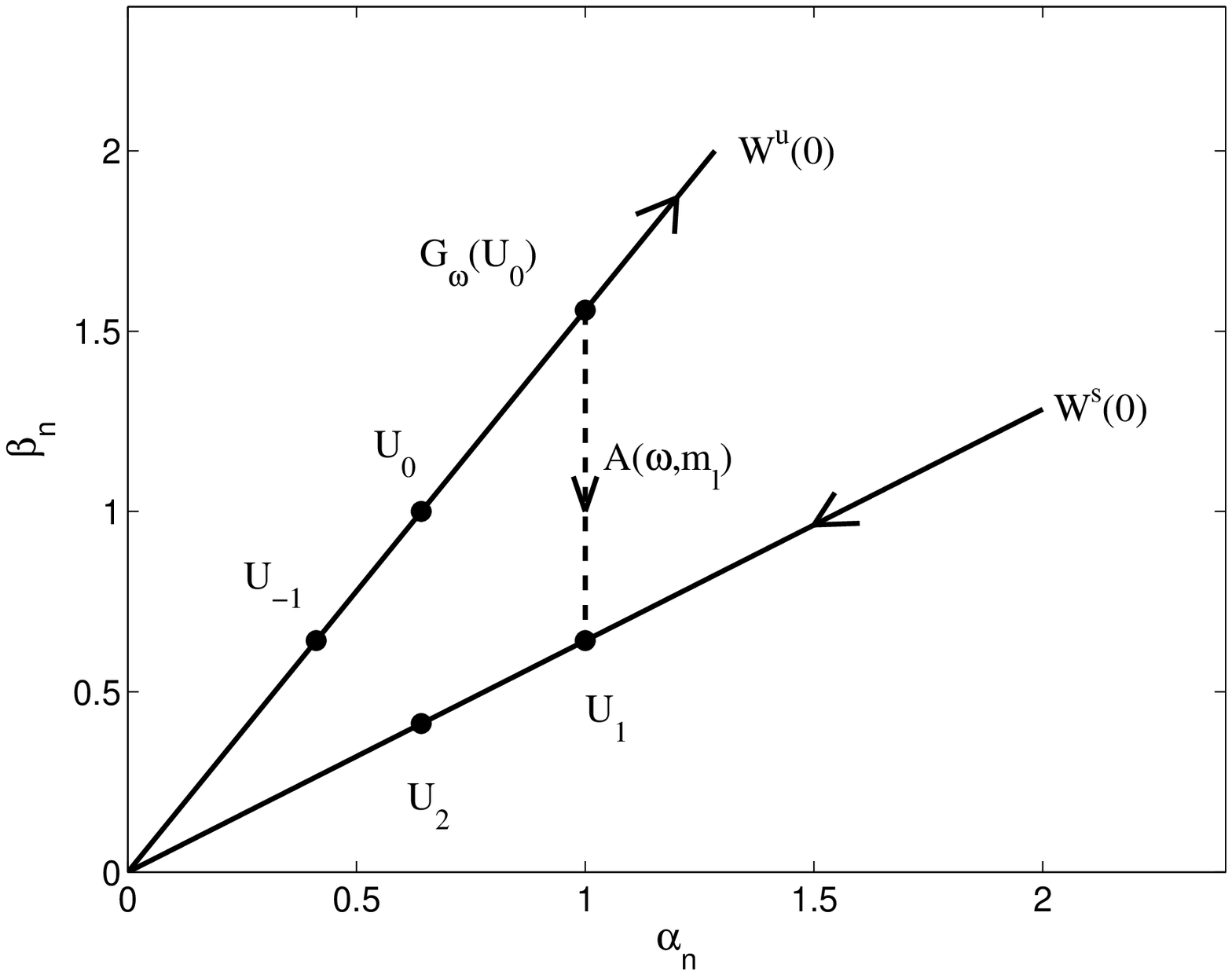}
\caption{Linear case ($B=0$). (Left panel) Stable manifold (in the half plane $\alpha >0$)
and images of the unstable manifold by $A(\omega,m_0)$ for $m_0=0.005$,
$m_0=m_l$ and $m_0=m_l+0.005$. (Right panel) Homoclinic orbit to 0
for $m_0=m_l$. In both panels we have fixed $\Omega = 10$, $\omega=9.99$, which implies
$m_l=0.0092$.}\label{splitting}
\end{center}
\end{figure}

Now keeping fixed $m_0 > 0$, condition (\ref{condlin}) can be rewritten
$\omega =\omega_l (m_0)$, where (for $\omega < \Omega$)
$$
\omega_l^2 = \frac{1}{1-m_0^2}[\, \Omega^2+2-(4+m_0^2 \Omega^2 (\Omega^2+4))^{1/2}  \, ],
\ \ \ m_0 \neq 1,
$$
$$
\omega_l^2 = \frac{1}{2}(\Omega^2+2)- \frac{2}{\Omega^2+2},
\ \ \ m_0 = 1.
$$
The solutions of (\ref{redequalin}) homoclinic to $0$ are spanned by
$\beta_n =\sigma^{|n|}$, and the corresponding solutions of (\ref{dfpu}) 
in the linear case read
$y_n(t)=\beta_n \cos{t}$ with $\omega =\omega_l (m_0)$.
One recovers a classical result, i.e.
if mass is increased at the defect then the linear localized mode frequency
lies below the phonon band and its frequency is given by $\omega_l$.

Now let us consider the effects of nonlinear terms.
For this purpose we start with the simplest case of a hard potential, i.e. $B>0$.
The situation when $B<0$ is far more complex and will be investigated later.

\subsubsection{\label{defmode}Nonlinear defect modes for hard on-site potentials}

If $B>0$, $W^s(0)$ and $W^u(0)$ do not intersect (except at the origin)
for $0< \omega <\Omega$.
Indeed, one can show by induction that $|\beta_n | > |\beta_{n-1}| >0$
for any nontrivial orbit on $W^u(0)$, which implies
$W^u(0)$ lies inside the sector formed by the lines $\alpha = \beta$ and $\alpha =0$.
In the same way, $W^s(0)$ lies inside the sector formed by the lines $\alpha = \beta$ and 
$\beta =0$ hence it does not intersect $W^u(0)$.
The above property also implies that $W^u(0)$ can be defined (globally)
as the graph $\alpha = g(\beta )$ of an increasing function $g$, and
the same holds true for $W^s(0)=R\, W^u(0)$ on which $\beta = g(\alpha )$.

\begin{figure}[!h]
\begin{center}
\includegraphics[width=6.0cm]{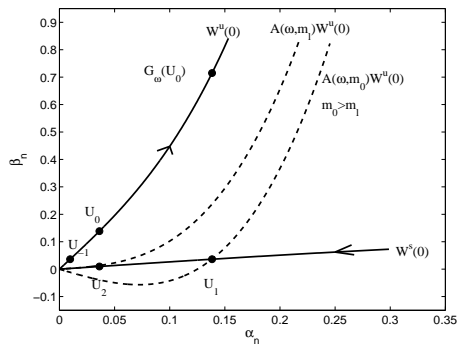}
\caption{Case $B>0$ and $\omega<\Omega$. Stable and unstable
manifolds (in the half plane $\alpha >0$), 
and image of the unstable manifold by $A(\omega,m_0)$ for
$m_0=m_l$ and $m_0=0.05 > m_l$. We have fixed $\Omega = 10$
and $\omega=9.9$.} \label{splitting2}
\end{center}
\end{figure}

For fixed $\omega \in (0,\Omega )$, the local unstable manifold can be
approximated by $\alpha = g(\beta )=\sigma \beta + b\, \beta^3 + O(|\beta |^5 )$,
with $b=\sigma^2 (\sigma^2 -\sigma^{-2})^{-1}B <0$
(this coefficient can be computed by a classical identification procedure,
using the fact that $W^u(0)$ is invariant under $G_\omega$). 
Consequently, 
$W^s(0)$ and $W^u(0)$ have the local shape represented in figure \ref{splitting2}.
The same situation occurs in the limit  $\omega \approx \Omega$ 
(one can locally approximate the map $G_\omega$
up to any order in $U , \mu$ using the time-one map of an integrable flow \cite{arrow},
which allows to determine the shape of $W^s(0)$ and $W^u(0)$
close to $U_n =0$).

In the case when $m_0 \leq 0$, the curves $W^s(0)$ and $ A(\omega ,m_0)W^u(0)$ do not intersect
($ A(\omega ,m_0)W^u(0)$ remains inside the sector formed by the lines 
$\alpha = \beta$ and $\alpha =0$).
However, $W^s(0)$ and $ A(\omega ,m_0)W^u(0)$ intersect if $m_0 >0$ provided
\begin{equation}
\label{inthard}
m_0 > m_l (\omega ),
\end{equation}
which means there exists an orbit homoclinic to $0$ for equation (\ref{redequapp2}).
This property is clear for $m_0 \approx m_l$ where there exists a unique
intersection point (in the half plane $\alpha >0$) close to $U=0$
due to the local shape of $W^s(0)$ and $W^u(0)$.
Moreover, we
numerically find a unique intersection
point for all values of $m_0$ satisfying (\ref{inthard}).

Condition (\ref{inthard}) is equivalent to
\begin{equation}
\label{inthard2}
\omega_l < \omega < \Omega.
\end{equation}
The amplitude of the homoclinic orbit is $O(\sqrt{\omega -\omega_l})$
as $m_0$ is fixed and $\omega \rightarrow \omega_l$
(at the limit $ A(\omega ,m_0)W^u(0)$ and $W^s(0)$ become tangent at the origin),
and its spatial decay rate $\sigma$ is given by equation (\ref{decay}).
This homoclinic orbit corresponds for equation (\ref{dfpu}) to
a nonlinear analogue of the above mentioned linear localized mode.
This solution can be approximated by
$y_n (t) \approx \beta_n \cos{t}$ for $m_0 \approx 0$ and the
frequency $\omega$ varies with amplitude contrarily to
the linear case. Note that the existence of this nonlinear
defect mode can be obtained in a standard way from equation
(\ref{dfpu}) in the small amplitude limit, 
using the Lyapunov centre theorem (in its
infinite-dimensional version). The Lyapunov family of 
periodic orbits exists when no multiple of $\omega_l$ lies
in the phonon band $[\Omega , (\Omega^2+4)^{1/2}]$, 
which is true in our parameter regime
where $\Omega > 2 / \sqrt{3}$ and $m_0 \approx 0$.

Lastly, let us notice that the above homoclinic orbit possesses
the symmetry $\beta_{-n}=\beta_n$, or equivalently
$R\, U_{-n+1}=U_n$. It suffices to check the latter relation
for $n=0$ to prove it for any $n$, since both solutions $R\, U_{-n+1}$ and
$U_n$ coincide if they satisfy the same initial contidion.
Since $U_0$ lies on the unstable manifold we have $\alpha_0 = g(\beta_0 )$,
and in the same way  $\beta_1 = g(\alpha_1 )$ since $U_1$ lies on the stable manifold.
Since by definition $\alpha_1 = \beta_0$, this implies $\alpha_0 = \beta_1$ and thus
$R\, U_{1}=U_0$.
Using the properties
$U_1 = G_\omega (U_0)-\omega^2 m_0 (0,\beta_0)^T$ and $R\, U_{1}=U_0$ we also deduce
the relations
\begin{eqnarray}
\label{relat1}
2\alpha_0 &=& [2+\Omega^2 -\omega^2 (m_0 +1)]\beta_0 + B\, \beta_0^3 \\
\label{relat2}
2\beta_1 &=& [2+\Omega^2 -\omega^2 (m_0 +1)]\alpha_1 + B\, \alpha_1^3,  
\end{eqnarray}
which are useful in particular
for the numerical computation of $U_0 , U_1$.

\subsubsection{Nonlinear defect mode with algebraic decay}

In the situation of section \ref{defmode} ($B>0$), the case when
$m_0$ is fixed and
$\omega \rightarrow \Omega$ deserves a special attention. Indeed,
the homoclinic orbit $(\alpha_n ,\beta_n)$ converges in this limit towards a solution having
an algebraic decay as $n\rightarrow \pm \infty$. 

\begin{figure}[!h]
\begin{center}
\includegraphics[width=6.0cm]{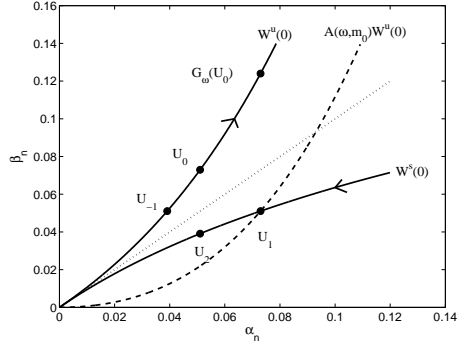}
\caption{Case $B>0$ and $\omega=\Omega$. Stable and unstable
manifolds (in the half plane $\alpha >0$), 
and image of the unstable manifold by $A(\omega,m_0)$ for
$m_0=0.05$. We have fixed $\Omega =10$ in this example.
} \label{splitting3}
\end{center}
\end{figure}

More precisely, if $\omega =\Omega$ and $m_0 >0$ then
$W^s(0)$ and $ A(\omega ,m_0)W^u(0)$ intersect at a
unique point $(\alpha_1 ,\beta_1)$ in the half plane 
$\alpha >0$ (see figure \ref{splitting3}).
This can be checked analytically for
$m_0 \approx 0$ and $(\alpha , \beta )\approx 0$, 
since the unstable manifold can be locally parametrized by
$\alpha = \beta  - (B/2)^{1/2} \beta^2 + O(|\beta|^3)$
in the half plane $\alpha >0$
(this expansion follows from a classical identification procedure).
Using this relation for $(\alpha_0 , \beta_0)$ in conjunction
with (\ref{relat1}) we find as $m_0 \rightarrow 0$
\begin{equation}
\label{ampli}
\beta_0 = {\Omega^2}(2B)^{-1/2} m_0 + O(m_0^2).
\end{equation}
Note that in this non-hyperbolic case,
the function $g$ having the unstable manifold as its
graph is not $C^2$ at $\beta =0$
(in the half plane $\alpha <0$, one has
$\alpha = \beta  + (B/2)^{1/2} \beta^2 + O(|\beta|^3)$
on the local unstable manifold).
Far from the small amplitude limit,
we have also checked numerically
the existence and uniqueness of 
$W^s(0)\, \cap \, A(\omega ,m_0)W^u(0)$
in the half plane $\alpha >0$.

Consequently, there exists
a solution of (\ref{redequapp2}) homoclinic to $0$
for $\omega = \Omega$ and $m_0>0$. This solution 
has an algebraic decay due to the fact that the origin is
not any more an hyperbolic fixed point for $\omega = \Omega$.
One can approximate the solution profile for
$m_0 \approx 0$, using the fact that (\ref{redequapp2}) 
admits at both sides of $n=0$ a continuum limit.
Indeed, setting 
\begin{equation}
\label{ansatz}
\beta_n \approx m_0 \beta (x), \ \ \ x = m_0 n,
\end{equation}
one obtains
the following differential equation 
$$
\frac{d^2\beta}{dx^2}=B\beta^3,\ \ \ x \in (-\infty , 0) \mbox{ or } (0,+\infty),
$$
from which we deduce (multiply by $\beta^\prime$ and integrate)
\begin{equation}
\label{continuum}
\frac{d\beta}{dx}=-\mbox{sign}(x)\, (B/2)^{1/2}\, \beta^2
\end{equation}
since $\beta (x)\rightarrow 0$ as $x\rightarrow \pm\infty$.
Using (\ref{continuum}) and (\ref{ampli}) one obtains the
following approximation of the homoclinic solution 
for $m_0 \approx 0$ 
\begin{equation}
\label{approx1}
\beta_n \approx m_0 \, \sqrt{\frac{2}{B}}\,  ({m_0 |n|+\frac{2}{\Omega^2}})^{-1}.
\end{equation}

This yields an approximate solution $y_n (t) \approx \beta_n \cos{t}$ of (\ref{dfpu}),
corresponding to a breather with an algebraic decay and a frequency $\omega = \Omega$
at the bottom of the phonon band.

\subsubsection{\label{csp}Case of soft on-site potentials}

Now let us make some considerations on the case $B<0$ (soft on-site potential $V$)
which is far more complex.

\ve

For $m_0 =0$, $\mu <0$ and $B<0$, equation (\ref{redequapp2}) 
possesses homoclinic solutions to $0$. This case has been analyzed
in several references with different viewpoints and for different parameter ranges, 
see e.g. \cite{hennig,alfimov,pankov,wein,james2}.

\begin{figure}[h!]
\begin{center}
\includegraphics[width=6.0cm]{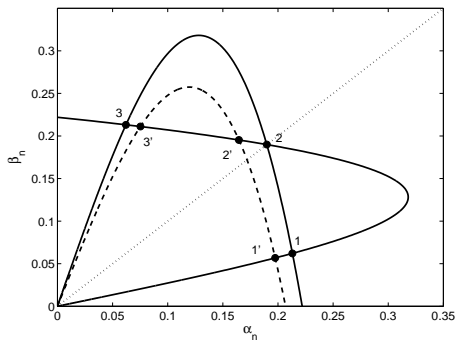}
\caption{First intersection points  between the stable and unstable
manifolds for parameters $\omega=9.9$ ($\mu=-1.99$) and $B=-75$. The
dashed line depicts the image of the unstable manifold by the linear
shear $A(\omega,m_0)$ for $m_0=0.005$. \label{fig7}}
\end{center}
\end{figure}

The dynamics of the map $G_\omega$ is rather complex due
to the fact that the stable and unstable manifolds of the origin
intersect transversally in general (see figures \ref{fig7} and \ref{fig9}).
This implies the existence of an invariant Cantor set
on which some iterate $G_{\omega}^{p}$ is topologically conjugate
to a full shift on $N$ symbols \cite{arrow}, which yields a rich variety
of solutions and in particular
an infinity of homoclinic orbits to $0$.

Among these different homoclinic orbits one can point out two particular ones
$U_n^i = (\alpha_n^i , \beta_n^i)^T$ ($i=1,2$), 
corresponding for the Klein-Gordon chain
to breathers solutions with a single hump near $n=0$
(site-centered or bond-centered). 
These solutions have been described in lemma \ref{exislemma1}
and theorem \ref{existence1} in the small amplitude limit.
The corresponding homoclinics $U_n^1 , U_n^2$ are reversible,
i.e. they satisfy $R\, U_{-n+1}^2=U_n^2$ ($\beta_{-n}^2=\beta_n^2$)
and $R\, U_{-n+2}^1=U_n^1$
($\beta_{-n+1}^1=\beta_n^1$).
In figure \ref{fig7},
the point with label $2$ lying on the axis $\alpha = \beta$ 
corresponds to $U_1^1$, and the points with labels $3,1$ 
correspond to $U_0^2, U_1^2$ respectively.
Obviously any translation of $U_n^i$ generates a breather
solution of (\ref{dfpu}) having its maximal amplitude near 
a different site.

Now let us consider the situation when $\omega$ is kept fixed
and a small mass defect $m_0$ is introduced in (\ref{redequapp2}).
As illustrated in figure \ref{fig7}, each of the above solutions is
structurally stable.
 
For example, let us consider in figure \ref{fig7} the intersection
points $1,2,3$ between $W^u(0)$ and $W^s(0)$.
Each of these intersections persists (points $1',2',3'$ in figure \ref{fig7})
as the linear shear $A(\omega ,m_0)$ is applied to $W^u(0)$
for $m_0 \approx 0$ (dashed line in figure \ref{fig7}).
Let us examine the corresponding solutions of (\ref{red11bis})
and the related breather solutions of the Klein-Gordon model.

We denote by $\tilde{U}_n^2 = (\tilde\alpha_n^2 , \tilde\beta_n^2)^T$ the solution
of (\ref{red11bis}) with initial data $\tilde{U}_1^2$ at the point $1'$.
This solution is homoclinic to $0$ according to lemma \ref{lemint}.
Repeating an argument of section \ref{defmode}, one can show that
$R\, \tilde{U}_{-n+1}^2=\tilde{U}_n^2$, i.e. $\tilde\beta_{-n}^2=\tilde\beta_n^2$.
Consequently, $\tilde{U}_n^2$ corresponds to an (approximate) breather solution
of (\ref{dfpu}) centered at the defect site $n=0$. This solution is 
a small deformation of the site-centered breather $y_n^2$
of theorem \ref{existence1}.

Now let us denote by $\tilde{U}_n^1$ the 
homoclinic solution of (\ref{red11bis}) with initial data $\tilde{U}_1^1$ at the point $2'$.
It corresponds to an (approximate) breather solution of (\ref{dfpu}),
whose profile is a small deformation of the breather $y_n^1$ centered
between $n=0$ and $n=1$ (see theorem \ref{existence1}).
Since $\tilde{U}_1^1$ does not belong to the line $\alpha = \beta$ 
(it lies at a distance $O(|m_0 |)$), the corresponding breather solution
is not symmetric any more, which was expected since the atomic masses 
at $n=0,1$ are different.

Lastly we note $\tilde{U}_n^3 = (\tilde\alpha_n^3 , \tilde\beta_n^3)^T$ the 
homoclinic solution of (\ref{red11bis}) with initial data $\tilde{U}_1^3$ at the point $3'$.
Since $\tilde{U}_1^3$ is $O(|m_0 |)$-close to $U_0^2$
(point with label $3$), $\tilde{U}_n^3$ is a small deformation of 
the solution $U_{n-1}^2$ existing for $m_0=0$.
In other words, $\tilde{U}_1^3$ corresponds to a small deformation of
the breather $y_{n-1}^2$ centered at $n=1$. The mass defect at $n=0$
breaks the mirror symmetry of the solution, since its amplitude
has only the imperfect symmetry
$\tilde\beta_{-n+1}^3-\tilde\beta_{n+1}^3 = O(|m_0 |)$ for $n\neq 0$.

\ve

A more delicate question concerns the continuation and the possible
bifurcations of the above homoclinic solutions as $m_0$ is further
varied. The evolution of $\tilde{U}_n^1, \tilde{U}_n^2 , \tilde{U}_n^3$ 
depends on the structure of the homoclinic windings near 
${U}_1^1, {U}_1^2 , {U}_0^2$. Numerically we find that the lobes formed
near these points by the stable and unstable manifolds have the
structure shown in figure \ref{fig7}. These manifolds windings
can be analytically approximated as explained in reference
\cite{hennig} (section 3.5) or \cite{hennig2} (section 4).

At a critical value $m_0 =m_c (\omega ) >0$, the points with label
$2'$ and $3'$ on $W^s(0)\, \cap \, A(\omega ,m_0)W^u(0)$ collide
as $W^s(0)$ and $ A(\omega ,m_0)W^u(0)$ become tangent.
Consequently the solutions $\tilde{U}_n^3$ and $\tilde{U}_n^1$
disappear through a tangent bifurcation above this critical
value of $m_0$. Obviously, since we consider the truncated
map (\ref{red11bis}) instead of the full recurrence relation 
(\ref{redequanh}), these solutions only correspond to approximate
breather solutions of (\ref{dfpu}). However we shall check
numerically  that the corresponding tangent bifurcation occurs 
for nearby breather solutions of (\ref{dfpu}), at a critical value
close to $m_c (\omega )$ (see section \ref{numerical}).

In what follows we give a simple method to estimate $m_c (\omega )$,
which is based on a simple approximation of $W^u(0)$. 
Let us consider a cubic approximation $W^u_{app}$
of the local unstable manifold of figure \ref{fig7}, 
parametrized by $\beta = \lambda\, \alpha - c^2 \, \alpha^3$.  
The coefficient $c$ depends on $\mu$ and $B$ and
need not be specified in what follows
(a value of $c$ suitable when $\lambda$ is large is
computed in \cite{hennig2}, equation (60)). We note 
$\lambda = \sigma^{-1}
=1-{\mu}/{2}+{\sqrt{\mu^2-4\mu}}/{2}$ the unstable eigenvalue. 
We have 
\begin{equation}
\label{awu}
\beta = \lambda_0 \alpha - c^2 \, \alpha^3
\end{equation}
on the curve $ A(\omega ,m_0)W^u_{app}$,
where $\lambda - \omega^2 m_0 = \lambda_0$.
By symmetry we can approximate the local stable manifold using
the curve $W^s_{app}$ parametrized by 
\begin{equation}
\label{ws}
\alpha = \lambda\, \beta - c^2 \, \beta^3 .
\end{equation}
The curves $ A(\omega ,m_0)W^u_{app}$ and $W^s_{app}$ become tangent
at $(\alpha , \beta )$ when in addition
\begin{equation}
\label{tan}
(\lambda  - 3 c^2 \, \beta^2 ) (\lambda_0  - 3 c^2 \, \alpha^2 )=1.
\end{equation}

In order to compute $m_0 =m_c$ as a function of $\omega$, 
or, equivalently, the corresponding value of $\lambda_0$ as a function of $\lambda$, 
one has to solve the nonlinear system
(\ref{awu})-(\ref{ws})-(\ref{tan}) with respect to
$\alpha$, $\beta$, $\lambda_0$, which yields
a solution depending on $\lambda$. 
Instead of using $\lambda$ it is practical to
parametrize the solutions by $t=\beta / \alpha$. 
This yields
$$
\alpha = \frac{1}{c\sqrt{2}}\, (t+\frac{1}{t^3})^{1/2},
\ \ \
\beta = \frac{t}{c\sqrt{2}}\, (t+\frac{1}{t^3})^{1/2},
$$
$$
\lambda_0 = \frac{3}{2} t + \frac{1}{2 t^3},
\ \ \
\lambda = \frac{3}{2 t}  + \frac{1}{2}\, {t^3}.
$$
Since $\mu = 2 - \lambda - \lambda^{-1}$ 
and $m_0 = (\lambda - \lambda_0 )\, (\Omega^2 + \mu )^{-1}$ it follows
\begin{equation}
\label{eqmut}
\mu = 2 - \frac{3+t^4}{2t} -  \frac{2t}{3+t^4},
\end{equation} 
\begin{equation}
\label{mcapp}
m_0 = \frac{1}{2} (t-\frac{1}{t})^3 (\Omega^2 + \mu )^{-1}.
\end{equation} 
Given a value of $\mu \in (-\Omega^2 , -1/2 )$, 
one can approximate $m_c $ by the value of
$m_0$ given by equations (\ref{eqmut})-(\ref{mcapp}).

For example, in the case numerically studied
in figure \ref{fig7} we have $\omega=9.9$ and $\mu=-1.99$.
Consequently $\lambda\approx 3.721$, $t\approx 1.7935$
and $\lambda_0 \approx 2.777$, which yields 
$m_c \approx 0.009632$. A numerical study of the map
yields $m_c \in (0.00963,0.00964)$, and consequently
our approximation works very well in this parameter regime.
Moreover, the approximation is 
extremely close to the actual value of $m_0$ at which
a tangent bifurcation occurs between the corresponding breather
solutions of the Klein-Gordon system (numerically we again find 
$m_0 \in (0.00963,0.00964)$,
see section \ref{numerical} for more details).

Despite it gives precise numerical results in a certain parameter
range, the approximation (\ref{eqmut})-(\ref{mcapp}) is not
always valid. Indeed, the parameter regime $\mu > -1/2$
is not described within this approximation. 
Moreover, one can check that $W^u_{app}$ intersects
$W^s_{app}$ on the line $\alpha = \beta $ with an angle depending solely on
$\lambda$, and not on the coefficient $B$
(in particular, $W^u_{app}$ and $W^s_{app}$ become tangent 
for $\lambda = 2$). This problem
could be solved by adding a quintic term $d\, \alpha^5$
in equation (\ref{awu}).

\vspace{1ex}

The intersection point with label $1'$ between
$W^s(0)$ and $A(\omega ,m_0)W^u(0)$ persists for 
$0<m_0 <m_l (\omega )$, or equivalently $0<\omega < \omega_l (m_0 )$,
and consequently the reversible homoclinic solution
$\tilde{U}_n^2$ exists within this parameter range.
At $\omega = \omega_l$, this solution disappears through a
pitchfork bifurcation with the symmetric solution $-\tilde{U}_n^2$
(the amplitude of the homoclinic orbit is $O(\sqrt{\omega -\omega_l})$
as $m_0$ is fixed and $\omega \rightarrow \omega_l$).
This homoclinic orbit corresponds for equation (\ref{dfpu}) to
a nonlinear analogue of the linear localized mode
of section \ref{linc}. As noticed in section \ref{defmode},
the existence of small amplitude exact solutions of this type 
(with $\omega \approx \omega_l$)
can be obtained in a classical way using the Lyapunov centre theorem, 
when $\omega_l$ does not resonate with the phonon spectrum.
For $m_0 \approx 0$ and $\omega \approx \omega_l$,
the breather solution of (\ref{dfpu}) can be approximated by
$y_n (t) \approx \beta_n \cos{t}$, and the
frequency $\omega$ varies with amplitude and lies below $\omega_l$.

\vspace{1ex}

More generally, the evolution 
of the set $A(\omega ,m_0)W^u(0)\, \cap \, W^s(0)$ as $m_0$ varies
is very complex, due to the complex shape of the stable and
unstable manifolds and the complicated structure of 
their intersection (see figure \ref{fig9}).
In section \ref{numerical}
we shall give some additional examples of breather bifurcations
which can be deduced from the fine structure of the stable and
unstable manifolds.

\vspace{1ex}

Note that previous studies have examined,
for certain families of reversible two-dimensional maps,
how parameter changes modify the intersections between 
the stable and unstable manifolds of the origin and
the associated set of homoclinic solutions \cite{bountis,carr}.
These (autonomous) maps are directly obtained from the discrete nonlinear Schr\"odinger equation
or generalized versions (due to their phase invariance), 
as one looks for oscillatory solutions with a single Fourier component. 
Althought we obtain similar types of tangent bifurcations as defects strength
are varied, our situation is quite different since we are concerned with a
nonautonomous map, where the impurity leads to consider
a linear shear of the unstable manifold.

\subsection{\label{many}Case of finitely many defects}

This section generalizes the analysis of the above section
to the case when equation (\ref{KG}) admits a finite
number of inhomogeneities. More precisely, we assume 
in equation (\ref{dfpu})
$\eta_n =\gamma_n =\kappa_n =\epsilon_n = 0$ if
$|n| \geq n_0 +1$, for a given integer $n_0 \geq 0$.
Note that this assumption allows one to cover the case of an odd number
of defects as well as an even number.

\ve

The situation is more complex than in section \ref{single},
because studying homoclinic solutions of 
(\ref{redequapp}) leads to finding the intersections of
the stable manifold with the image of the unstable manifold
under a {\em nonlinear} transformation (see lemma \ref{lemint2} below).
However, one can recover the linear case if one replaces
the relevant spatial map by a suitable one, both being equal
at leading order only. 
In what follows we shall develop this leading order theory,
considering as higher order terms
all terms being $o( \| (\alpha_n , \beta_n ) \|^3)$.

\ve

Equation (\ref{redequapp}) reads
\begin{equation}
\label{appr}
\beta_{n+1}-2\beta_{n}+\beta_{n-1}=
(\theta_n - \mu )\beta_n - \kappa_n \beta_{n-1}
+B\, \beta_n^{3},
\end{equation}
where 
$\theta_n =\Omega^2 ( \eta_n  + \eta_n \gamma_n + \gamma_n ) - \omega^2 \epsilon_n + \kappa_n$,
$\omega^2 = \Omega^2 + \mu$.
In the sequel we shall note
$\varepsilon = \| \, \{ \theta \} \, \|_{\ell_\infty (\mathbb{Z}) } + 
\| \, \{ \kappa \} \, \|_{\ell_\infty (\mathbb{Z}) }  $.

Setting $\beta_{n-1}=\alpha_n$ and
$U_n =(\alpha_n ,\beta_n)^T$, equation (\ref{appr}) can be rewritten
\begin{equation}
\label{redappr}
U_{n+1}=F_n (U_n),
\end{equation}
\begin{equation}
\label{defmapappr}
F_n (\alpha ,\beta )
=
\left(
\begin{array}{c}
\beta  \\
- (1+\kappa_n )\alpha + (2+\theta_n -\mu )\beta +B \beta^3
\end{array}
\right).
\end{equation}
Noting $F=G_\omega$ for simplicity (see definition (\ref{defmap})), 
one can observe that
\begin{equation}
\label{lintrans}
F_n = (I+T_n)\, F + O(|\kappa_n | |\beta |^3),
\end{equation}
$$
T_n = 
\left(
\begin{array}{cc}
0 & 0 \\
\theta_n + (\mu - 2)\, \kappa_n & \kappa_n
\end{array}
\right) .
$$
Note that higher order terms are absent from equation
(\ref{lintrans}) if $\kappa_n =0$.

\ve

Since $F_n = F$ for $|n| \geq n_0 +1$ one has the following property.

\begin{lemma}
\label{lemint2}
Fix $\mu <0$ and denote by $W^s(0)$, $W^u(0)$ the
stable and unstable manifolds of the fixed point $U=0$ of $F$.
Consider the nonlinear map
$G = F_{n_0}\, \circ \, F_{n_0-1}\, \circ \, \cdots \, \circ \, 
F_{-n_0 } \, \circ \, F^{-2n_0 -1}$.
Equation (\ref{redappr})
possesses an homoclinic orbit to $0$ if and only if $W^s(0)$ and 
$G ( W^u(0) )$ 
intersect.
\end{lemma} 

Lemma \ref{lemint2} is hard to use for analyzing homoclinic
solutions since it involves a nonlinear transformation $G$ instead
of a linear one as in lemma \ref{lemint}.
However one can recover the linear case when replacing $F_n$ by
a suitable approximation $\hat{F}_n$, equal to $F_n$ up to higher order
terms. This is possible thanks to property (\ref{prophat})
of lemma \ref{lemsimp} below.
In the sequel we note
$$
L= DF(0) =
\left(
\begin{array}{cc}
0 & 1 \\
-1 & 2-\mu 
\end{array}
\right) .
$$

\begin{lemma}
\label{lemsimp}
Consider the collection of maps $\hat{F}_n$ 
($ -n_0 \leq n \leq n_0$) defined by
\begin{equation}
\hat{F}_n = A_n \, F\, \circ \, A_{n-1}^{-1},
\end{equation}
where $A_{-n_0 -1}=I$ and for $n \geq -n_0$
\begin{equation}
\label{defmatam}
A_n = L_n \, L_{n-1} \, \cdots \, L_{-n_0}\, L^{-n-n_0-1},
\ \ \
L_n = (I+T_n) L.
\end{equation}
The map $\hat{F}_n$ is a leading order approximation of $F_n$, i.e.
$\hat{F}_n = F_n + O( \varepsilon \, \| (\alpha , \beta ) \|^3 )$.
Moreover one has the property 
\begin{equation}
\label{prophat}
\hat{F}_{n_0}\, \circ \, 
\hat{F}_{n_0-1}\, \circ \, \cdots \, \circ \, \hat{F}_{-n_0}=
A \, F^{2n_0 +1},
\end{equation}
where $A=A_{n_0}$ reads
\begin{equation}
\label{defmata}
A = L_{n_0} \, L_{n_0 -1} \, \cdots \, L_{-n_0}\, L^{-2n_0-1}= 
I+O(\varepsilon ).
\end{equation}
\end{lemma} 

\proof

First we note that the sequence $A_n$ satisfies 
$A_{-n_0}=I+T_{-n_0}$ and 
\begin{equation}
\label{reca}
A_{n+1}=(I+T_{n+1})\, L\, A_n \, L^{-1}
\end{equation}
for all $n \geq -n_0 - 1$.
It follows for $ -n_0 \leq n \leq n_0$
\begin{equation}
\label{expf}
\hat{F}_n =
(I+T_{n})\, L\, A_{n-1} \, L^{-1} \, 
F\, \circ \, A_{n-1}^{-1}.
\end{equation}
Now let us note that $A_n = I + O(\varepsilon )$.
Moreover, the following identity holds true
for any parameter-dependent
matrix $M \in M_2 (\mathbb{R})$ with $\| M \| = O(\varepsilon )$
\begin{equation}
\label{commute}
F\, \circ (I+M) = L\, (I + M)\, L^{-1} \, F 
+O( \varepsilon \, \| (\alpha , \beta ) \|^3 ).
\end{equation}
Consequently one has also
$$
F = L\, (I + M)\, L^{-1} \, F \, \circ (I+M)^{-1}
+O( \varepsilon \, \| (\alpha , \beta ) \|^3 ).
$$
Using this property in equation (\ref{expf}) leads to
$$
\hat{F}_n =
(I+T_{n})\, F +O( \varepsilon \, \| (\alpha , \beta ) \|^3 ).
$$
Using (\ref{lintrans}) this yields 
$\hat{F}_n = F_n + O( \varepsilon \, \| (\alpha , \beta ) \|^3 )$,
therefore $\hat{F}_n$ is a leading order approximation of $F_n$.
Property (\ref{prophat}) follows directly from the definition
of $\hat{F}_n$.
$\fin$

It is worthwhile stressing that $A=DG(0)$, where $G$ is the nonlinear
transformation introduced in lemma \ref{lemint2}.

Now we fix in addition $\hat{F}_n= F = F_n$ for $|n| \geq n_0 +1$.
According to lemma \ref{lemsimp} we have also
$\hat{F}_n = F_n + O( \varepsilon \, \| (\alpha , \beta ) \|^3 )$
for $|n| \leq n_0$. In the sequel we
approximate system (\ref{redappr}) by the new one
\begin{equation}
\label{redappr3}
U_{n+1}=\hat{F}_n (U_n).
\end{equation}
Property (\ref{prophat}) implies the following result, 
since $W^u(0)$ is invariant under $F^{2n_0 + 1}$.

\begin{lemma}
\label{lemint3}
Fix $\mu <0$ and denote by $W^s(0)$, $W^u(0)$ the
stable and unstable manifolds of the fixed point $U=0$ of $F$.
Equation (\ref{redappr3})
possesses a solution $U_n$
homoclinic to $0$ if and only if $W^s(0)$ and 
$A ( W^u(0) )$ intersect, 
where the matrix $A= I + O(\varepsilon )$ is defined in
lemma \ref{lemsimp}. The intersection point corresponds to
$U_{n_0 +1}$.
\end{lemma} 

Consequently, as in section \ref{single}
one recovers the problem of finding the intersection of
$W^s(0)$ with the image of $W^u(0)$ under the (near-identity)
linear transformation $A$. 
Note that $A=I+T_0$ in the single defect case $n_0$=0.

Here we shall not attempt to relate the bifurcations
of breather solutions of (\ref{dfpu})
with the properties of the inhomogeneities, via
an analysis of homoclinic solutions of (\ref{redappr3}).
This question will be considered in future works using
the simplification provided by lemma \ref{lemint3}.
As for a single defect, for $B<0$ one can expect multiple
tangent bifurcations between (deformations of)
site-centered and bond-centered breathers as inhomogeneities
are varied, due to the winding structure of $W^u(0)$ and
$ W^s(0)$.

\vspace{1ex}

It is now interesting
to compute the leading order contribution
of the sequence of inhomogeneities to the matrix $A$. This is the object of
the following lemma.

\begin{lemma}
\label{simplm}
The matrix $A$ of
lemma \ref{lemsimp} takes the form
$A=I+M + O(\varepsilon^2 + \varepsilon |\mu |)$, where
$$
M = 
\left(
\begin{array}{cc}
M_{11} & M_{12} \\
M_{21} & M_{22}
\end{array}
\right) ,
$$
\begin{eqnarray*}
M_{11}&=&\sum_{n=0}^{2n_0}{n(n+1)\rho_{n_0-n}-n\, \kappa_{n_0-n}},\\
M_{12}&=&\sum_{n=0}^{2n_0}{-n^2\rho_{n_0-n}+n\, \kappa_{n_0-n}},\\
M_{21}&=&\sum_{n=0}^{2n_0}{(n+1)^2 \rho_{n_0-n}-(n+1)\, \kappa_{n_0-n}},\\
M_{22}&=&\sum_{n=0}^{2n_0}{-n(n+1)\rho_{n_0-n}+(n+1)\, \kappa_{n_0-n}},
\end{eqnarray*}
$\rho_n =\Omega^2 ( \eta_n  + \gamma_n - \epsilon_n )$.
\end{lemma}

\proof

Since $T_n = O(\varepsilon )$ it follows from definition (\ref{defmata})
\begin{equation}
\label{approxa1}
A=I + \sum_{n=0}^{2n_0}{L^n\, T_{n_0 -n}\, L^{-n}} + O(\varepsilon^2 ).
\end{equation}
Now we use the expansions
$$
T_n = M_n + O(\varepsilon^2 + \varepsilon |\mu | ), \ \ \
M_n = 
\left(
\begin{array}{cc}
0 & 0 \\
\rho_n -  \kappa_n & \kappa_n
\end{array}
\right) ,
$$
$$
L = L_c + O(|\mu |), \ \ \ L_c = 
\left(
\begin{array}{cc}
0 & 1 \\
-1 & 2
\end{array}
\right) ,
$$
to obtain
$$
A=I +M  + O(\varepsilon^2 + \varepsilon |\mu | ),
\ \ \
M=
\sum_{n=0}^{2n_0}{L_c^n\, M_{n_0 -n}\, L_c^{-n}},
$$
where
$$
L_c^n =
\left(
\begin{array}{cc}
-n+1 & n \\
-n & n+1
\end{array}
\right) .
$$
Then simple computations lead to the coefficients of $M$ provided above. 

$\fin$

Interestingly, lemma \ref{simplm} shows that the influence
of the inhomogeneities on the set of homoclinic solutions
depends at leading order (via the matrix $I+M$)
on algebraically-weighted averages of
$\{\kappa \}$ and $\{\rho \}$.

\ve

Now let us return to the original parameters 
$m_n , d_n , a_n , k_n $ describing the lattice
inhomogeneities (see equations (\ref{KG}) and (\ref{dfpu})),
with $m_n = d_n = a_n =0$ for $|n| \geq n_0 +1$,
$k_n = 0$ for $n \leq -n_0 -1 $ and $n \geq n_0$.
Let us note
$\tilde\varepsilon = \| \{ m_n / m \} \|_{\ell_\infty} 
+ 
\| \{ d_n / d \} \|_{\ell_\infty } 
+
\| \{ a_n / a \} \|_{\ell_\infty }
+
\| \{ k_n / k \} \|_{\ell_\infty }.
$
One obtains
$$
\kappa_n = \frac{k_{n-1}-k_n}{k}+O(\tilde\varepsilon^2 ),\ \ \
\rho_n =r_n +O(\tilde\varepsilon^2 ),
$$
where
$$
r_n =  \Omega^2 \, ( \frac{d_n}{d} + 2 \frac{a_n}{a} - \frac{m_n}{m})
$$
is a linear combination of the
on-site potential and mass defect impurities.
Some coefficients of $M$ can be simplified since
$$
\sum_{n=0}^{2n_0}{\kappa_{n_0 -n}}=O(\tilde\varepsilon^2 ), \ \ \
\sum_{n=0}^{2n_0}{-n\, \kappa_{n_0 -n}}=
\frac{1}{k}\, \sum_{n=-n_0}^{n_0 -1}{k_n} + O(\tilde\varepsilon^2 ).
$$
Noting
$$
I_k = \sum_{n=-n_0}^{n_0}{n^k \, r_n}, \ \ \
J_0 = \frac{1}{k}\, \sum_{n=-n_0}^{n_0 -1}{k_n},
$$
one finally obtains
$A=I +\tilde{M}  + O(\tilde\varepsilon^2 + \tilde\varepsilon |\mu | )$ with
$$
\tilde{M} = 
\left(
\begin{array}{cc}
\tilde{M}_{11} & -\tilde{M}_{11} +n_0 I_0 - I_1 \\
 \tilde{M}_{11} +(n_0 +1) I_0 - I_1 & -\tilde{M}_{11}
\end{array}
\right) 
$$
and $\tilde{M}_{11} = n_0 (n_0 +1)I_0 - (2n_0 +1) I_1 +I_2 + J_0$.

Consequently, the matrix $A$ depends (at leading order in $\tilde\varepsilon$ and $\mu$) on
the average values $I_0 , J_0$
of $r_n$, $k_n /k$, and on the
weighted averages $I_1 , I_2$ of $r_n$ (with linear and quadratic 
weights respectively).
Since $\mbox{Tr}(\tilde{M})=0$, it follows 
$\mbox{Tr}(A)=2+ O(\tilde\varepsilon^2 + \tilde\varepsilon |\mu | )$
and $\mbox{Det}(A)=1+ O(\tilde\varepsilon^2 + \tilde\varepsilon |\mu | )$.
As a consequence, in order to study the spectrum of $A$ for $\tilde\varepsilon ,\mu  \approx 0$
(and determine to which type of linear transformation it corresponds)
it would be necessary to compute the quadratic terms in $(\tilde\varepsilon ,\mu )$
in its expansion.

\section{\label{numerical} Numerical results}

We have performed numerical computations in order to check
the range of validity of the analysis of section \ref{single}, and
in particular if discrepancies appear for large amplitude solutions 
or if parameters $(m_0,\omega)$ are moved away from $(0,\Omega )$.
More precisely, we have computed breather solutions of the Klein-Gordon lattice
\begin{equation}\label{K-G}
  \omega^2 (1+m_n) \frac{d^2 y_n}{d t^2}+\Omega^2
  V'(y_n)=y_{n+1}-2y_n+y_{n+1}
\end{equation}
with a single mass defect $m_n=m_0 \delta_{n,0}$ and
periodic boundary conditions $y_{-N}(t)=y_N (t)$.
In general we have used a lattice with $101$ particles, except
for the computations of breathers with algebraic decay
(case $\omega=\Omega$) where $401$ particles have been considered. The computations
have been compared with homoclinic orbits to 0 of the 
two-dimensional map (\ref{red11bis}). 
For the numerical computations we have always fixed $\Omega=10$
(recall $\Omega$ is the lower phonon band edge for the infinite
system). This can be done taking, for instance, $k=0.01$, and $d=1$ 
in the original problem (\ref{dfpu}).
For the potential $V$ we have chosen
a polynomial of degree $4$ with $V^{''}(0)=1$.

\subsection{\label{hardp}Hard potentials}
    
To start we have considered the simplest case of a hard potential,
i.e. a potential with a strictly positive
hardening coefficient $B$ (see definition (\ref{defbnh})).
We have chosen
\begin{equation}\label{potential}
  V(x)=\frac{x^2}{2}+\frac{x^4}{4} \, ,
\end{equation}
for which $B=75$.

\begin{figure}[h!]
\begin{center}
\includegraphics[width=6.0cm]{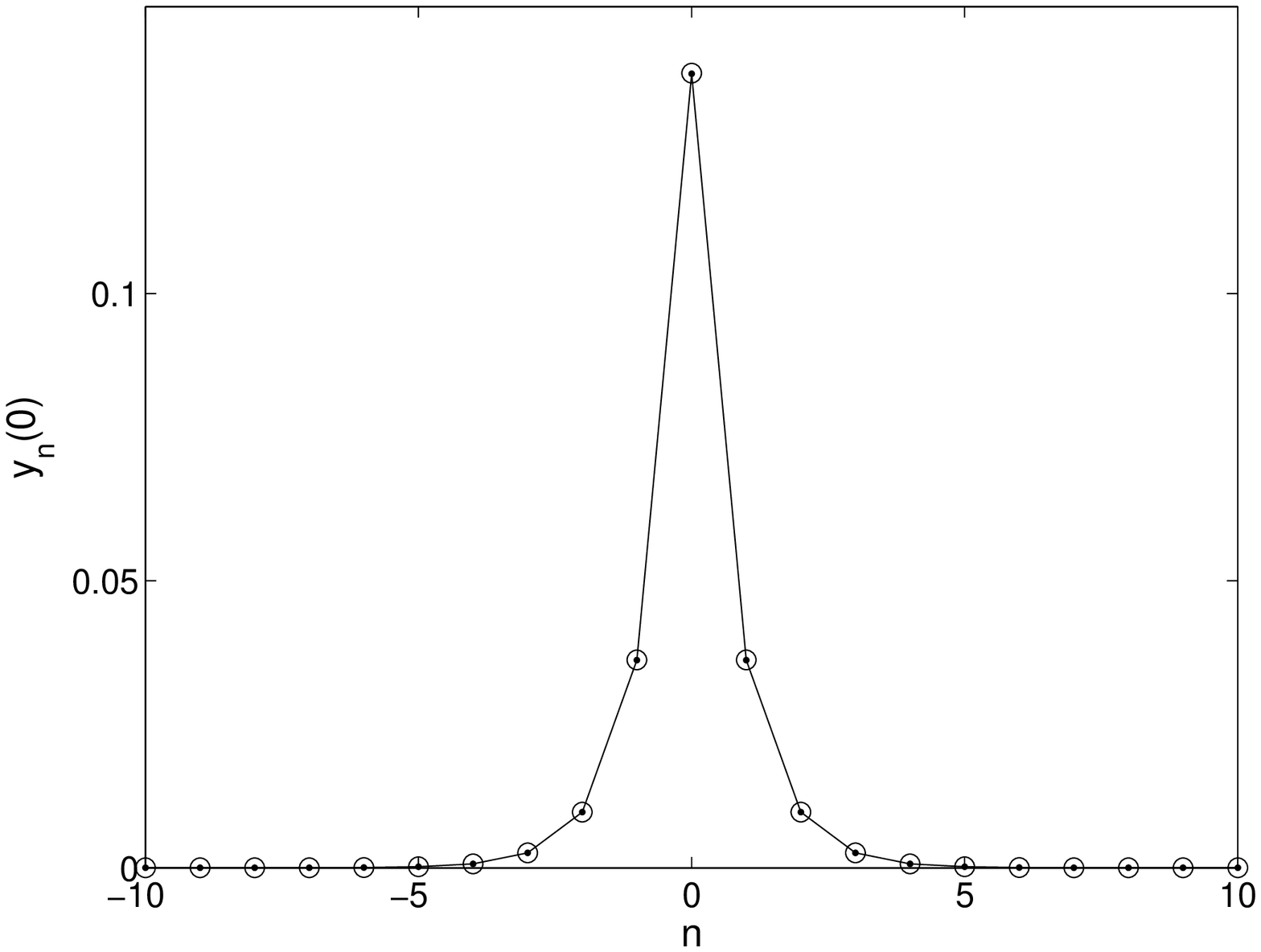} 
\includegraphics[width=5.8cm]{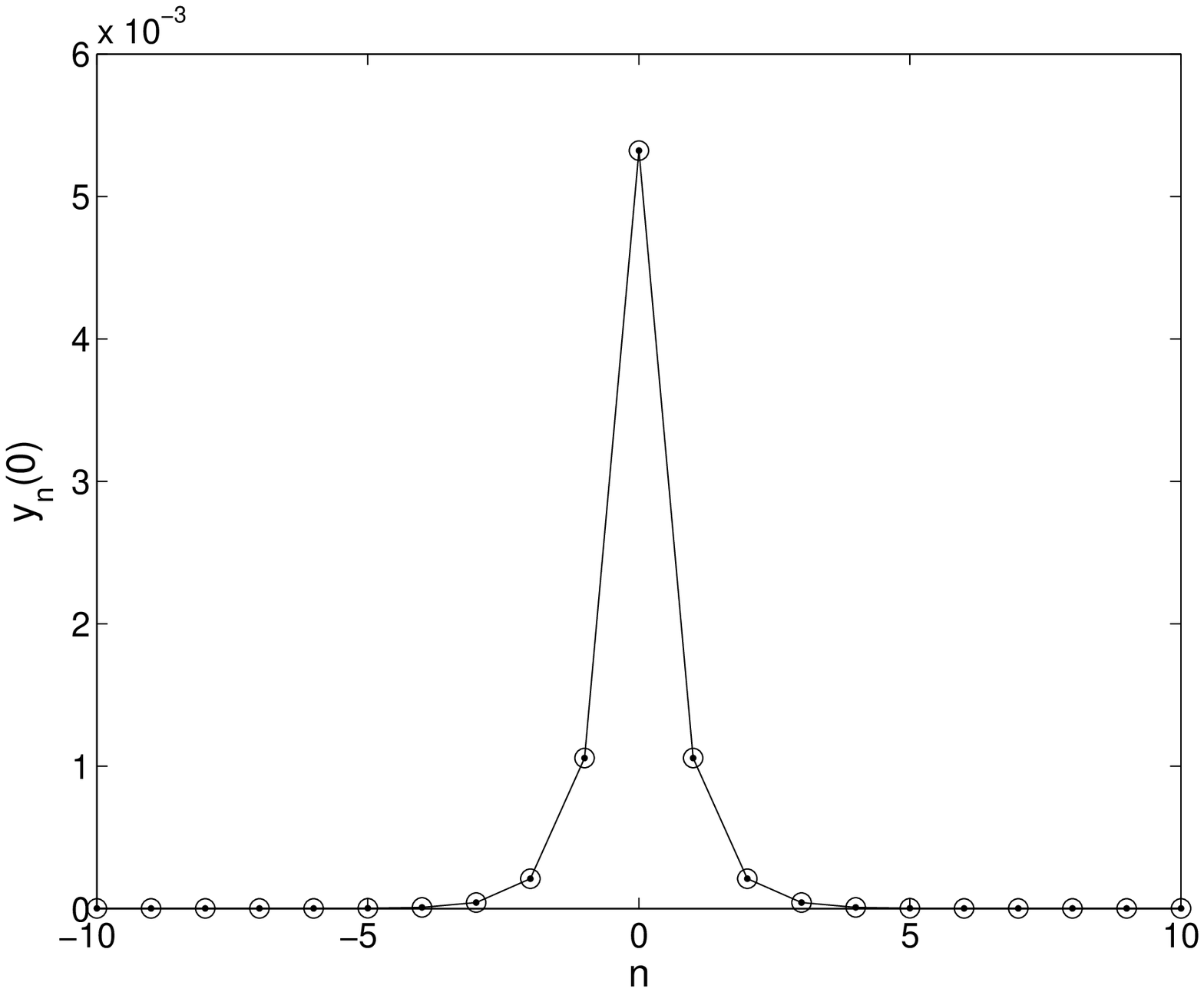}
\caption{Comparison between the profile
of a breather solution (continuous line) of the Klein-Gordon system
(\ref{K-G}) with hard potential (\ref{potential}) and the approximate solution
$y_n = \beta_n \cos{t}$
(circles) constructed with the homoclinic orbit of (\ref{red11bis}). 
We have considered a mass defect $m_0=0.05$.
In the left panel we have chosen
a frequency $\omega=9.9$ ($\mu=-1.99$). 
In the right panel we have fixed
$\omega=9.837$ ($\mu=-3.23$) very close to $\omega_l$
(note the change of scale for the vertical axis).
\label{fig2}}
\end{center}
\end{figure}

In this case, the reduced map
(\ref{red11bis}) possesses a unique orbit homoclinic to 0  in the
sector $\alpha>0, \beta>0$, for $m_0>0$ and
$\omega_l<\omega<\Omega$. An example of this homoclinic orbit is
shown in figure~\ref{splitting2} for a frequency $\omega=9.9$
($\mu=-1.99$) and a mass defect $m_0=0.05$. In figure~\ref{fig2} (left panel) we
compare the approximate solution $y_n = \beta_n \cos{t}$ obtained with
this homoclinic orbit (circles) with the exact breather
profile computed with the standard numerical method based on the
anti-continuous limit \cite{marin} (continuous line). 
The agreement is excellent even if the solution profile is 
very localized. Indeed, as one computes the eigenvalues $\sigma , \sigma^{-1}$
(equation (\ref{decay})) of the linearized map
(\ref{redequalin}) with $m_0 =0$, one obtains $\sigma \approx 0.27$,
which implies a strong spatial localization visible in figure \ref{fig2}.
The accuracy of the
centre manifold reduction (a priori expected for $\sigma \approx 1$) is
surprisingly good in this parameter regime.

    The breather solution can be continued for decreasing frequencies up
to $\omega_l\approx 9.8369$, which is the frequency of the linear
defect mode at which the breather solution bifurcates.
Figure~\ref{fig2} (right panel) compares again the numerically computed breather profile 
and the approximate solution obtained with the homoclinic orbit, but now very close
to this bifurcation point (at $\omega=9.837$, i.e. $\mu=-3.23$). 
We still observe an excellent agreement.
Note that the oscillations amplitudes are very small, but the solution is still strongly
localized.

\begin{figure}[h!]
\begin{center}
\includegraphics[width=6.0cm]{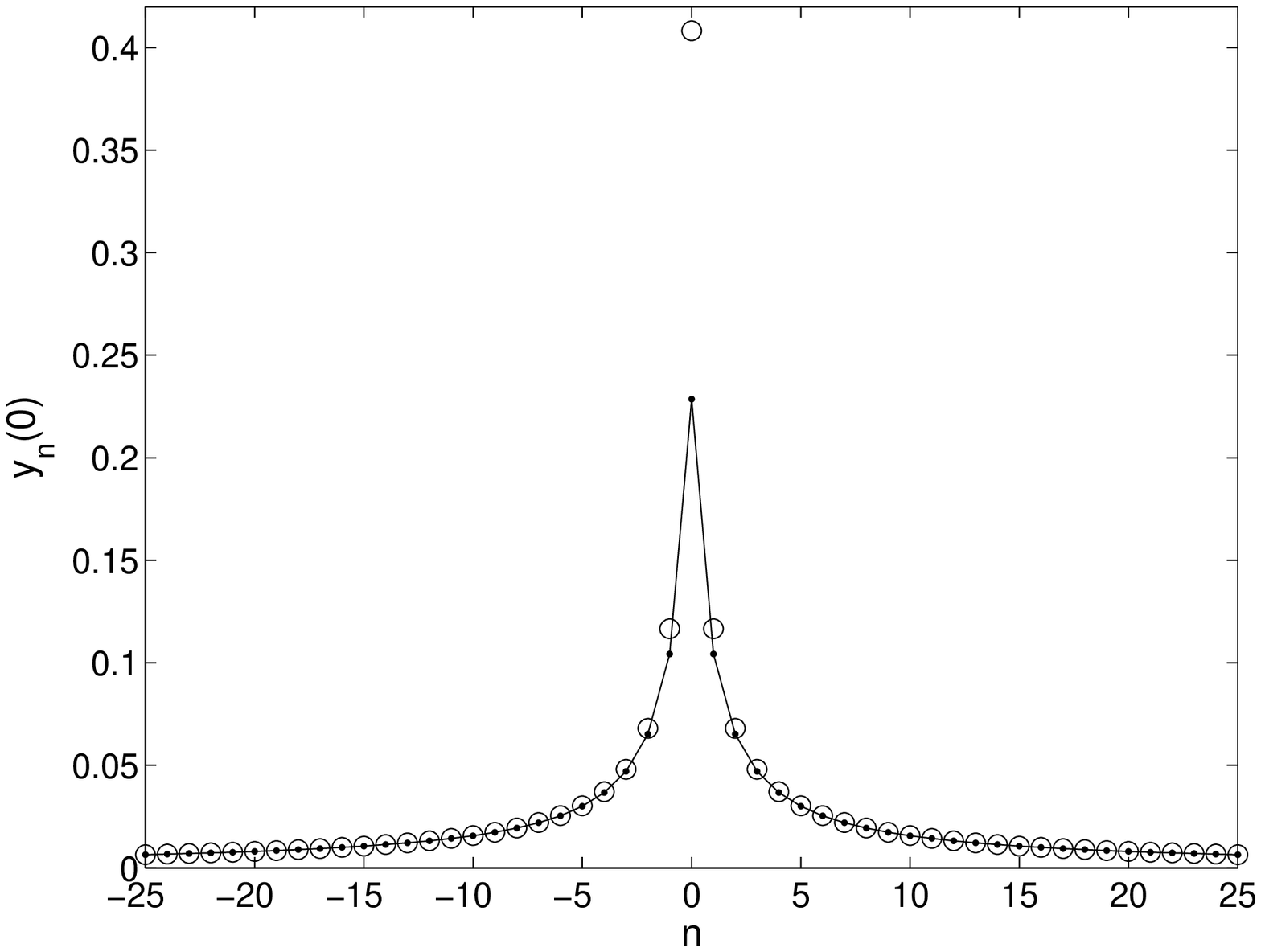} 
\includegraphics[width=6.0cm]{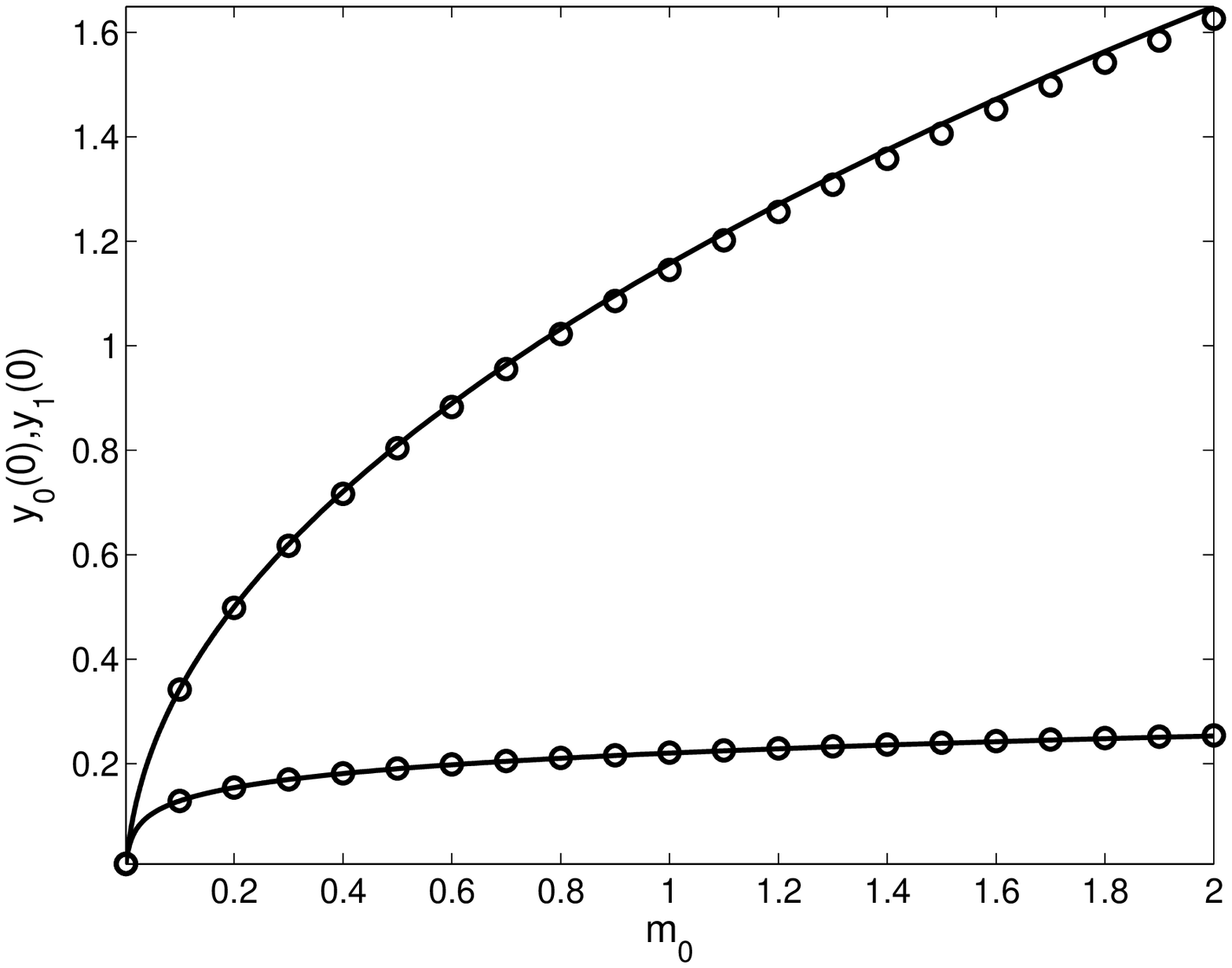}
\caption
{Left panel~: breather solution
at the lower edge of the phonon band $\omega=\Omega$ ($\mu=0$) for a
mass defect $m_0=0.05$ and
the symmetric potential
$V(x)={x^2}/{2}+{x^4}/{4}$.
The continuous line corresponds to the
numerically computed breather solution.
The circles represent approximation
(\ref{approx1}) of the homoclinic orbit that fits very well the
algebraic decay of the breather tails.  
Right panel~: the continuous lines represent the
amplitude of the breather solution at $n=0$ (upper curve)
and $n=1$ (lower curve)
versus mass defect. The circles correspond to the
homoclinic solution of the nonlinear map (\ref{red11bis})
(the upper plot represents $\beta_0$ and the lower plot $\beta_1$).
\label{fig4}}
\end{center}
\end{figure}

    For increasing frequencies the continuation path ends up at
the lower edge of the phonon band $\omega=\Omega$
($\mu=0$).
For this particular frequency value the breather solution (see
continuous line in figure~\ref{fig4}, left panel) presents an algebraic decay
which is very well described by approximation (\ref{approx1}).
This approximation fails to describe the maximum amplitude of the
oscillation $\beta_0$ for these parameter values. 
This is not surprising since 
$\beta_0$ is not small, and $\beta_n$ varies rapidly near
$n=0$, hence $m_0$ should be further decreased to attain the
domain of validity of the ansatz (\ref{ansatz}) 
near the solution centre.

However, the value of $\beta_0$ obtained from the
exact homoclinic orbit of (\ref{red11bis})
fits very well the maximum amplitude of the
breather solution, as it is shown in figure~\ref{fig4}, right panel. Note that
the agreement is very good even for very large amplitudes or very
large mass defect i.e. far away from the parameter range in
which the centre manifold reduction and normal form analysis
are valid in principle.

\ve

\begin{figure}[h!]
\begin{center}
\includegraphics[width=6.0cm]{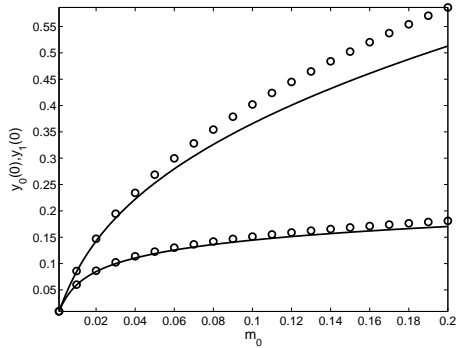} \caption{
Same computation as in figure \ref{fig4}, right panel, 
but now for the asymmetric potential
$V(x)={x^2}/{2}+x^3/6+{x^4}/{4}$.
\label{fig6}}
\end{center}
\end{figure}

    It is interesting to remark that the accuracy of this fit
depends on the symmetry of the potential $V(x)$ we have chosen.
Figure~\ref{fig6} shows what happens if we add to the polynomial
potential (\ref{potential}) a cubic term $x^3/6$ that breaks its
symmetry. The range of validity of our leading order 
approximation reduces significantly. A similar result was obtained
in reference \cite{PRB} for breather solutions in spatially
homogeneous Fermi-Pasta-Ulam lattices.
Obviously the agreement would be improved by taking into 
account the Taylor expansion of the reduction function $\phi$
(see theorem \ref{reductionnh}) and computing the normal form at a
higher order.

    Finally we have numerically checked that all breathers solutions in the gap
$\omega_l<\omega<\Omega$ are spectrally stable, at least for the value
of the frequency parameter $\Omega=10$ we have considered. This
result is in agreement with the stability analysis performed in
reference~\cite{MST} for Klein-Gordon chains with an impurity
and at low coupling $k$.

\subsection{\label{softp}Soft potentials}

In the case of soft potentials 
(when the coefficient $B$ defined by
(\ref{defbnh}) is strictly negative), 
the situation is far more complex due to the much more intricate structure of the
intersections between the stable and unstable manifolds. Therefore
one expects a richer bifurcation scenario as parameters 
(breather frequency, mass defect) are varied.
Our computations have been performed with the symmetric potential
\begin{equation}\label{soft}
  V(x)=\frac{x^2}{2}-\frac{x^4}{4} \, ,
\end{equation}
for which $B=-75$.

Let us recall some basic features of the analysis
performed in section \ref{csp}, in order to compare
the results with numerical computations.
For the (truncated) reduced mapping (\ref{red11bis}) with $m_0=0$,
figure~\ref{fig7} shows some intersections of stable and
unstable manifolds emanating from the saddle point at the origin,
for a frequency value $\omega=9.9<\Omega$. Iterating the map
with an initial condition $U_1$ at the homoclinic point with label 1, 
we obtain an homoclinic orbit which corresponds to a one-site breather 
centered at $n=0$. With an initial condition $U_1$ at the
homoclinic point with label 2, the corresponding breather is
a two-site breather with maximal amplitude at $n=0$ and $n=1$.
An initial condition $U_1$ at the homoclinic point with label 3
(symmetric of point 1 respect to the line $\alpha=\beta$)
corresponds to a one-site breather
centered at site $n=1$. 

\begin{figure}[h!]
\begin{center}
\includegraphics[width=6.0cm]{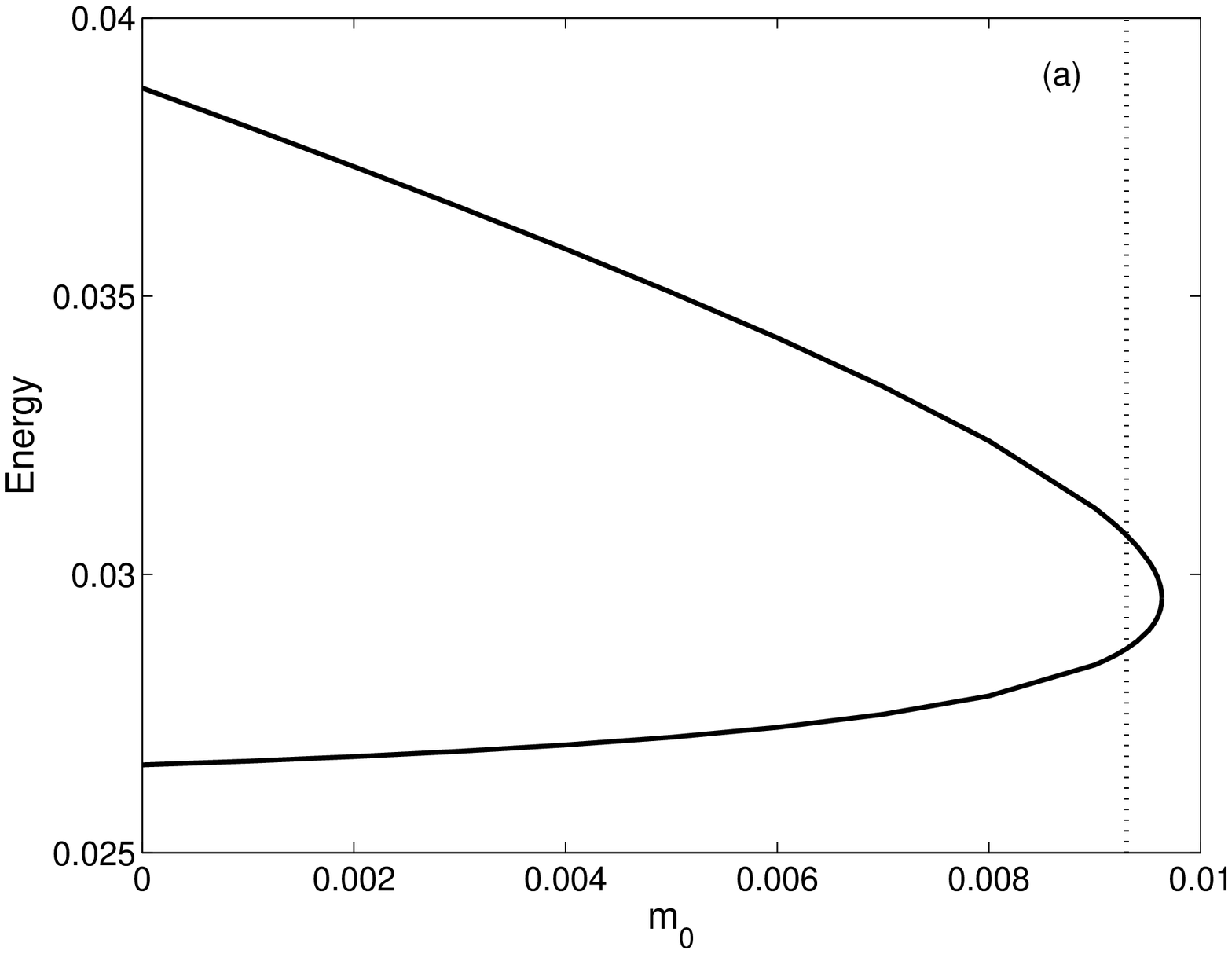}
\includegraphics[width=6.0cm]{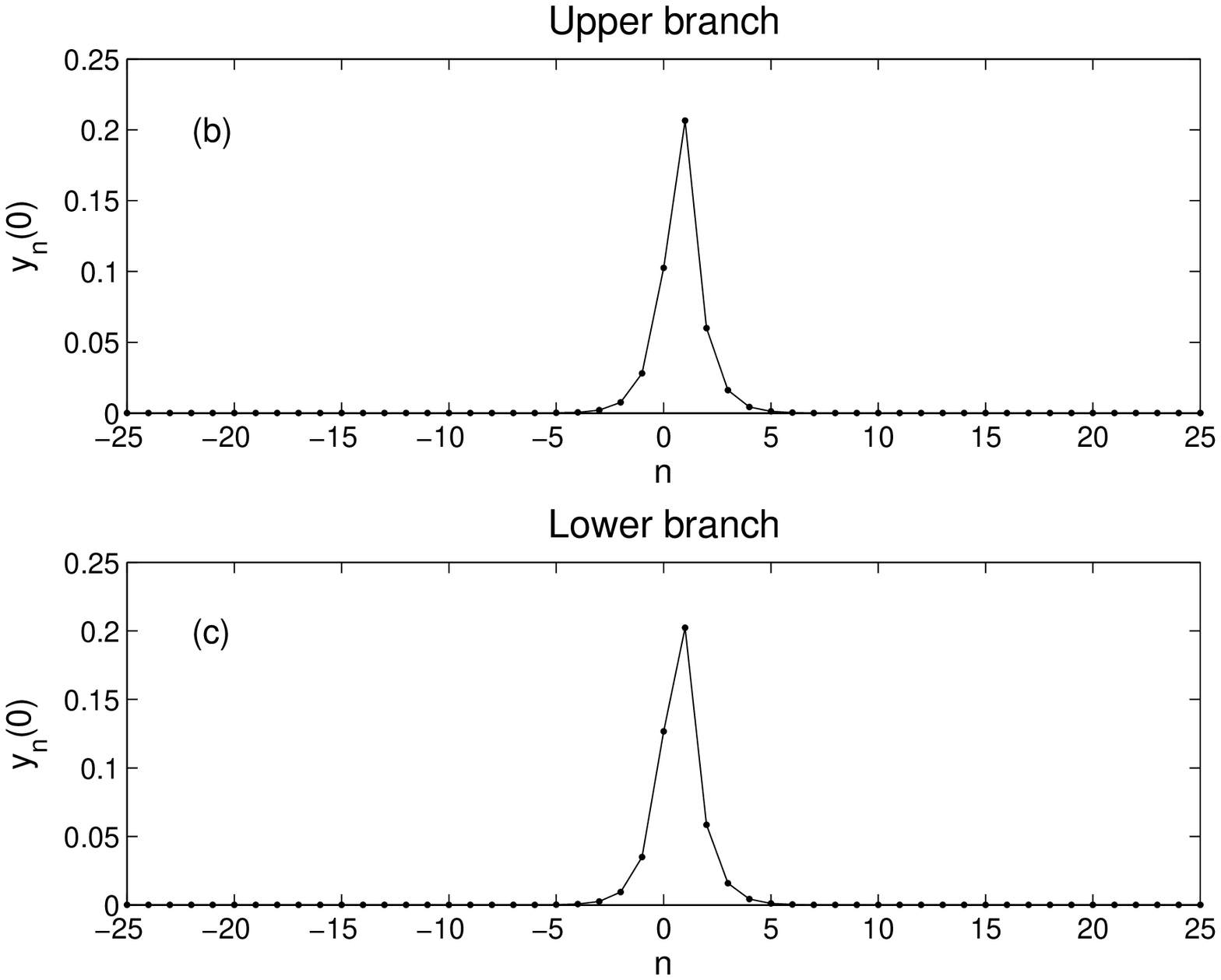}
\caption{Tangent bifurcation between breather solutions numerically
computed in a Klein-Gordon chain with a soft potential.
The chain presents a mass defect $m_0$ at $n=0$, and the bifurcation
occurs as $m_0$ is increased. 
In the left panel, the breathers energies
$E=\sum_{n\in\mathbb{Z}}{\Omega^2 V(y_n (0))+(y_{n+1}(0)-y_n (0))^2 /2}$
are depicted versus $m_0$ (the breathers are even in $t$ with 
frequency $\omega = 9.9$).
For $m_0 \approx 0$, the upper branch represents a
two-site breather centered between sites $n=0$ and $n=1$. 
The lower branch
represents a one-site breather centered at $n=1$. The breathers profiles
close to the bifurcation point are plotted in the
right panels (the value of $m_0$ is marked with a dashed line in the
left panel). \label{fig8}}
\end{center}
\end{figure}

The dashed line of figure~\ref{fig7} depicts the image of the
unstable manifold by the linear shear $A(\omega,m_0)$ for
$m_0=0.005$. As $m_0$ increases $A(\omega,m_0)\; W^u(0)$ moves
further down so that intersection points 2' and 3', corresponding to
homoclinic orbits of the inhomogeneous problem, becomes closer and
closer. So there exists a critical value of $m_0$ for which these
intersection points collide and then disappear. 
In fact we have
checked numerically 
that this tangent bifurcation occurs at a critical value
$m_0\in(0.00963,0.00964)$ for problem (\ref{red11bis}). 
This critical value can be approximated 
using equations (\ref{eqmut})-(\ref{mcapp}), which yields
$m_0 \approx 0.009632$ in the present case. These results
correspond very precisely to 
a breather bifurcation numerically observed in the Klein-Gordon chain 
(at a critical value $m_0\in(0.00963,0.00964)$)
and depicted in figure~\ref{fig8}.

    The upper branch of figure~\ref{fig8}(a) represents the
energy of a breather solution corresponding to point 2'. For
$m_0 \approx 0$, the breather has a maximal amplitude at sites 
$n=0,1$. A profile of this breather for $m_0=0.0093$, close to the bifurcation
point, is shown in figure~\ref{fig8}(b), where the amplitude is now
much larger at $n=1$. 
The lower branch of figure~\ref{fig8}(a) represents
a one-site breather centered at $n=1$ and
corresponds to point 3'. Its profile for $m_0=0.0093$ is shown in
Figure~\ref{fig8}(c). According to analytical results of
reference~\cite{MST}, at low enough coupling $k$ the solutions on the lower branch are 
spectrally stable whereas the solutions on the upper branch are unstable.
We have checked this result numerically for the parameter values of figure~\ref{fig8}
by computing the Floquet spectra of the two families of breather solutions.

As in section \ref{hardp}, we have also computed one-site breathers centered 
at the mass defect, corresponding to point 1' in figure~\ref{fig7}. Again we
have found an excellent agreement between the numerically computed
breather profiles and the approximate solutions obtained using the map (\ref{red11bis}).
As expected from the analysis of section \ref{csp}, these breathers survive up to 
$m_0 = m_l (\omega)$, i.e. up to a much higher value of $m_0$ than the families 2', 3' described above.


\begin{figure}[h!]
\begin{center}
\includegraphics[width=6.0cm]{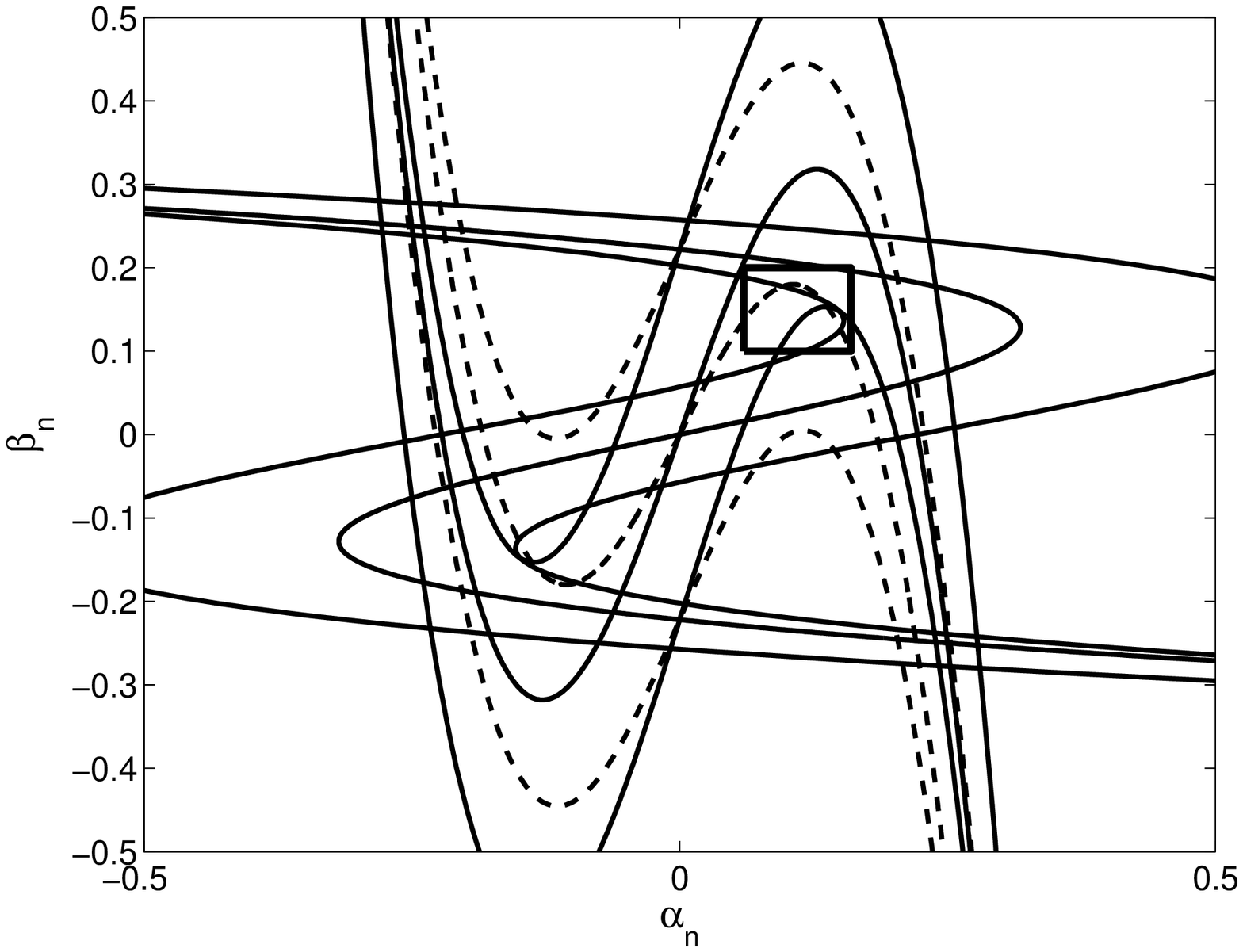}
\includegraphics[width=6.0cm]{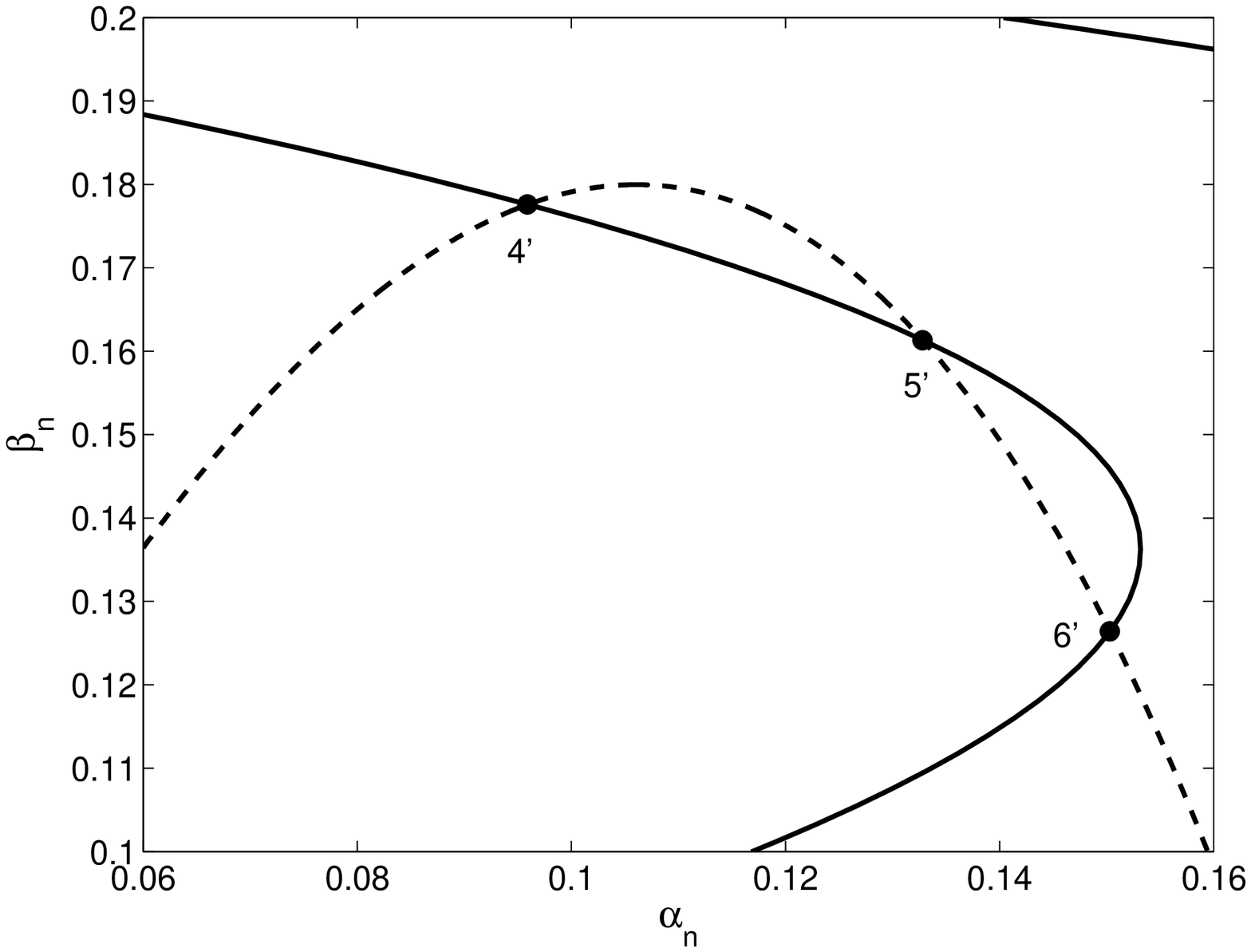}
\caption{Emergence of new intersection points between 
$A(\omega,m_0)W^u$ (dashed curve) and $W^s$
(drawn with a full line) as the mass defect is increased. 
The figure corresponds to $m_0=0.012$ and
$\omega = 9.9$. The right panel
shows a zoom of the left panel
over the region marked with a rectangle. 
The new homoclinic points 5' and 6'
correspond to new breather
solutions of the Klein-Gordon lattice. 
The point with label 4' corresponds to a two-site breather, which
exists in the homogeneous lattice and persists for $m_0 \leq 0.012$.
\label{fig9}}
\end{center}
\end{figure}

A part of the intersecting stable and unstable manifolds is shown in 
the left panel of figure~\ref{fig9}. Due to their complicated windings,
new intersections points appear between $A(\omega,m_0)\; W^u(0)$ and
$W^s(0)$ as $m_0$ is chosen in certain windows of the parameter space,
giving rise to new homoclinic solutions of (\ref{redequapp2}).

An example is shown in the region marked with a rectangle
(see the details in the right panel of figure~\ref{fig9}).
For some value of $m_0\in(0.01064,0.01065)$, a new intersection
point between $A(\omega,m_0)\; W^u(0)$ and $W^s(0)$ appears.
As $m_0$ is further increased, this inverse tangent
bifurcation gives rise to two new homoclinic points 5' and 6'.  
Correspondingly, we have numerically checked that an inverse tangent bifurcation occurs
in the Klein-Gordon chain at a critical value of
$m_0\in(0.01064,0.01065)$, giving rise to new breather solutions which
do not exist in the homogeneous chain.

The point 4' in figure~\ref{fig9} also exists for $m_0 =0$. Returning to
figure \ref{fig7}, it is obtained by applying the inverse map $G_\omega^{-1}$
to the point with label 2. In the homogeneous limit $m_0 =0$, this
homoclinic point corresponds consequently to a two-site breather centered between
$n=1$ and $n=2$. As figure~\ref{fig9} shows, an increase
of the mass defect $m_0$ moves point 5' against point 4' until
they collide and disappear through a new tangent bifurcation.
This tangent bifurcation is also numerically found
in the Klein Gordon chain at critical value of the
mass defect very close to the theoretical one
(in both cases one obtains $m_0 \approx 0.01268$).

\begin{figure}[h!]
\begin{center}
\includegraphics[width=6.0cm]{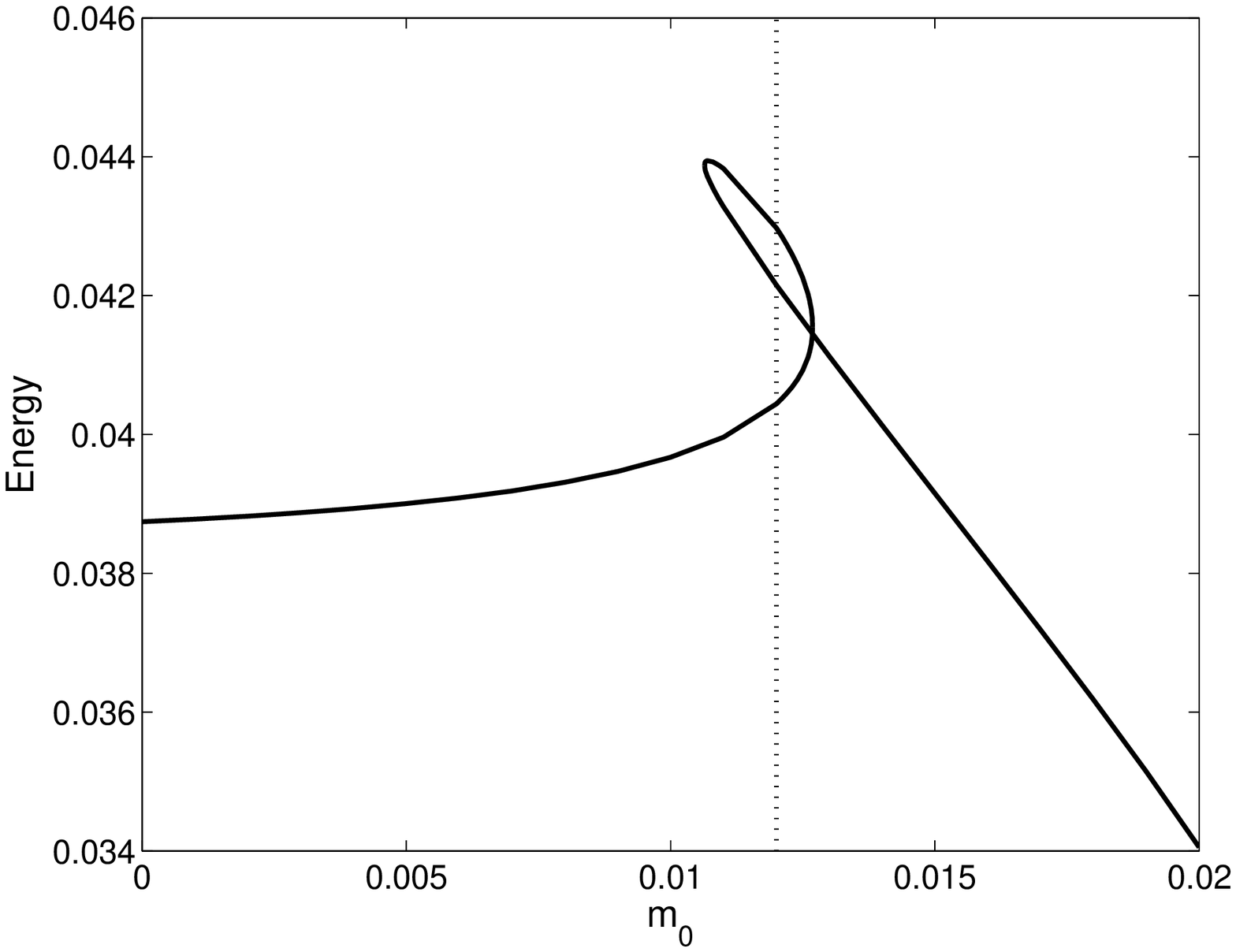}
\includegraphics[width=6.0cm]{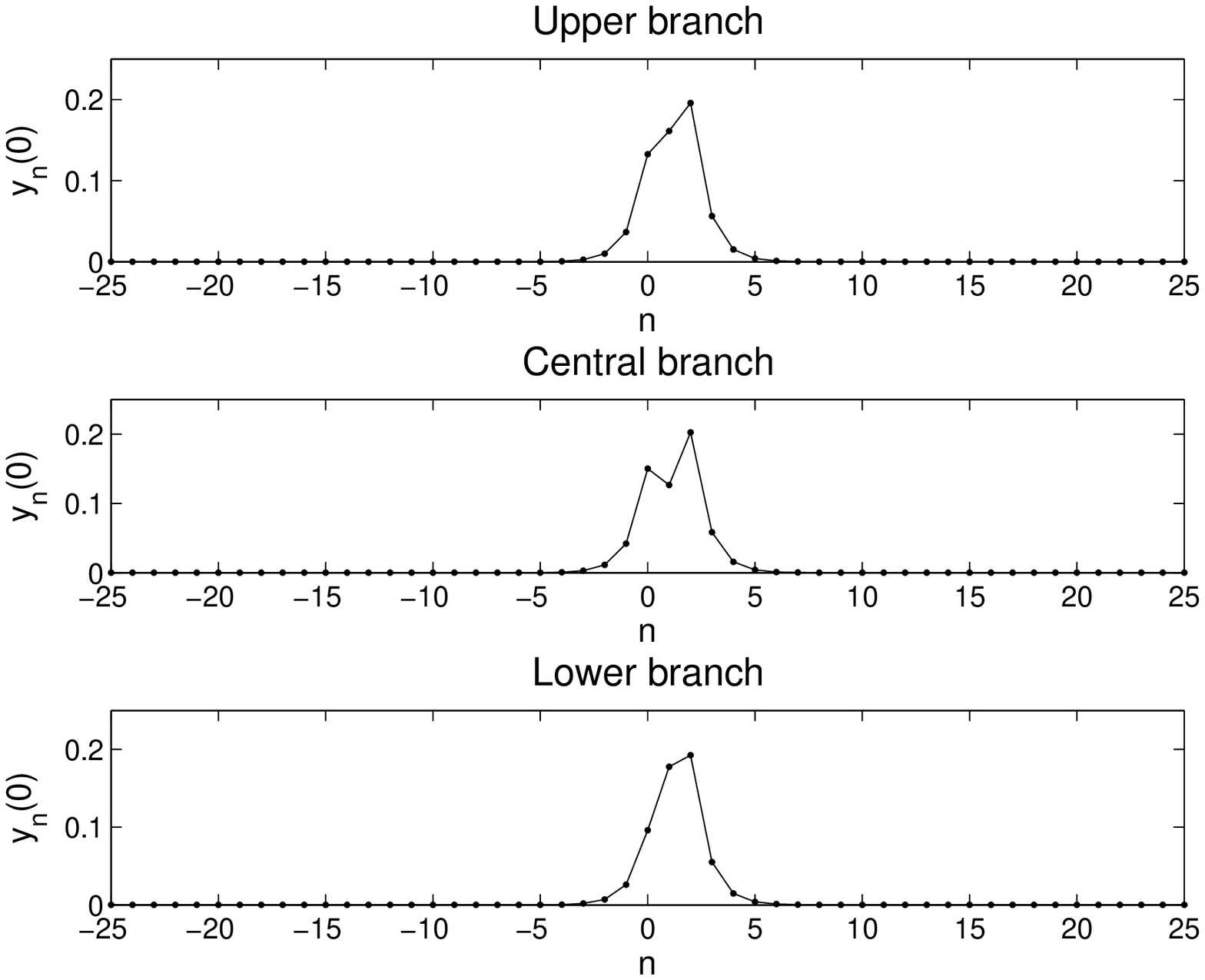}
\caption{Bifurcation diagram of breather solutions numerically
computed in the Klein-Gordon chain, with a soft potential
and a mass defect $m_0$ at $n=0$.
In the left panel the breathers
energies $E$ are depicted versus $m_0$
(see the definition of $E$ in the caption of figure \ref{fig8}).
The breathers frequency is
$\omega = 9.9$.
The lower branch at the left of the
vertical line corresponds
to a two-site breather centered between
$n=1$ and $n=2$.
The right panel shows the
profiles of the three breathers for $m_0=0.012$,
when all of them coexist (the value of $m_0$ is
marked with a vertical line in the left panel).
\label{fig10}}
\end{center}
\end{figure}

Figure~\ref{fig10} shows the bifurcation diagram of the
numerically computed breathers corresponding to homoclinic points
4', 5', 6' (left panel),
and gives their profiles for a given value of $m_0$
in the right panel. A numerical computation of Floquet spectra
shows that all these breathers are unstable.

As a conclusion, we have seen that the truncated normal
form (\ref{redequapp2}) allows one to predict with a high
precision certain breather bifurcations in the Klein-Gordon chain,
which occur as the mass defect $m_0$ is varied. 
These bifurcations depend on the fine structure of the windings of 
the stable and unstable manifolds of the origin, computed 
on the truncated normal form without defect.

\vsp{3}

\noindent
{\it Acknowledgements.}
This work has been supported by
the French Ministry of Research through
the CNRS Program ACI NIM (New Interfaces of Mathematics).
G.J. wishes to thank Michel Peyrard for
initiating this research, and is grateful to R. MacKay
for pointing out interesting bibliographical references. 
B.S-R. and J.C. acknowledge sponsorship by the 
Ministerio de Educaci\'on y Ciencia, Spain, project
FIS2004-01183.
B.S-R. is grateful to the Institut de Math\'ematiques
de Toulouse (UMR 5219) where a part of this work
has been carried out during a visit in Sept.-Oct. 2006. 

\vsp{3}

\end{document}